\documentclass[a4paper,final,onecolumn]{article}
\usepackage[a4paper,total={7.24in,10.51in}]{geometry}
\usepackage[T1]{fontenc}
\usepackage[utf8]{inputenc}
\usepackage[english]{babel}

\usepackage{csquotes}
\usepackage[bibstyle=nature,date=year,backend=bibtex,citestyle=numeric-comp,sortcites=true,sorting=none,mincrossrefs=1]{biblatex} 
\usepackage{lineno,hyperref}
\usepackage{amssymb}
\usepackage{mathtools}
\usepackage{stmaryrd}
\usepackage{bm}

\usepackage{miller}
\usepackage{graphicx}
\usepackage{subfig}

\usepackage[binary-units=true]{siunitx}
\DeclareSIUnit{\atpercent}{at.\%}

\usepackage[font=scriptsize]{caption}
\usepackage{cleveref} 

\usepackage{xcolor}

\usepackage{booktabs}
\usepackage{tabularx}

\usepackage{algorithm}
\usepackage[noend]{algpseudocode}
\def\BState{\State\hskip-\ALG@thistlm}
\algdef{SE}[DOWHILE]{Do}{doWhile}{\algorithmicdo}[1]{\algorithmicwhile\ #1}

\usepackage{pifont}

\usepackage{authblk}
\usepackage[draft]{changes}
\definechangesauthor[name={MarkusKuhbach}, color=red]{MK}
\colorlet{Changes@Color}{red}

\newcommand{\autoreporter}{paraprobe-autoreporter}
\newcommand{\parmsetup}{paraprobe-parmsetup}
\newcommand{\transcoder}{paraprobe-transcoder}

\newcommand{\ranger}{paraprobe-ranger}
\newcommand{\distancer}{paraprobe-distancer}
\newcommand{\surfacer}{paraprobe-surfacer}
\newcommand{\nanochem}{paraprobe-nanochem}
\newcommand{\intersector}{paraprobe-intersector}
\newcommand{\tessellator}{paraprobe-tessellator}
\newcommand{\spatstat}{paraprobe-spatstat}
\newcommand{\clusterer}{paraprobe-clusterer}
\newcommand{\crystalstructure}{paraprobe-crystalstructure}
\newcommand{\paraprobe}{paraprobe}

\newcommand{\nomad}{NOMAD}

\newcommand{\cxx}{C/C++}

\newcommand{\python}{Python}

\newcommand{\cgal}{CGAL}

\newcommand{\voroxx}{Voro$++$}
\newcommand{\hdf}{HDF5}

\newcommand{\paraview}{Paraview}
\newcommand{\jupyter}{jupyter notebook}
\newcommand{\jupyters}{jupyter notebooks}

\newcommand{\ivas}{IVAS}

\newcommand{\tetgen}{TetGen}
\newcommand{\inspico}{Inspico}
\newcommand{\neofourj}{neo4j}
\newcommand{\cypher}{cypher}
\newcommand{\pqp}{GammaUNC/PQP}
\newcommand{\apsuite}{AP Suite}
\newcommand{\blender}{Blender}

\newcommand{\nexus}{NeXus}

\newcommand{\ashapes}{$\alpha$-shapes}
\newcommand{\ashape}{$\alpha$-shape}

\newcommand{\swisswatch}[2]{$\SI[mode=math]{#1}{\hour}$:$\SI[mode=math]{#2}{\minute}$}

\newcommand{\gammaxx}{$\gamma^{\prime\prime}$}
\newcommand{\gammax}{$\gamma^{\prime}$}

\newcommand{\cameca}{CAMECA/AMETEK}

\newcommand{\talos}{TALOS}

\title{On Strong-Scaling and Open-Source Tools for High-Throughput Quantification of Material Point Cloud Data: Composition Gradients, Microstructural Object Reconstruction, and Spatial Correlations}
\author[1,2]{Markus K\"uhbach}
\author[3]{Vitor Vieira Rielli}
\author[3]{Sophie Primig}
\author[4]{Alaukik Saxena}
\author[4,5]{David Mayweg}
\author[5]{Benjamin Jenkins}
\author[4]{Stoichko Antonov}
\author[7]{Alexander Reichmann}
\author[7]{Stefan Kardos}
\author[7]{Lorenz Romaner}
\author[1]{Sandor Brockhauser}
\affil[1]{Consortium FAIRmat, Humboldt-Universit\"at zu Berlin, Zum Gro{\ss}en Windkanal 2, D-12489 Berlin, Germany.}
\affil[2]{Structure Research \& Electron Microscopy Group, Department of Physics, Humboldt-Universit\"at zu Berlin, Newtonstra{\ss}e 15, D-12489 Berlin, Germany.}
\affil[3]{School of Materials Science \& Engineering, UNSW Sydney, Kensington, 2052 NSW, Australia.}
\affil[4]{Max-Planck-Institut f\"ur Eisenforschung GmbH (MPIE), Max-Planck-Stra{\ss}e 1, D-40237 D\"usseldorf, Germany.}
\affil[5]{Chalmers University of Technology, Department of Physics, Division of Microstructure Physics, Fysikgr\"and 3, SE-412 96 G\"oteborg, Sweden.}
\affil[6]{Department of Materials, University of Oxford, 16 Parks Road, Oxford, OX1 3PH, United Kingdom.}
\affil[7]{Department of Materials Science, Montanuniversit\"at Leoben, Franz Josef-Stra{\ss}e 18, A-8700 Leoben, Austria.}

\date{}
\begin{document}
\maketitle

\begin{abstract}
Characterizing microstructure-material-property relations calls for software tools which extract point-cloud- and continuum-scale-based representations of microstructural objects. Application examples include atom probe, electron, and computational microscopy experiments. Mapping between atomic- and continuum-scale representations of microstructural objects results often in representations which are sensitive to parameterization; however assessing this sensitivity is a tedious task in practice.

Here, we show how combining methods from computational geometry, collision analyses, and graph analytics yield software tools for automated analyses of point cloud data for reconstruction of three-dimensional objects, characterization of composition profiles, and extraction of multi-parameter correlations via evaluating graph-based relations between sets of meshed objects. Implemented for point clouds with mark data,
we discuss use cases in atom probe microscopy that focus on interfaces, precipitates, and coprecipitation phenomena observed in different alloys. The methods are expandable for spatio-temporal analyses of grain fragmentation, crystal growth, or precipitation.
\end{abstract}

\newpage
\section{Introduction}
Microscopy techniques, such as electron microscopy (EM) \cite{Williams1996,Kirkland2020}, atom probe microscopy (APM) \cite{Miller2000,Larson2013,Lefebvre2016,Gault2021Nat}, and even computational microscopy \cite{Stukowski2010,Zepedaruiz2020,Casalino2021}, are essential tools for characterizing the atomic architecture of materials. Collecting quantitative evidence in support of or against a set of research hypotheses is the purpose of microscopy research. Microscopy data are processed into descriptors that can be used in physically- or artificial-intelligence-based surrogate models to decode material properties and develop understanding as to how microstructures change with processing and also during service.

In most cases, descriptors are of two kinds: Either they are quantities at the continuum-scale, like (spatial) statistics of crystal defect or atom ensembles, such as dislocation density, or they are spatially-detailed three-dimensional descriptors for the static arrangement or oftentimes even dynamics of crystal defect ensembles. Examples include descriptions of point \cite{Huber2018} and line defects \cite{Stukowski2010a,Ghamarian2020,Zepedaruiz2020} or of grains and crystals of different thermodynamic phases and descriptions for the junctions in the network of crystal defects of a microstructure \cite{Lazar2015,Lazar2017,Spencer2017}. 

Such a scale-bridging encoding of point clouds into microstructural objects requires assumptions and models which have often adjustable parameters which control the shape, extent, and topology of the objects' representation. The significance of this parameter sensitivity should not only be explored but ideally quantified in detail within a research study. Enabling and supporting researchers with this quantification is one key role of software tools. In the condensed-matter physics as well as the computational materials science community, a substantial number of simulation tools are not only open-source \cite{Momma2011,Jain2013,Draxl2019} but are especially build to reduce barriers with management and comparison of results according to the FAIR data stewardship principles \cite{Wilkinson2016,Draxl2020,Kuehbach2021MM}. These activities enable and drive the development of programmatically usable scientific software around existing (software) tools and instruments to explore and assess the parameter sensitivity of descriptors, their uncertainty, and the vast set of materials they describe.

For the materials-science-branch of especially the EM community sophisticated open-source tools have been developed (e.g. \cite{Hyperspy2021,LiberTem2021,AbTem2021}) but sophisticated platforms for sharing experimental data remain to be developed. For the atom probe microscopy community, the situation with respect to data sharing platforms is similar \cite{Kuehbach2021MM} but fewer options of general enough open-source software exist which could enable domain scientists to take advantage of specialist tools' and algorithms of other scientific communities such as computational geometry. Many microscopes are operated with proprietary software which combines metadata management, data acquisition, and analysis services in a single application through a graphical user interface (GUI). Provisioning instrument calibrations, abstracting hardware layers, and offering intuitive GUIs are advantages of these tools for scientists with routine analysis needs.

However, such software can decouple scientists from the capabilities of open-source scientific software, unless vendors interface their software via advanced programming interfaces (APIs) or scripting solutions. Here we report a set of tools which can complement the vendors' efforts to equip scientists with a diverse set of tools, which should all ideally be made interoperable. Our proposal is focused on automating analyses, motivating a more detailed understanding of the functioning of numerical algorithms and the significance of parameter sensitivity. With this we serve the improvement of research processes by reducing needs for eventually unnecessarily ineffective or restrictive tools which currently are barriers to FAIR-compliant research \cite{Scheffler2022}.

Atom probe scientists are one community who face this situation \cite{Kuehbach2021MM}. The two main techniques they use, atom probe tomography (APT) and field-ion microscopy (FIM), both indirectly measure positions of atoms. They both use controlled field evaporation to characterize a needle-shaped nanoscale specimen. This is done by holding the specimen in an electric field before superimposing a laser or high-voltage electric pulse to achieve controlled evaporation. For APT, the atoms are measured as ions which are evaporated via the controlled pulsing and successively measured via position-sensitive time-of-flight-resolved mass spectrometry. For FIM, imaging gas atoms are used which field-ionize specifically above individual surface atoms and are then accelerated along the electric field lines towards a position-sensitive detector \cite{Mueller1956a}. Exploitation of the fact that the field strength can be changed to enforce also the evaporation of the surface atoms has blurred the boundary between APT and FIM experiments \cite{Katnagallu2018,Morgado2021}. Data from such experiments can be processed into three-dimensional tomographic reconstructions which are point cloud models of the evaporated specimen volume with ion and atom type information. Despite differences in the position and ion-type resolution between APT and FIM, these reconstructions offer a unique combination of isotopic and sub-nanometer spatially-resolved information about the atomic architecture of materials.

Like every experiment and computational model, the technique and the reconstruction faces limitations \cite{Oberdorfer2015,Gault2021}: Limited detector efficiencies cause that not every atom is detectable. Limited mass-to-charge-resolving power causes that not every ion can be decomposed into its atoms. The reconstruction process can result in regions of the dataset which have a lower positional accuracy and precision as compared to electron microscopy (EM) experiments or especially to (atom trajectory) data available with molecular dynamics simulations \cite{Brehm2020,Casalino2021}. These difficulties have motivated efforts towards using APM and EM correlatively \cite{Xu2015,Kelly2021} surplus taking advantage of computational methods \cite{Ceguerra2013}.

In effect, atom probe microscopy has developed into an experimental technique whose capabilities of measuring a statistically significant number of ions delivered a tool for not only accurate and precise composition analysis \cite{Gault2021Nat} but also for studies of the spatial arrangement of atoms and how they are organized as an ensemble of crystal defects at different length scales \cite{Kontis2018,Gault2021Nat}.
Characterizing these so-called microstructural objects geometrically and developing efficient methods for assessing the parameter sensitivity of these geometrical descriptions is the focus of this work.

Our work is a continuation of open-source software development efforts to support the scientific community \cite{APTTC2020}. Seminal previous work in this regard with relevance for the characterization of microstructural objects were the introduction of iso-surface-based methods \cite{Miller1986,Hellman2000a} for revealing grain and phase boundaries (interfaces), signed-distance-based composition quantification, via so-called proxigrams \cite{Hellman2000,Hellman2003}, computational-geometry-based methods \cite{Karnesky2007a,Felfer2012b,Felfer2013,Felfer2015b} and connecting these with composition analyses and interfacial excess mapping \cite{Krakauer1993,Felfer2015a,Peng2019}, taking advantage of clustering methods for segmenting precipitates or dislocations \cite{Hyde2000,Stephenson2007,Marquis2017,Ghamarian2020}, and recently the introduction of artificial intelligence methods \cite{Sarker2020,Zhou2021a}. Most of the associated software tools are proof-of-concept implementations. Their source code is shared openly. The implementation and maintenance is in most cases decoupled from the commercial tools.

The more frequently the proprietary and open-source tools were used the more apparent became the sensitivity of the descriptors they delivered as a function of the parameter settings \cite{Torres2011,Martin2015}. This revealed unsolved challenges: Some are conceptual, like the assumption that using a single threshold value suffices to segment a dataset into useful iso-surfaces. As this work supports this is an often inadequate \cite{Hornbuckle2015,Barton2019,Jenkins2020b}. Other challenges, like data accessibility restrictions in commercial software, pose practical barriers with respect to how completely computational geometry, concentration field, and detailed metadata are exportable \cite{London2017,Kuehbach2021MM}; and thus how APM research studies are (programmatically) comparable to one another and can become FAIR-compliant. Suggestions were made by individual researchers how data from the GUI of commercial software can be extracted into spreadsheets \cite{Theska2019,Rielli2020}. As far as tasks like composition, interface-based analyses, and microstructural-object-centric analyses are considered, we found only few cases where tools other than the commercial ones were used \cite{Haley2018b,Keutgen2020,Heller2021} (apart from the case of interfacial excess quantification \cite{Felfer2015a,Zhou2022}.)

To improve the situation we want to substantiate the assumption that numerically more efficient and better, in the sense of more automated and more functionally capable, tools can be developed by embracing open-source tools from different scientific communities and blending them into a set of complementary open-source tools that can be coupled into complex workflows and customized without barriers.

In previous studies \cite{Kuehbach2021NPJ,Kuehbach2021JAC}, we substantiated the validity of this assumption for methods including spatial statistics \cite{Kuehbach2021NPJ} and crystallography \cite{Kuehbach2021JAC}. In continuation of this research, we exemplify that it is also possible to develop efficient parallelized tools which are programmatically automatable for analyzing point cloud data for composition and object-based geometrical analyses. Specifically, in this work we blend tools from computational geometry, robotics, game engine, and graph analytics communities. We exemplify their benefit when processing point cloud data - here using different datasets from atom probe microscopy. Specifically, solutions for the following data analysis tasks are developed:

\begin{enumerate}
    \item Different methods are implemented for representing the edge of datasets and compare these representations geometrically and quantitatively. The methods are useful for reducing bias through detection of object-edge-intersections. These results are useful for quantifying parameter sensitivity and detection of mesh properties such as closure.
    \item Methods for distance-based segmentation of datasets are implemented which accept generic triangulated surface meshes and/or input from clustering analyses, obtained from other software tools, to segment datasets volumetrically into regions and yield corresponding volume composite or surface meshes of three-dimensional regions.
    \item We implement an open-source tool for delocalization, i.e. smoothing and discretizing point cloud data and subsequently executed iso-surface-based analyses on the such discretized datasets. We equip this tool with methods which enable automated computational-geometry-based characterization of surface facets and three-dimensional objects with respect to object closure, shape, size, and composition (ionic and/or elemental).
    \item We revisit previous work of \cite{Felfer2015a,Peng2019} on methods for creating automated triangulated surface patches of interfaces and show how a more robust numerical protocol can be used to make such analyses even more automated and FAIR-compliantly documented.
    \item We implement algorithms which automate the placing and aligning of an arbitrary number of regions-of-interest (ROIs) at facets of triangulated surface meshes. Specifically, we study three important cases: Namely, interfaces about closed objects (precipitates), surface patches between regions with preferential composition gradients to exploit, and surface patches which reveal themselves in APM measurements only via their decoration with solutes. We implement procedures which fully automate the characterization of composition profiles and interfacial excess mapping for these ROIs. These keep automatically track of all metadata and probe the composition profiles consistently where this is numerically possible.
    \item We implement a tool which enables to logically relate an arbitrary number of such object-based results via graph analytics to formulate robust methods for characterizing coprecipitation. We use these tools for studying the sensitivity of object representations as a function of parameterization.
    \item We explore multithreading options for these tools. All of them are applied to analyze experiments using a laptop. Finally, we document the scalability of our solution when processing among the nowadays largest known reconstructions where a total of one billion ions is computed on a single computing node of a computer cluster.
\end{enumerate}

These solutions are implemented as modified or additional tools respectively into the paraprobe-toolbox. This is a modularized software toolbox with efficient tools \cite{Kuehbach2021NPJ,Kuehbach2021JAC} for programmatic analyses of point cloud data and inscribed sets of other geometric primitives.

\section{Methods}
\subsection{The paraprobe-toolbox is modularized}
\label{Methods}
\begin{figure}[h!]
\centering
    \includegraphics[draft=false,width=1.0\textwidth]{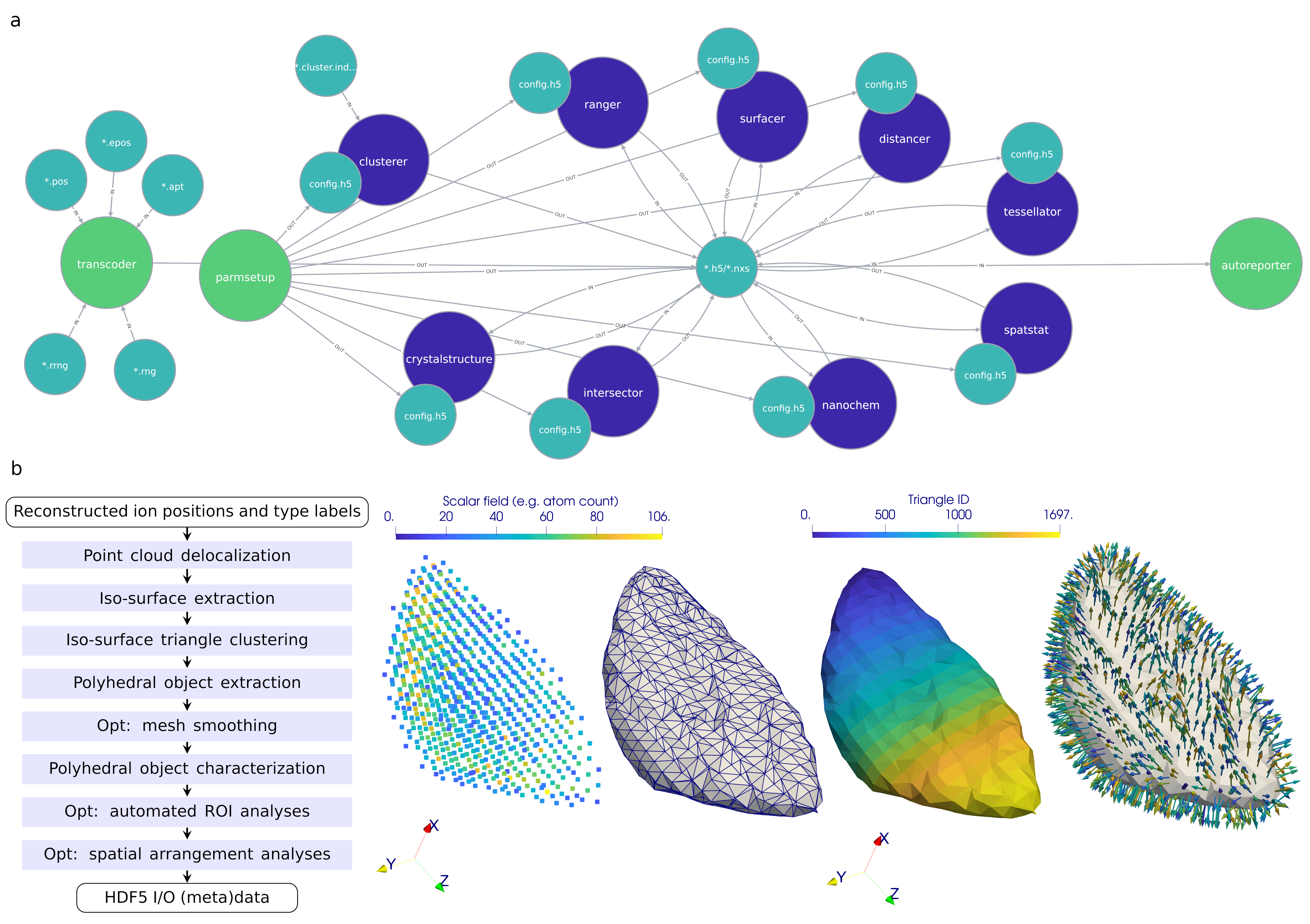}
\caption{Sub-figure a) summarizes the collection of tools (bluish nodes) which the paraprobe-toolbox includes to solve specific sets of analyses for three-dimensional point cloud data with ion or atom type marks and geometrical primitive sets. The tool interfaces with commercial APM software and tools of the community via \transcoder{}. This tool reads community file formats (turquoise nodes) as input. Each run of a tool has a dedicated configuration file (the config.h5 nodes) which complies with a \nexus{} application definition. The results of the tools, numerical data and metadata, are stored as a set of specifically-formatted \hdf{} binary files. Sub-figure b) details the workflow which \nanochem{} implements, representing an open-source code route of performing classical iso-surface-based analyses. The reconstructed volume is discretized into scalar fields of element-specific counts, composition, or concentration, respectively via ion- or atom-type-specific kernel density estimation. These fields can be exported. Subsequently, triangulated iso-surfaces can be approximated and post-processed to identify which triangles are connectable into triangle clusters. Applying subsequently polygon mesh processing on these clusters enables to detect which triangle clusters represent surface meshes of closed polyhedra and which represent isolated surface patches. Finally, ROIs can be placed and processed automatically for local composition analyses. All surface mesh and patch data can be exported. The spatial arrangement of meshes can be analyzed with \intersector{}. Supplementary files are written for basic visualization via the eXtensible Data Model and Format (XDMF).}
\label{FigMethods}
\end{figure}

\paragraph{Context}
This work continues activities on building strong-scaling, open-source, and FAIR-compliant software tools for processing point cloud data with associated mark data \cite{Kuehbach2020MSMSE,Kuehbach2021NPJ,Kuehbach2021JAC,Kuehbach2021MM} (the so-called paraprobe-toolbox). In this work we improved existent tools and implemented two new ones (\nanochem{} and \intersector{}). Figure \ref{FigMethods}a summarizes the tools. An analysis is scripted in \python{} as a workflow using a \jupyter{} \cite{Kluyver2016}. This removes the need for routine users to deal with the C/C++ backend and can serve as a starting point for formulating workflows via workflow management systems like pyiron \cite{Janssen2019}. The individual paraprobe-tools, displayed as blue nodes in Fig. \ref{FigMethods}a, are instructed through an \nexus{}/\hdf{} configuration file, represented by the  config.h5 nodes in turquoise of Fig. \ref{FigMethods}a. Internally the tools batch process a list of analysis tasks. Each tool and run returns an \hdf{} file which gives unrestricted access to all numerical data and metadata. Python convenience functions are available on the input (\parmsetup{}) and the post-processing side (\autoreporter{}) to support users with creating configuration files and with extracting commonly used quantities from the eventually deep data tree in the \hdf{} file. Most of the tools implement strong-scaling multithreading via Open Multi-Processing (OpenMP). Strategies for process parallelism have been explored \cite{Kuehbach2021NPJ}. 


\paragraph{Tool modifications} First, we refactored the tools of \cite{Kuehbach2021NPJ} to improve their modularity and replaced similar code portions with more generally applicable functions. Unnecessary restrictions in \cite{Kuehbach2021NPJ} were also removed which makes possible now studies with arbitrarily complex molecular ions \cite{Kuehbach2021MM}, realized via an adjustable maximum number of atoms per molecular ion (currently 32) to handle complex ions \cite{Elzoka2020,Felfer2021IonList} and a processing of selected elements or isotopes in molecular ions (atomic decomposition). Analyses can now be restricted via applying spatial and/or ion attribute filters. Ions can be filtered based on type, hit multiplicity, or evaporation ID. Spatial filters can be combined with sub-sampling each $n$-th ion of the dataset. Spatial filters are implemented via set operations for geometric primitive ROI(s) $\mathcal{M} \in {\mathcal{R}}^3$ such as oriented bounding boxes (OBBs), rotated cylinders, or spheres. These can be combined. Polyhedra could be implemented in the future. Robust point inclusion tests \cite{Eberly2020,CGAL2021} for each respective primitive were implemented.

\paragraph{Importing reconstructions and ranging definitions}
Due to the availability of many community codes (e.g. \cite{Felfer2022b}) for performing reconstruction and creating ranging definitions and the proprietary nature of reconstruction protocols in commercial software, \paraprobe{} does not implement an own tool for building reconstructions so far. Instead, each analysis after having made a new measurement starts with using the \transcoder{} tool to load a dataset as a pair of tomographic reconstruction and ranging definitions. Data from third-party software like \apsuite{}, \inspico{}, or other APM community software \cite{APTTC2020}, see Fig. \ref{FigMethods}a, are supported. Community interaction with vendors would enable to make this interface even more sophisticated in the future. Paraprobe-transcoder then creates a \hdf{} file with ion positions, mass-to-charge-state-ratio values, and a set of ranging definitions. Once transcoded, existent ranging definitions are applied using \ranger{} which computes an ion type array. Multiple mappings can be stored in the same \hdf{} file to reflect that analyses in atom probe are often sensitive to ranging definitions and reconstruction parameters which calls for a pedantic versioning  \cite{Kuehbach2021MM}.

\paragraph{Geometrical models for the edge of a dataset and distance-based filtering}
In previous work \cite{Kuehbach2021NPJ,Kuehbach2021MM} we discussed the importance of characterizing the edge of the reconstructed point cloud because atom probe specimens can contain incompletely measured (truncated) microstructural objects such as precipitates or patches of grain or phase boundaries. Edge models can be created and characterized with \surfacer{}, \distancer{}, or \tessellator{}. 

We modified the \surfacer{} tool to support convex hull computations and added diagnostics for \ashapes{} \cite{Edelsbrunner1994,Kai2021} such as returning the geometry of interior tetrahedra for given $\alpha$-values, mesh closure tests, and the option to characterize the set of possible \ashapes{} as a function of $\alpha$-value, also to offer a bridge to related work on \ashapes{} for cluster analysis \cite{Still2021}. Paraprobe-surfacer can store the geometry of all created objects as triangle and tetrahedra sets. 

The \distancer{} tool implements functionalities for computing the shortest Euclidean distance of points to an arbitrary inputted set of (non-degenerate) triangles. Practical applications are segmentations of datasets into ions within a certain distance range to microstructural objects or said mesh of an edge model. 



\paragraph{Tessellations and spatial statistics with distance-based filtering}
Furthermore, we modified the \tessellator{} tool compared to \cite{Kuehbach2021NPJ} to support loading of optional input from \distancer{} computations with the aim to identify interfaces between adjacent Voronoi cells with specific cell attributes. First, \tessellator{} computes a Voronoi tessellation of the entire dataset. Second, the tool optionally reports facets of Voronoi cells only for these cells whose associated distance attribute value is above $d_{ero}$ to those cells whose neighboring cells' distance is below $d_{ero}$. This enables to compose surfaces similarly like it was reported for visualizing atomically-resolved slip planes within molecular dynamics simulations \cite{Stukowski2010,Zepedaruiz2020}.

Also the \spatstat{} tool \cite{Kuehbach2021NPJ} for computing spatial statistics was modified. Specifically, the tool was equipped with a functionality which enables users to load two optional distance values per ion which can be used for filtering ions and thus restrict the computation to customizable arbitrarily shaped regions in the dataset in addition to the above-mentioned spatial filters. A possible application are spatial statistics for all ions within a (signed) maximum distance to a set of microstructural objects combined with the filtering of ions for their distance to the edge of the dataset to reduce bias. 


\paragraph{Point-in-primitive inclusion tests and primitive intersection analyses}
The relative spatial arrangement of individual points to geometric primitives (sphere, rotated cylinder, oriented bounding box, or polyhedron) is evaluated with algorithms of the computational geometry community: Inclusion tests are used to identify if a point $p \in {\mathcal{R}}^d$ with $d = {2, 3}$ is lying inside (including the edge of the primitive) or outside given primitives. Self-intersections of polyhedra \cite{Loriot2021} are tested for and reported. Paraprobe can handle polyhedra with convex and non-convex surface patches. The shortest Euclidean distance between a point and a triangle is computed via a function from the Geometrytools \cite{Eberly2020}. The shortest Euclidean distance between a point $p_A$ on a triangle $A$ to its closest point $p_B$ on a triangle $B$ is computed via a function of the \pqp{} library \cite{Gottschalk1996,Larsen2000,Larsen2020}.

\paragraph{Import cluster analysis results from third-party tools}
We saw the need for a utility tool, \clusterer{}, with which results from cluster analyses of commercial or other APM tools can be loaded into \paraprobe{}. The tool takes definitions of clusters from currently \ivas{}/\apsuite{} as input and disentangles the commercial encoding that all ions of a cluster are stored with the same artificial mass-to-charge-state-ratio value. Paraprobe reconstructs from these values the unique cluster IDs surplus recovers the evaporation ID of each ion by comparing ion positions. 

\paragraph{Interface modeling}
A set of algorithms was implemented in \nanochem{} which enables a computation of triangulated surface meshes to a set of points in ${\mathcal{R}}^3$. These points are the reconstructed positions of the solute ions. Ions are atomically decomposed, i.e. points duplicated with respect to their multiplicity within the ion. Most of the steps use functionalities of the \cgal{} library, specifically algorithms offered by the polygon mesh processing package \cite{cgal:lty-pmp-22a, Botsch2010}. First, a set of points is filtered from the dataset. These are the locations of the selected segregating species (eventually further spatially filtered) which guide the locating of the interface. This subset of the entire point cloud is interpolated via principal component analysis (PCA) \cite{cgal:ap-pcad-22a}. The resulting plane is taken to slice the axis-aligned bounding box to the dataset via a polyline whose interior area is subsequently triangulated to yield an initial interface model.

Next, an iterative algorithm is used in which course the mesh is isotropically refined, smoothen, consistent facet and vertex normals computed, and evolved via DCOM \cite{Felfer2015a}. During DCOM the barycenter of all solute ions within a control sphere about each vertex is computed. The triangle vertices are then translated in the direction of their vertex normal towards the respective barycenter positions. Vertices without ions in the control sphere are not touched. After each DCOM iteration the mesh is inspected for self intersections. The evolution of the mesh is documented with writing all meshes to the \hdf{} results file. Automated placing and analyzing of ROIs follows the procedures described in the automated composition profiling paragraph below.



\paragraph{\nexus{}/\hdf{} data models and enabling cloud computing}
Another key modification to previous work is the I/O handling. Specifically, the configuration files for each tool follow an \nexus{} application definition which defines what each parameter conceptually represents, which datatype and unit it has. Each configuration file is stored as a \nexus{}-compliant \hdf{} file \cite{Koennecke2015}. These application definitions build on recent work in APM research data management \cite{FAIRmatNeXusProposal2022} to work towards making interoperable input and output data between different software tools via a community-developed set of open application definitions, glossary terms, and eventually ontology \cite{Collins2021, Goerzig2022}. Thanks to all modifications to the toolbox, it has now become possible to package the toolbox in a Docker container which allows its efficient integration into FAIR data management platforms, like NOMAD \cite{Draxl2019}. Here, we can take advantage of \paraprobe{} already because of the open data model and open file format. We encourage the atom probe community to contribute to these efforts by exchanging ideas at meetings and conferences and making suggestions through the online documentation \cite{FAIRmatNeXusProposal2022}. If fostered by the community that could lead to the formulation of a common data exchange model to make atom probe microscopy analyses more interoperable and reproducible. 

\subsection{High-throughput composition and object analyses}
\paragraph{Open-source deconvolution, iso-surfaces, and microstructural object reconstruction}
Paraprobe-nanochem implements multiple computational geometry algorithms which, when used in combination, enable users to programmatically instruct delocalization tasks, iso-surface-based analyses, and subsequent computational geometry processing to reconstruct microstructural objects and automated ROI-based composition, concentration, proxigram, and interfacial excess analyses. This work focuses on three-dimensional objects, such as triangulated surface meshes of precipitates. Unique is that the user can access all intermediate results and objects, including their geometry. Figure \ref{FigMethods}b summarizes the individual algorithmic steps of the \nanochem{} tool for quantifying microstructural objects. Key details of these algorithms are summarized in the following paragraphs.
 


\paragraph{Delocalization}
This is a strategy for smearing ion positions into the continuum to enable subsequent approximating of topologically simpler and smoother features via iso-surfaces \cite{Larson2013}. Delocalization can be achieved with defining first a discretization volume (3D voxel grid) and superimposing it on the point cloud. Second, the voxels are scanned with a delocalization kernel which evaluates how strongly each ion contributes signal intensity to each voxel. 

So far, \paraprobe{} implemented \cite{Kuehbach2021NPJ} a naive approach whereby an ion was binned into the voxel that covered its position using rectangular binning. More sophisticated methods were proposed \cite{Hellman2003,Ulfig2009b,Larson2013} but these have so far been accessible almost exclusively via commercial software \cite{Keutgen2020} where exporting the scalar fields is, to the best of our knowledge, not practically possible. Given that delocalization settings are often not reported in the literature in sufficient detail or many users rely on default settings \cite{Kuehbach2021MM}, it is often difficult to judge the sensitivity of iso-surface-based results that were reported in the literature. However, as Larson et al. pointed out \cite{Larson2013}, applying delocalization always calls for making a compromise between how strongly one smears positions in an effort to achieve blunter surfaces and smoother composition gradients but to smear not too strongly to render very different or even questionable numerical results. Several authors reported in fact a strong parameter sensitivity of iso-surfaces \cite{Martin2015,Hornbuckle2015,Barton2019}, and suggested improvements but these remained, with few exceptions \cite{Haley2018b}, at the level of accepting the results from commercial tools oftentimes paired with very tedious manual inspection. To close this gap was our incentive to implement an open-source alternative. Paraprobe-nanochem implements a multi-threaded kernel density estimation which uses  an anisotropic 3D Gaussian delocalization kernel with variances $\sigma_x = \sigma_y := \sigma \in \mathcal{R}$, $\sigma > 0$, and $\sigma_z = 0.5\sigma_x$. Ion types are decomposed into isotopes of elements and accounted for via element-specific scalar fields. Users can instruct multiple delocalization analyses with different settings to study the sensitivity of iso-surfaces to delocalization (see case study \ref{CaseStudy5} specifically). Further technical details are reported in the supplementary material.


\paragraph{Iso-surface extraction}
After each delocalization run, \nanochem{} activates another internal batch queue for approximating triangulated iso-contour surfaces (ion-count-, composition-, or concentration-based). This enables studies with a customizable set of iso-surface threshold values $\{\varphi\}$. Already computed delocalization results are reused rather than recomputed. Multiple methods have been reported in the computational geometry community for approximating a continuum description for surfaces. We decided to use an open-source implementation of a topologically more robust marching cubes (MC) algorithm \cite{Lewiner2003}. The reader is referred to the literature for a detailed overview of the functioning, the history, and differences between MC implementations and alternatives \cite{Lorensen1987,Newman2006,Lorensen2020,Engwirda2020,Kazhdan2006}.

The result is a triangle soup, i.e. triangles without connectivity information, representing a complex (set) of iso-surfaces. Although MC has frequently been applied in the atom probe literature, mostly via its implementation in commercial software, few atom probers have discussed that the implementation of the topological rule set can differ between MC implementations \cite{Zhou2022}. These differences can result in eventually significant effects on the local topology and closure of the iso-surface, in particular when there is a strong sensitivity on the threshold value $\varphi$. 

\paragraph{Microstructural object (feature) reconstruction}
Iso-surfaces serve the quantification of a key methodology in material science which is to coarse-grain specific atomic arrangements into objects, or features, for which descriptors, like curvature tensor, line or surface energy, can be attributed at the continuum scale. Continuing with the triangle soup, \nanochem{} performs first a proximity clustering of the triangles, using a modified DBScan algorithm \cite{Ester1996} to cluster nearby triangles. Once clustered, polygon mesh processing is used, including combinatorial steps, to identify which of the triangle cluster represent individually closed surface meshes of polyhedra and which cluster represent open or free-standing triangle patches. Shortest Euclidean triangle-to-triangle distances were computed via the \pqp{} library \cite{Larsen2020}. Polygon mesh processing uses functionalities of \cgal{} \cite{CGAL2021,Loriot2021}. Triangles are clustered sequentially, polygon mesh processing uses multithreading.

\paragraph{Object characterization}
After clustering the triangles yet another internal queue can be processed which analyzes further each identified closed polyhedron: These analyses can be configured according to user needs offering the volume of the polyhedron, an inspection if triangles of each surface mesh intersect with the mesh of the edge of the dataset, a computation of outer unit normal vectors for each triangle facet, and a shape analysis of the object via computing an (approximate) oriented bounding box (OBB) \cite{Katrioplas2021} or rotated ellipsoid \cite{Fischer2021} for the object respectively. Furthermore, the tool inspects which ions, i.e. which evaporation IDs, are located inside or on the surface mesh of each closed polyhedron \cite{CGAL2021}. An exact algorithm for computing the optimal OBB has been reported \cite{ORourke1985,Melchior2018}. However, its cubic numerical complexity in the number of points renders it impractical. Therefore, we opted to interface the tool with a faster approximating algorithm \cite{Barequet2001,Chang2011,Katrioplas2021} implemented in \cgal{} \cite{CGAL2021}. Point-in-polyhedron intersection tests are evaluated also via \cgal{}. All object characterization uses multithreading.

\paragraph{Automated composition profiling}
As another optional step, \nanochem{} implements algorithms for placing customizable ROIs at each triangle facet of an object or of a free-standing surface mesh and evaluating projected signed distances of the ion relative to the signed interface facet normal (1D composition profiles, Fig. \ref{FigMendezMartinCarbide}e, \ref{FigMaywegOverview}b). Using the outer unit normal and barycenter of each triangle facet enables to place and align each ROI. The cylinder-triangle intersection algorithm of \cite{Eberly2010} enables to identify if ROIs lie completely inside the dataset or not to detect bias or profile truncation. The result is a set of ROI mesh geometrical metadata and at least a collection of ion and/or element-specific sorted lists of projected distances, eventually preprocessed already into cumulated composition profiles. If desired, such analyses can be performed for every run to study sensitivity on delocalization, iso-surface, and object parameterization. This automates a set of tasks which would otherwise very likely not be performed manually because of the extremely time consuming number of thousands if not millions of ROI analyses necessary through GUI interaction. Paraprobe documents all ROI metadata (location, orientation, dimensions) to offer numerical exact reproduction, which is another clear advantage over hitherto reported manual procedures. ROIs are processed via multithreading.

\subsection{Spatial correlation analyses}
\paragraph{Motivation}
These functionalities enable scientists to collect detailed ensembles of sets of mesh data which is useful for parameter sensitivity studies, uncertainty quantification, and planning more targeted or detailed manual analyses commercial or community software. Despite its value for scripting analyses, we found that the amount of generated data can be overwhelming because the results report, depending on how the analyses were configured, the sensitivity on multiple sets of parameters from the geometrical, the surface approximation, and delocalization, and eventually even ranging and reconstruction. Being able to investigate this wealth of information is the strength of high-throughput tools.

Therefore, we felt we should develop an automated post-processing tool to support scientists with quantifying the sensitivity of an object's representation as a function of parameters used - \intersector{} is the result of this - the second key novelty of our work. The tool enables generic spatial correlation and location analyses of ensembles of sets of three-dimensional triangulated surface meshes. Tools for such a characterization have been developed in the biological structure characterization and fluorescence microscopy communities \cite{Manders1992,Costes2004,Bolte2006,Zinchuk2008,Gilles2017} but these work with discretized objects, i.e. voxel grids. Known as so-called colocalization analysis tools in those communities, these tools have not found a particular recognition or application in the materials science community though yet. Our implementation works for objects in continuum space. Here, assuring numerical robustness is a more difficult challenge than for discretized objects. 

\paragraph{Methods}
With the \intersector{} tool of \cite{Kuehbach2021JAC} now becoming replaced by \crystalstructure{} \cite{Kuehbach2021JAC} we changed the scope of the \intersector{} tool and reimplemented it. Now \intersector{} takes a collection $\mathcal{S}$ of sets ${\mathcal{M}}_k$ of three-dimensional triangulated surface meshes ${\mathcal{T}}^l_k \in {\mathcal{M}}_k$ of polyhedra $l$. Application on the set $\mathcal{S} = \{{\mathcal{M}}_k\}$ with $k, l \in \mathcal{N}$ quantifies a number of spatial location, mesh composition, and collision analysis tasks by pair-wise comparing objects ${\mathcal{T}}^a_i, {\mathcal{T}}^b_j$ (with $i, j$ values of $k$ and $i = j$ allowed and $a, b$ values of $l$ with $a = b$ allowed). Objects can be compared for different $k$ or $l$. Specifically, the tool provides algorithms for answering the following questions:

\begin{enumerate}
    \item With which objects does an object intersect? We term this collision analyses.
    \item If two objects collide, what is the volume of the intersection (${\mathcal{T}}^a_i \bigcap {\mathcal{T}}^b_j$)? Details of the volume computation of these, so-called intersection analyses, are reported in the supplementary material.
    \item If an object does not collide/intersect (volumetrically) with another object, does it have other objects within a threshold distance $d_{prx}$ (shortest Euclidean)? We term this proximity analyses.
    \item We store individual collisions and proximity relations as directed graphs per set $k$. In these graphs, nodes represent said objects (polyhedron $l$ in set $k$). Edges represent collision- or proximity-type unidirectional relations between node pairs. This enables to logically relate objects even though they are assigned eventually different identifiers throughout the analyses.
    \item Using Voronoi-tessellation methods (see \cite{Kuehbach2021NPJ}), we construct composite objects from Voronoi cells which belong to specific reconstructed ion positions inside specific objects. Resulting in composite meshes of sets of individual Voronoi cells, where each cell has object ID and evaporation ID attribute data, selected nearest and higher-order nearest neighbor cell adjacencies are evaluated. These topological analyses enable a segmentation of the cell set into three-dimensional regions about objects and a representation of interfaces between objects as 3D meshed entities. An example are ions within the interface between two or more coprecipitating second-phase precipitates. The approach is equivalent to describing the three-dimensional structure of grain boundaries via Voronoi cells \cite{Lazar2015,Lazar2017}.
    \item Using attribute data for each object, such as polyhedron volume or element-specific composition values (from \nanochem{}), we post-process the resulting directed graphs along the direction $k$, while traversing forward and backward, to characterize the evolution of shape and properties of objects as a function of their representation for different $k$. Essentially this graph traversal yields results that remind of Matryoshka dolls which enables the desired quantification how volume and composition of specific objects change as a function of iso-surface value $\varphi$ for instance. The traversal copes with the problem that iso-surface analyses are independent and thus objects will get different identifiers assigned throughout the high-throughput analyses. Paraprobe-intersector essentially identifies the relations between these identifiers. Evidently, the variable $k$ denotes a general coordinate in parameter or phase space, respectively, whose interpretation depends on the use case: Another possible one for which the tool could be used is spatio-temporal tracking of triangulated surface meshes of crystals from e.g. crystal plasticity simulations \cite{Kuehbach2020MSMSE}. In such a case, $k$ could be chosen as equivalent to the time step $t$, strain step $\epsilon$, or iteration counter. Then, a comparison of mesh ${\mathcal{T}}^a_t$ (at time $t$) with mesh ${\mathcal{T}}^b_{t\pm1}$ is equivalent to the re-identification (tracking) of the crystal (or its fragments) between the previous ($t-1$, backward tracking) or the next time step ($t+1$, forward tracking) respectively. 
\end{enumerate}

Further details are given in the supplementary material. These include also details about the implementation, the computers that were used, and the scheduling of the jobs.





\section{Results and discussion}
\subsection{Comparative study of methods for modeling the edge of a dataset}
\label{CaseStudy1}
Although an atom probe reconstruction captures often a statistically significant number of ions, it represents a point cloud within a finite region. Consequently, datasets often contain incompletely measured but relevant microstructural features. Tightly-fitting triangulated surface meshes to the edge of the point cloud \cite{Haley2009} can be a useful tool in this regard for quantifying such edge effects and studying the sensitivity of engineering-relevant descriptors (e.g. number of precipitates or atoms per unit volume).

In the first case study we compare how quantitatively different such results and descriptors are for different edge meshing strategies. A mesh-based description of microstructural features offers further opportunities. One is to create distance-based or even tessellation-based volumetrical segmentations of datasets. Another one is to inspect eventual collisions of meshes, or portions of meshes, and evaluate the proximity of meshes. With these functionalities the first case study also documents how spatial object analyses, like those reported in \cite{Karnesky2007,Karnesky2007a}, can be  generalized for meshes. Finally, we use the first case study also as a verification of the functionalities for the implemented \distancer{}, \surfacer{}, \tessellator{}, and \intersector{} tools.

As an illustrative example we reanalyze a reconstructed dataset of an neutron-irradiated steel specimen. The dataset was measured, reconstructed, and ranged originally by Jenkins and coworkers. Compositional information, irradiation and experimental conditions \cite{Jenkins2020}, and details to the heat treatment of the material \cite{Jenkins2020c} were reported in the literature. The specimen contains a dispersion of clusters, which are richer in copper, silicon, nickel, and manganese than the matrix. The authors characterized these clusters with the core-linkage algorithm \cite{Stephenson2007,Jenkins2020}. Sharing the original reconstructed ion positions and mass-to-charge-state-ratio values surplus which ions were assigned a cluster, enabled our work. With a total of \SI{7.57e6}{} detected ions the dataset volume is small but instructive because the smaller a dataset is (in volume) the relatively more likely its ions lie close to the edge of the dataset. Especially in this case, a quantification of edge effects is more important than it is for larger datasets. We should also mention that acquiring substantially more than a few million ions is often problematic for materials which are prone to premature fracture due to stress on the specimen.


The reconstructed point cloud was transcoded and ranged. In this and subsequent case studies we perform atom-type-specific quantification by decomposing (molecular) ion types into atoms and computing respective multiplicity of each atom. Next, four different sets of edge models were computed: The first set of meshes are approximations of specific iso-contour surfaces or iso-surfaces for short. The here developed \nanochem{} tool was used to delocalize the position of each ion and compute the total number of atoms per voxel without distinguishing atom types. We refer to such iso-surfaces as iso-total-atom-count ones. The adjustable parameter is $\varphi$ the total number of atoms per voxel. The delocalization returns real-valued quantities. The second set of meshes was computed via the \surfacer{} tool which constructs a set of \ashapes{} \cite{Edelsbrunner1983,Felfer2015a,Kuehbach2021NPJ}. In principle these are generalizations of convex hulls with $\alpha$ as the adjustable parameter. The third edge model is the special case of $\alpha \shortrightarrow \infty$ which is equivalent to the convex hull \cite{Haley2009}. The fourth edge model was created by computing first the distance of each ion to an iso-surface from the first edge model (the one for $\varphi = 0.1$ atoms). Second this model was combined with ion-to-edge distance attribute data and a Voronoi tessellation of the entire dataset. With these pieces of information an edge model was instantiated which is composed of all facets between pairs of cells where one cell has a distance larger and the other one a distance that is smaller than an ion-to-edge distance threshold $d$. This threshold is the adjustable parameter.

\begin{figure}[h!]
\centering
	\includegraphics[draft=false,width=1.0\textwidth]{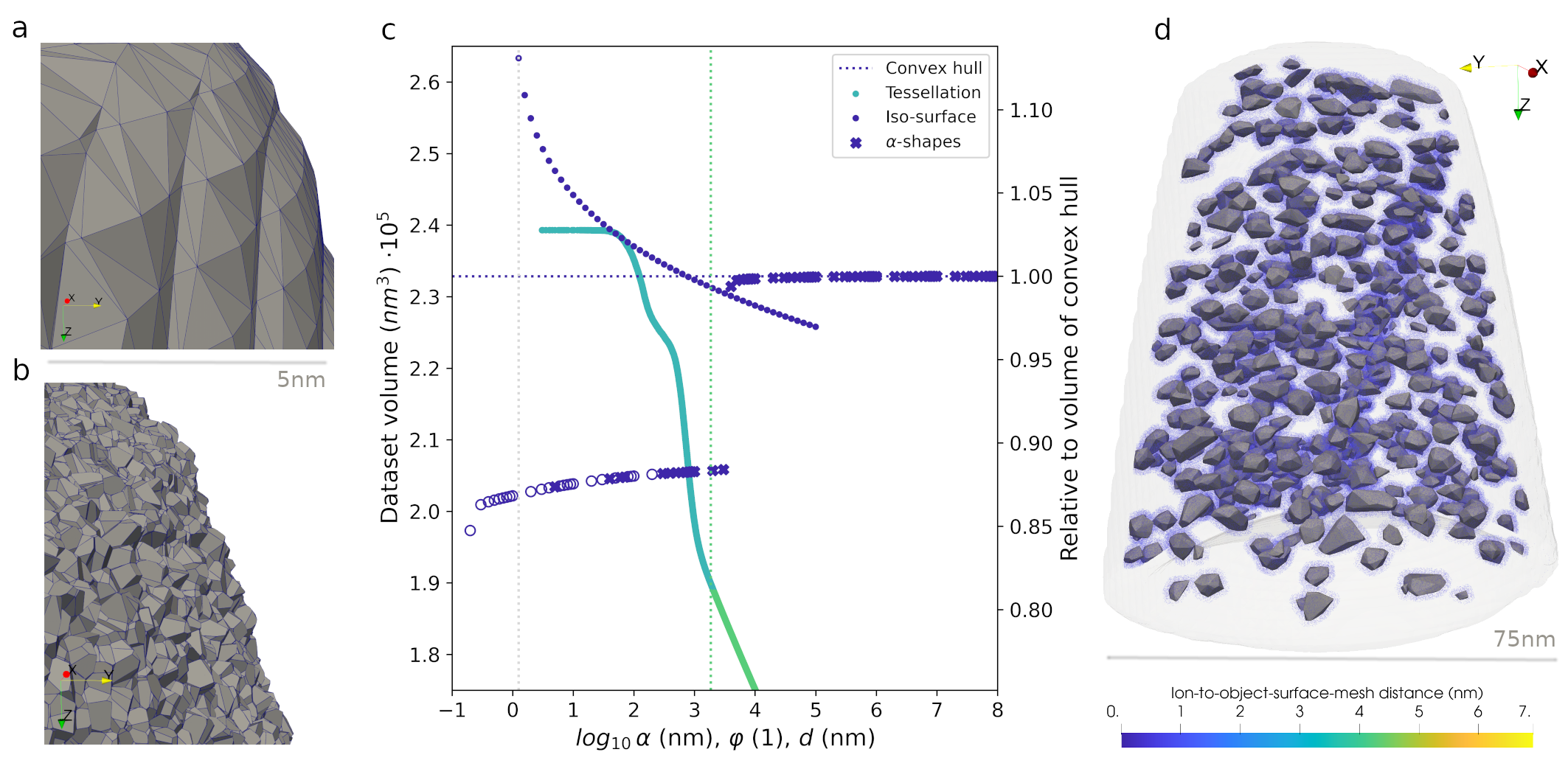}
\caption{We compare different methods for quantifying surface meshes of the edge to the dataset. Sub-figure a) shows an exemplar rendering of a portion of the iso-surface-based edge model ($\varphi = 0.1$, see also comments to d). Sub-figure b) shows an exemplar rendering of the same region as in a) for the tessellation-based edge model. Sub-figure c) compares the dataset volume as a function of the parameterization. Volume here refers to either the interior volume of the closed polyhedron (in the case of convex hull, tessellation, and iso-surfaces eventually), or the accumulated volume of interior tetrahedra (in the case of \ashapes{}). A numerical analysis of the triangle set to each iso-surface, tessellation, and the convex hull edge models confirmed these sets represent closed polyhedra. Using the same analysis on all \ashapes{}, though, identified that not all of these represent closed polyhedra. This is distinguished in the respective curve for \ashapes{} by plotting thick \ding{54} symbols for objects which are closed and thin open $\circ$ symbols for those objects which are not closed. The volume of the convex hull serves as the reference (dark-bluish-dotted lower horizontal line). Volume differences can be \SI{10}{\percent} and larger in particular when using \ashapes{}. The right vertical (green) dotted curve is associated with the dotted curve of accumulated volume of Voronoi cells (tessellation). The colour change from turquoise to green marks the distance beyond which cells are no longer affected by edge effects. Sub-figure d) shows a rendition of the dataset with the $\varphi = 0.1$ atoms iso-surface in light grey and the set of convex hulls around all clusters. The corresponding volume in sub-figure c) is marked with the light grey vertical line. The bluish halo about each cluster represents the positions of all those ions which lie within \SI{1.0}{\nano\meter} (as an example) to a surface mesh. The unit for the parameter $\alpha$ was plotted as $log_{10}(\alpha)$ only to remind the reader that $\alpha$ is a length. Plotting the logarithm of $\alpha$ enables a comparison between order of magnitude differences on the same x-axis for all three parameters $\alpha, \varphi, d$.}
\label{FigJenkinsLondon}
\end{figure}

Figures \ref{FigJenkinsLondon} summarize key results when comparing these edge models. Specifically, we inspect the surface meshes for closure and compare their interior volumes. The volume of the dataset is a key descriptor for normalizing many other descriptors of materials engineering relevance like the density of microstructural features. These engineering descriptors will be explored in more detail in the fifth case study. The results in Figs. \ref{FigJenkinsLondon} substantiate that parameter settings exist for which all four edge models yield watertight surface meshes. The respective interior volume, though, differs by \SIrange{2}{10}{\percent} between different models. Especially, the description of the volume based on \ashapes{} is sensitive. These results add quantitative substantiation to previous studies \cite{Felfer2016,Jenkins2020,Still2021,Kuehbach2021NPJ}.

Our study supports that edge models based on the convex hull have advantages. These are lower numerical costs than for the other edge models, no parameter sensitivity, and guaranteed watertightness. Returning in most cases a lower number of triangles compared to \ashapes{} obtained with $\alpha \ll \infty$ is an additional advantage when computing ion-to-edge distances. The key disadvantage of convex hulls is that the computed distances of ions in local concavities are inaccurate. The results for \ashapes{} document why using previously reported practices of downsampling should be used very carefully if at all. The results show that \ashapes{} which are computed for differently sampled input but the same $\alpha$-value have not only different shape but also a different interior volume evolution curve with changing $\alpha$.


Thanks to using the Computational Geometry Algorithms Library (\cgal{}) (see methods section), the numerical efficiency of using \surfacer{} for analyses as compared to those performed by Jenkins et al. \cite{Jenkins2020} is well one order of magnitude higher. The authors reported it took sequentially more than two hours to compute an \ashape{} of the entire dataset (using a package of the R programming language \cite{Jenkins2019,Lafarge2020}). Paraprobe-surfacer by contrast computed an \ashape{} for the entire dataset in \SI{4}{\minute} \SI{45}{\second} (sequential execution, setting $\alpha \xrightarrow[]{} \infty$). This substantiates that instead of using downsampling practices, i.e. take only each $n$-th ion, it is better to use tools which are numerically more efficient. We confirm that downsampling reduces indeed trivially the numerical costs. Tested here with repeating the authors' downsampling experiments confirmed the approximately the same order of magnitude faster processing so that minutes reduced to seconds. In effect, this enables \ashape{}-based analyses for a larger number of cases as it was so far considered in the APM literature. We expect that the practical benefit of our solution becomes even more important when working with larger datasets because algorithms for computing \ashapes{} scale worse than linearly. 


Frequently it is useful to segment a dataset into regions of different distance to microstructural features. Figures \ref{FigJenkinsLondon} verify the tools' capabilities to compute this segmentation. As an example, we study the same clusters which the authors identified via convex hull surface meshes. For this purpose a utility tool (\clusterer{}) was implemented which imports arbitrarily made definitions of clusters from third-party tools of the APM community into \paraprobe{}. Here, we exemplify for clusters computed with \ivas{}/\apsuite{}, which is the main commercial software. The figure visualizes all ions in the dataset within \SI{1.0}{\nano\meter} distance to these surface meshes (on either side). The method can be extended to segment also regions in the dataset with specific distance to spline-based line or tubular features.

Another purpose of the first case study is to verify the capabilities of the \intersector{} tool to correctly identify collisions between and proximity of individual or entire sets of triangles of surface meshes and/or surface patches. Furthermore, we test the tool functionality that computes Voronoi-tessellation-based volumetric segmentation. Specifically, each convex hull was tested for collisions on neighboring convex hulls in the set. Although this is an empirical strategy for verifying an implementation, it is expected that detecting eventual numerical differences could help to identify if severe implementation errors exist. We expect that each convex hull should collide only with its own copy and on neighbors eventually.

\begin{figure}[h!]
\centering
	\includegraphics[draft=false,width=1.0\textwidth]{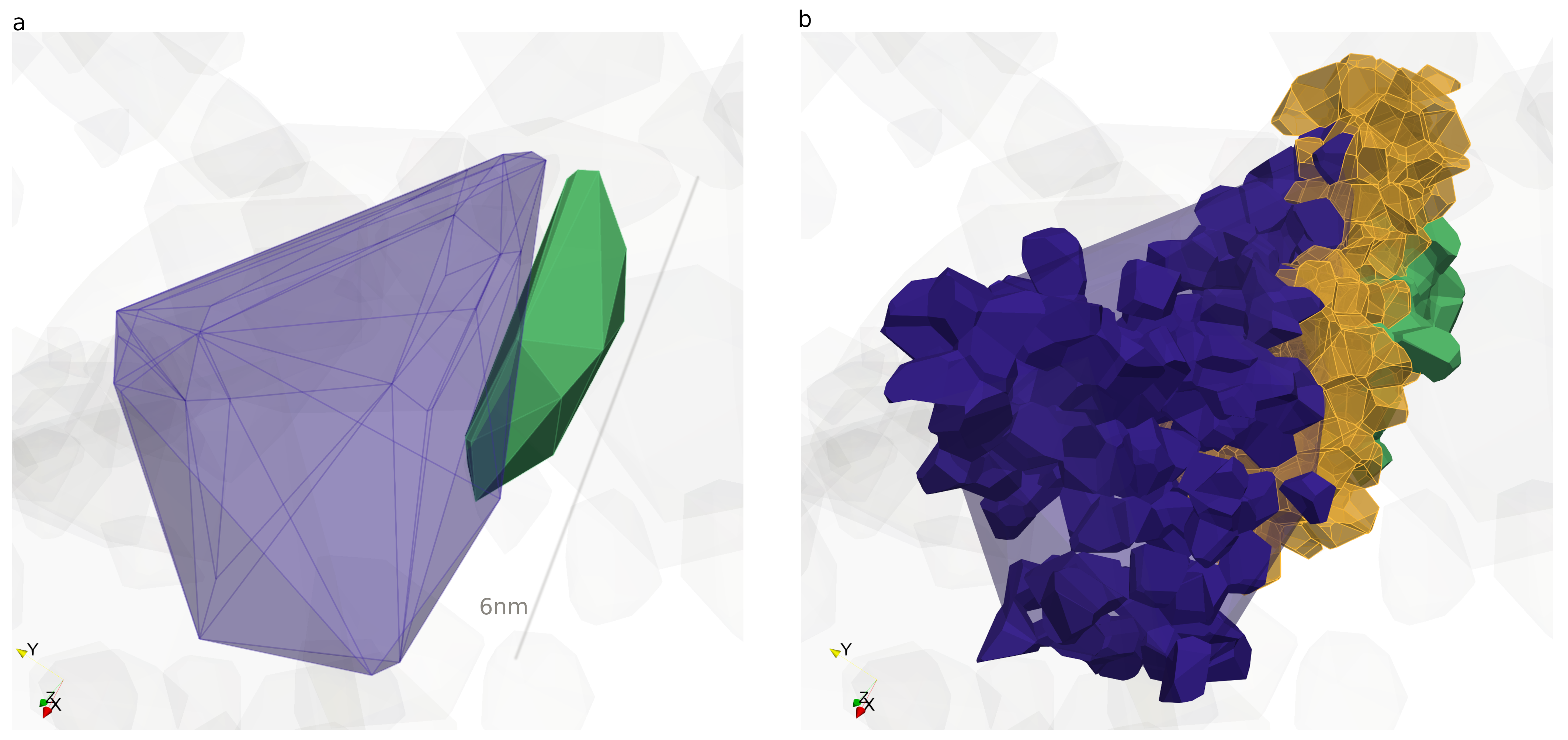}
\caption{With combining results from surface meshes, clustering, and tessellations, it is possible to three-dimensionally segment ROIs which represent portions of interfaces or junctions between interfaces. Sub-figure a) shows a rendition of the convex hull meshes for two exemplar clusters, object 130, in blue and object 133, in green, respectively. These objects intersect three-dimensionally. Convex hulls of neighboring clusters are shown in light grey in the background. The figure is a visual confirmation of the quantitative analysis and graph-based description which \intersector{} yields. Sub-figure b) shows the segmented 3D models of Voronoi cell composites. The orange composite are all Voronoi cells within nearest or second-nearest neighbour connections to a cell of either cluster. The composite can be used as a volumetric description of the region where the two objects intersect and help to uniquely assign each ion to either an object or the interface. The blue and green composites show the Voronoi cells to all ions supporting either cluster 130 (blue) or 133 (green), respectively.}
\label{FigJenkinsIntersection}
\end{figure}

Figure \ref{FigJenkinsLondon}d confirms all of these expectations. All meshes get detected as overlapping with themselves. Interestingly, also cases of additional collisions on neighboring convex hulls were detected. One random example is shown in Fig. \ref{FigJenkinsIntersection} for a collision pair. Using \clusterer{} it was possible to confirm that no ion was coincidentally a member of two clusters within the authors' original cluster analysis results. However, the point set which is closer, in an Euclidean closest distance sense, to all ions of a given cluster (the composite volume of all Voronoi cells to the points in the cluster) is different to the point set that is covered by the respective convex hull of that point set.

Using the meshes that \tessellator{} generated and connecting these with the collision analysis results of \intersector{} enabled to confirm the situation using \paraview{}. The tools deliver all data which enable scientists to inspect all collision situations. As an example we deliver the respective 3D models as supplementary material. This material documents that also all other cases of collisions on more objects than ones own mesh where cases like the one which is exemplarily shown in Fig. \ref{FigJenkinsIntersection}. This verification suggests that the \intersector{} can solve also a number of other collision analyses like those reported for iso-surface-based analyses in \cite{Kuehbach2021MM}.

\subsection{Automated composition profiling for closed objects}
\label{CaseStudy2}
The second case study aims to verify the \nanochem{} tool. Furthermore, we use it to present an automated method for composition profiling to support scientists with collecting reliable statistics of composition gradients across interfaces in an efficient manner. As an example, we process a reconstructed dataset of Cerjak et al. \cite{Cerjak2017} that was taken from a S690 steel specimen. The dataset contains \SI{23.8e6}{} ions in total. We work with the original reconstructed ion positions and ranging definitions of the authors \cite{Cerjak2017} for which data were made public and experimental procedures explained in previous work \cite{Kuehbach2021MM}. The dataset represents a material volume which contains a partially measured carbide located at a grain boundary segment which is decorated with phosphorus as solute atoms. We focus first on characterizing a single object, here the carbide, to familiarize the reader with object-based analyses before continuing with results of such studies for ensembles of objects and finally for ensembles of such ensembles and their parameter sensitivity. Specifically, we computed carbon iso-composition surfaces (${\varphi}_C = $ \SIrange{1.0}{30}{\atpercent} with \SI{1.0}{\atpercent} step) for a delocalization grid with cubic cells of $l = \SI{1.0}{\nano\meter}$ edge length. The variance of the kernel is $\sigma = \SI{1.0}{\nano\meter}$.

\begin{figure}[h!]
\centering
	\includegraphics[draft=false,width=1.0\textwidth]{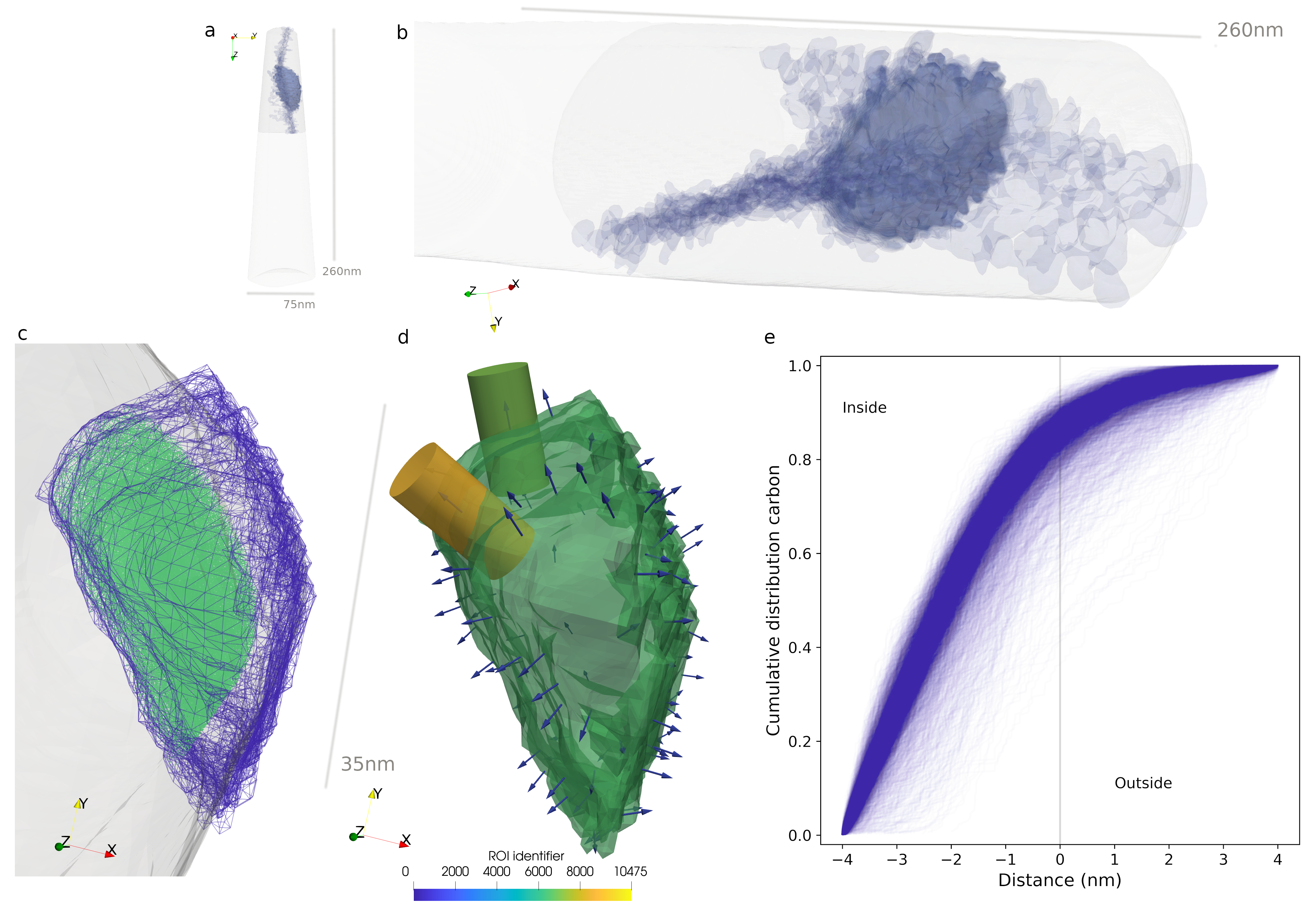}
\caption{Sub-figure a) documents that \nanochem{} can be used to analyze the entire dataset or portions of it. Automated inspections of objects via local ROIs to the surface mesh are possible. Sub-figures a) and b) give an overview of the entire dataset a) and a close-up b) of the carbide at a curved interface segment. Sub-figure c) and d) show a close-up of the carbide to document the computational geometry and composition analysis capabilities. Iso-surface-based analyses deliver triangulated surface meshes such as the blue wireframe mesh of the carbide in c). This sub-figure supports that also incompletely measured microstructural features in contact with the edge of the dataset can be analyzed. For each (closed) object the tool computes which ions are lying inside or on the surface of the object (green point cloud inside the bluish wireframe in c). Sub-figure d) shows the triangles of the surface mesh with individual outer unit normal vectors. These are used to automatically place and orient ROIs with the local facets (here exemplified for cylinders). To convey this message clearer, we rendered only a random sub-set of ROIs and their guiding normals. 3D models are available as supplementary material. Paraprobe-nanochem evaluates if ROIs intersect with the edge of the dataset. Thereby, automated analyses of composition profiles, like those shown in sub-figure e), can be controlled to detect eventually incomplete profiles. Here, all 1D cumulated carbon composition profiles are shown for all those ROIs which are fully embedded in the dataset. Sub-figures like the last one can be created with scripting using \autoreporter{}, thus offering tailored analyses.}
\label{FigMendezMartinCarbide}
\end{figure}

Figures \ref{FigMendezMartinCarbide} document that \nanochem{} yields detailed quantitative and spatially-resolved composition results. Specifically, Fig. \ref{FigMendezMartinCarbide}d verifies that objects can be meshed and ions inside these meshes detected correctly. The triangle facet normals have the correct orientation which is, according to our definition, positive when pointing out of a closed object. The 3D models in the supplementary material document also a correct detection of eventual inclusion of, or intersections between, individual ROIs and an edge model of the dataset Fig. \ref{FigMendezMartinCarbide}c to pinpoint for which ROIs edge effects have to be considered. Figures like \ref{FigMendezMartinCarbide}c and \ref{FigMendezMartinCarbide}d were creatable because \paraprobe{} stores all numerical data through \hdf{}. This enables efficient and programmatic figure creation, for instance to either explore composition gradients locally Fig. \ref{FigMendezMartinCarbide}e or to script models which consume and post-process the characterized data in the \hdf{} file according to user needs. In this example the results confirm there is a decreasing carbon gradient towards the matrix and local differences of the carbon concentration profile across the surface mesh.

It is important to mention that the only input from commercial software is the reconstructed dataset with the ranging definitions. The novelty is in the combination of giving unrestricted access to algorithms for delocalization, iso-surface extraction, subsequent polygon mesh processing, and automated ROI analyses. This offers an alternative for many analyses that would manually be very time consuming to perform with GUIs. Examples are given especially in the following case studies. Such high-throughput analyses were in the past often neither feasible nor reported by experimentalists across APM publications, but with now having access to also the 3D atom intensity fields as data arrays it is possible to compare results to those from other tools.  

\begin{figure}[h!]
\centering
    \includegraphics[draft=false,width=1.0\textwidth]{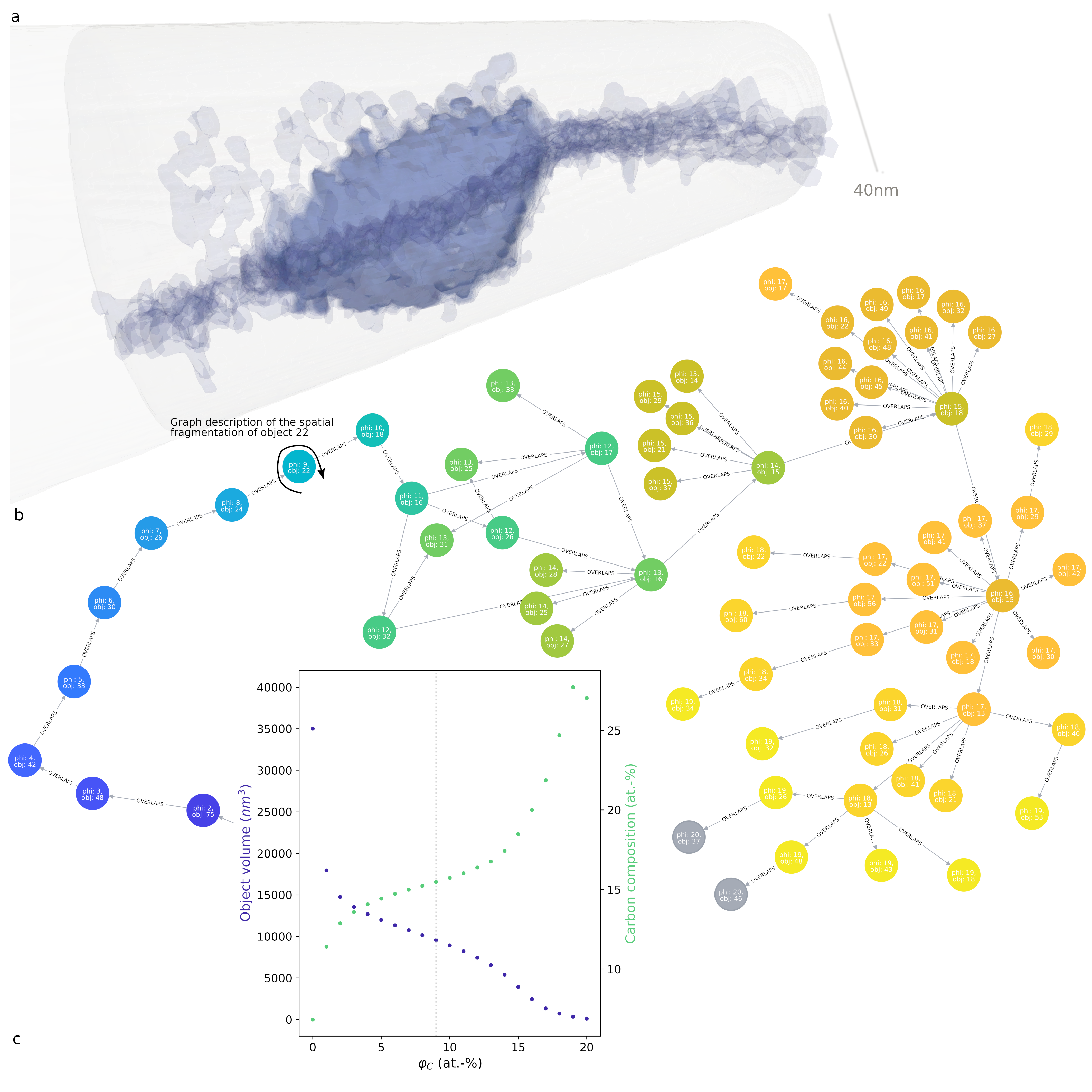}
\caption{Paraprobe-intersector is a tool for studying the parameter sensitivity of individual or ensembles of objects. Sub-figure a) shows a composite of 3D meshes of the carbide for different iso-surface settings. Reminding of Matryoshka dolls, the meshes for the carbide have different size and shapes. Paraprobe-intersector evaluates which meshes are logically connected via spatial collisions on other or proximity relations to meshes of neighboring objects. Sub-figure b) documents how this functionality yields automated graph network descriptions and visualization of the logical relations. These quantify the successive fragmentation of the iso-surface with increasing ${\varphi}_C$. Sub-figure c) documents that connecting results from different tools, here individual object properties for each iso-surface value from \nanochem{} (carbon ${\varphi}_C$ iso-composition), with graph analytics from \intersector{}, yields rigorous quantitative characterization how sensitive descriptors like volume or composition of an object are to parameterization. With scripting such diagrams could be created for each object of a high-throughput analysis in an automated manner.}
\label{FigMendezMartinGraph}
\end{figure}

Figures \ref{FigMendezMartinGraph} summarizes the tools' capabilities for quantifying the spatial arrangement between different objects. The result suggests an example how the sensitivity of an object's representation can be described as a function of the parameterization for the respective geometrical model. Specifically, Fig. \ref{FigMendezMartinGraph}a exemplarily shows how different choices for the iso-composition threshold $\varphi_C$ result in different object representations. Specifically, a composite of 3D meshes of the carbide is shown for different carbon iso-composition ${\varphi}_C$ values. Such a  qualitative characterization is an established prerequisite step during exploratory analyses of APM data \cite{Kuehbach2021MM} especially when using commercial software. Our strategy and built tool, though, enables quantitative studies via logically combined analyses as a function of different threshold values, plus the rigorous documentation of these. This strategy serves not only advanced sensitivity and uncertainty quantification but also repeatability and reproducibility because of the documentation of the steps. The carbide example documents a sensitivity of the shape and composition to ${\varphi}_C$. These findings confirm that a single threshold value is in most cases inadequate to distinguish objects via iso-contour approximation techniques \cite{Martin2015,Hornbuckle2015,Barton2019,Jenkins2020b}. Our work solves how these inherent limitations of iso-surfaces can at least be studied quantitatively in an exactly reproducible and fully automated manner.

The suggested delocalization method pinpoints a shortcoming of edge effect handling for microstructural objects in commercial software that is commonly faced, though, rarely discussed for atom probe data: We expect that delocalizing and discretizing a finite point cloud results in edge effects because the delocalization kernel extends beyond the edge of the point cloud. This affects the representation of surfaces. A more detailed discussion is offered in the supplementary material.


\subsection{Automated meshing of interfaces aided by compositional gradients}
\label{CaseStudy3}
Automated composition profiling as in the second case study requires orientable meshes to obtain consistently signed surface normals for each facet. For closed objects such a set of normal vectors can be computed as it was shown in the previous case study. In the case of polyhedra usually weighting schemes \cite{Baerentzen2005} are in use because otherwise discontinuous changes of surface normals are encountered at vertices and edges. A related atom probe study \cite{Keutgen2020} used these so-called pseudonormal vectors.

Many iso-surfaces, though, contain portions which are not closed but represent free-standing, so-called surface patches. In this case, the (global) context of the model and often further assumptions are needed to decide how the normals of the mesh primitives (here triangles) can be oriented consistently. 


One important application is quantification of composition variation in the vicinity of grain or phase boundaries. Here,  it is important that the normals point ideally in the same direction of an existent composition gradient instead of being flipped in an uncontrolled manner. The procedure which is used to compute normals in such situations should be well documented as to guide users where eventual inconsistencies exist. Given the variety of possible triangle configurations there can be inconsistencies when computing so-called proximity diagrams and studying their respective composition profiles \cite{Hellman2000,Keutgen2020}.

In this (third) case study, we test an automated protocol for orienting triangle normals by taking into account gradient information of the underlying composition field and formulate a quantitative quality descriptor which documents for which triangles this procedure is locally eventually not reliable. Exemplarily, the case study here reports local composition profiles across an interface protrusion and positions of eventual junctions between microstructural features (dislocations or patches of phase boundaries) at this protrusion. 

\begin{figure}[h!]
\centering
	\includegraphics[draft=false,width=1.0\textwidth]{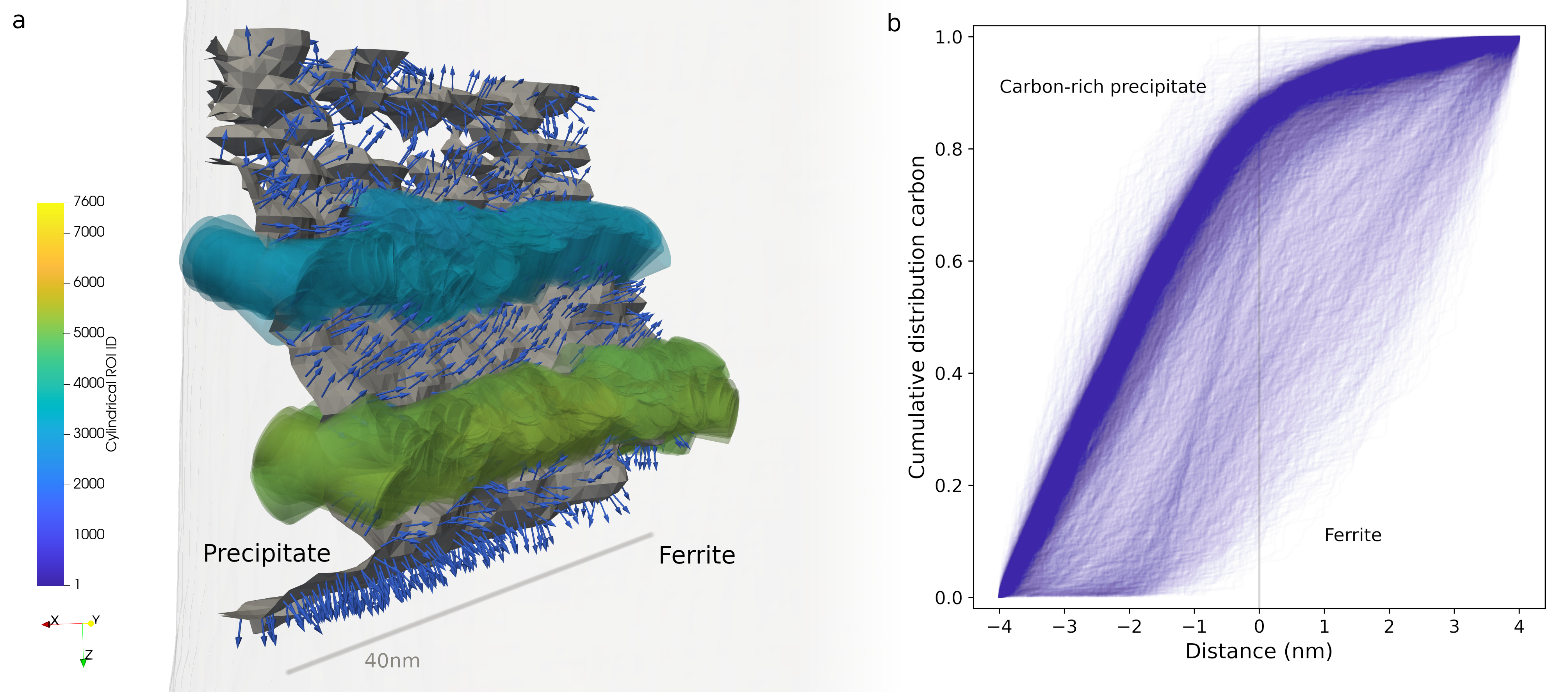}
\caption{The automated placing functionalities for ROIs and composition analyses of \nanochem{} can also be used for quantifying surface patches which are not part of closed objects. In this case the local triangle normal vector is oriented via an inspection of the gradient of the composition field $\nabla{\vec{c}}$. Sub-figure a) shows an exemplar free-standing surface patch with normal vectors which are aligned with the direction of decreasing carbon concentration $c_C$. Two exemplar sets of cylindrical ROIs are displayed in transparent colors to show how they overlap and align with the local normals. Sub-figure b) shows a collection of all 1D cumulated carbon composition profiles for all ROIs which do not intersect with the edge of the dataset. The results give access to substantial details of the composition gradients from the carbon-rich precipitate towards the ferrite with a lower carbon content. All profiles are stored in the \hdf{} result file to enable further programmatic analyses with \python{} scripting for example.}
\label{FigMaywegOverview}
\end{figure}

Specifically, the analysis is for a reconstruction of a 100Cr steel specimen from Mayweg et al. \cite{Mayweg2021a,Mayweg2021b}. The author shared the original dataset with \SI{30.7e6}{} ion positions and ranging definitions. A convex hull edge model was computed for the entire dataset. Subsequently, different carbon iso-composition surfaces ($l = \sigma = \SI{1.0}{\nano\meter}$ and ${\varphi}_C = $ \SIrange{5.0}{15}{\atpercent} with \SI{1.0}{\atpercent} step) were processed. Figure \ref{FigMaywegOverview}a displays the ${\varphi}_C = \SI{12}{\atpercent}$ iso-composition surface. To orient the normals of the triangles we first compute the gradient of the composition field $\nabla{\vec{c_C}}$. Second-order accurate central differences are computed for interior voxel and first order accurate one-sided (forward or backward) differences at the boundaries. Next, the voxel with the closest barycenter to the triangle barycenter is taken to guide the direction of the normal. Specifically, the magnitude of the gradient $\Vert{\nabla{\vec{c_C}}}\Vert$ at this voxel is taken as the quality descriptor to filter if the voxel can guide the eventual flipping of the triangle normal vector. As an example, we ignore triangles with associated voxels with lower than \SI{0.01}{\atpercent\per\nano\meter}. For all other triangles of the patch, the respective normals $\vec{n}$ are evaluated against the gradient direction and eventually flipped to assure $\vec{n}\cdot\nabla{\vec{c_C}} \geq 0$.

Note that the magnitude of the gradient and the cosine of the angle between the gradient vector and the triangle normal can serve as quality descriptors to identify local configurations where the gradient vector is nearly perpendicular to the proposed triangle normal. In such a case, the cosine evaluation yields values close to zero regardless from which side one approaches so that the projection is prone to sign fluctuations of the normal and therefore of lower quality than compared to a case of a near parallel alignment.


Figures \ref{FigMaywegOverview} summarize that this approach makes the triangle normals point in a direction which is consistent with the composition gradient that is here pointing into the ferrite with its lower carbon concentration. With these normals computed, automated ROIs are placed (similarly like in the second case study). An extension of this work and combination with results from \distancer{} could use these normals and quality descriptors to filter out those regions in the dataset with a specific proximity to the surface patch and exclude for instance those regions in the dataset where the above-mentioned approach of orienting the triangle normals delivered results with low quality.

\subsection{Automated meshing of interfaces aided by chemical decoration}
\label{CaseStudy4}
There are many datasets where iso-surface-based analyses of grain and phase boundaries, like those discussed so far, are unsuccessful. Facing imprecision of current reconstruction models, insignificant chemical gradients across the interface or lacking correlative results from electron microscopy methods means insufficient latent crystallographic information to identify interfaces in atom probe datasets. In some cases features within field desorption images can be used to inform the reconstruction of interfaces inside the reconstructed volume \cite{Wei2019}.

There are cases, though, where this is a tricky, if not a conceptually questionable approach: A grain or phase boundary is defined as the three-dimensional region between two adjoining crystals where the long-range crystallographic symmetry breaks. This region is often spatially correlated but not necessarily located exactly where a local segregation of a decorating solute is highest.

With the increased usage of site-specifically prepared atom probe specimens to probe spatial details of solute segregation at interfaces (see e.g. \cite{Raabe2014,Kontis2021}), there has been an interest to use this chemical decoration for supporting the construction of triangulated surface meshes to stand in as models of interfaces. Felfer et al. implemented computational geometry methods for this task \cite{Felfer2012b,Felfer2013,Felfer2015a}. Their so-called DCOM algorithm has influenced several authors \cite{Peng2019,Zhou2022,Felfer2022a}. Implemented in practice, these tools are semi-automated and may or not need manual mesh processing which is most conveniently performed with GUI-based tools like \blender{}. A noteworthy subtlety of DCOM is the stability of the algorithm when it gets applied iteratively. These and how users assure the creation of robust meshes when working manually has not been covered in substantial detail in the atom probe literature.

DCOM is an iterative algorithm which relocates the triangles of an initial interface model by moving vertices towards the barycenter of the local cloud of solute atoms about the respective triangle vertices. However, if left unconstrained, these mesh operations can result in mesh self-intersections. To support this discussion we specifically revisited the authors' implementation \cite{Felfer2022b} and investigated whether a complete automation of the algorithm and adding of established intermediate steps of polygon mesh processing and mesh quality inspection yields a more frequently applicable automation of DCOM. We were also interested whether this improves robustness or not. 

As an example, we applied our modified implementation first on an interface in a molybdenum-hafnium alloy from Leitner et al. \cite{Leitner2018}. The authors shared the original dataset with \SI{4.58e6}{} ion positions surplus the original ranging definitions. The dataset details a reconstruction with a single curved interface patch with boron and carbon segregation.

\begin{figure}[h!]
\centering
    \includegraphics[draft=false,width=1.0\textwidth]{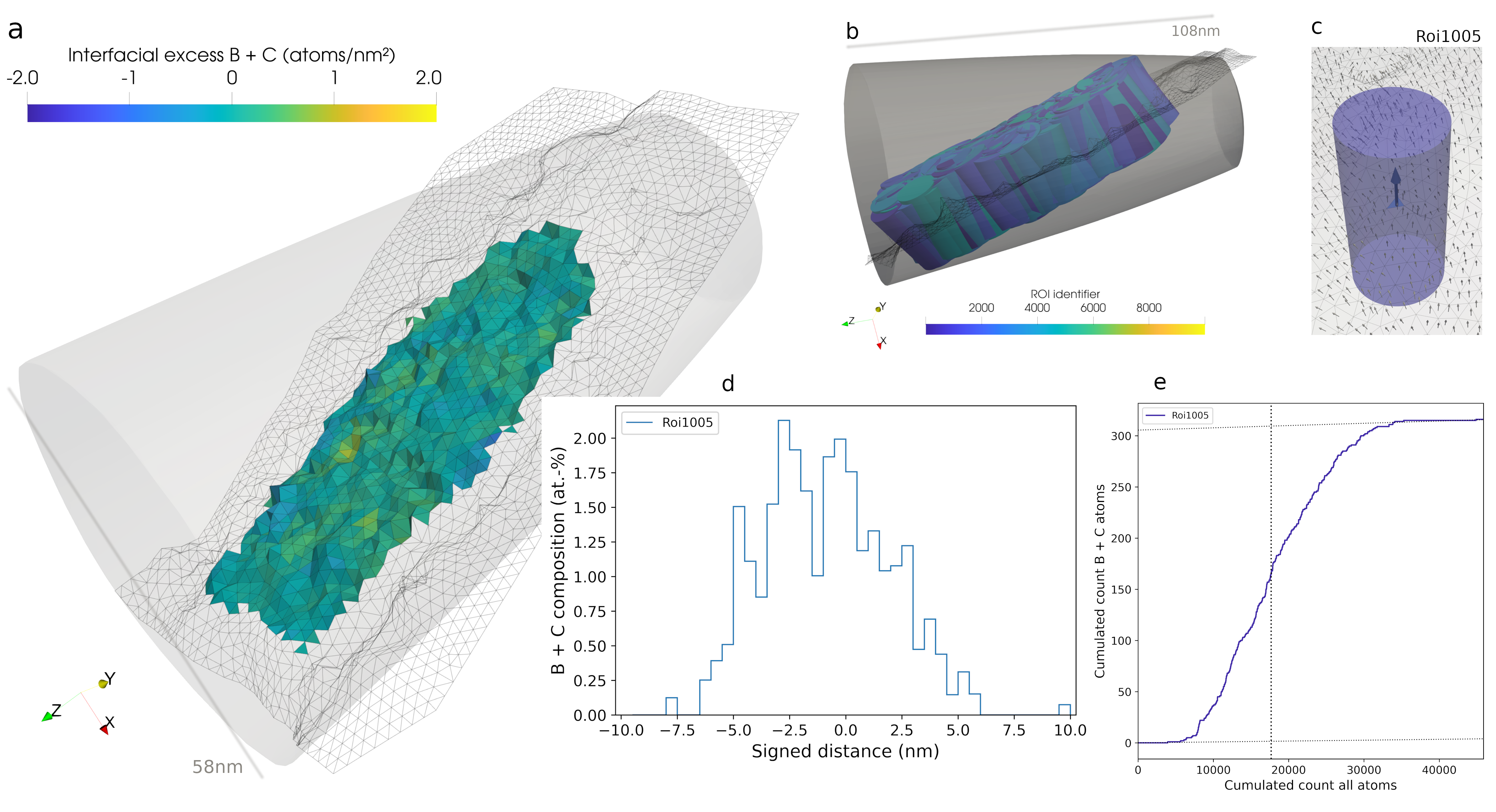}
\caption{Paraprobe-nanochem implements a set of robust and automated methods for creating triangulated surface mesh models of individual interfaces. This is useful in cases when a decoration of solute atoms is the exclusive experimental hint of the interface location. The tool combines this with a robust implementation of the DCOM method which detects if triangles self intersect during the iterative refinement of the interface model. Once identified, triangulated surface meshes are processed via fully-automated methods where ROIs are placed and aligned at each mesh triangle, if desired, to characterize composition, concentration, proximity diagrams, and interfacial excess mappings a), d), e). The light-gray contour about the dataset is a mesh of the convex hull. Intersections between ROIs and the convex hull, here representing the edge of the dataset, are detected in this process to control analysis bias by ignoring those ROIs that are only partially contained in the dataset. Sub-figure b) shows a composite of all those ROIs which lie inside the convex hull. For all these ROIs, of which one is shown exemplarily in c), the tool computes the projected distances of each ion and its respective atoms. The results are stored in the \hdf{} results file. This enables post-processing of different types of composition analyses via the \autoreporter{} Python tool. Exemplified for ROI 1005 these analyses yield e.g. composition d) or interfacial excess e) profiles. Sub-figure a) summarizes the results of such interfacial excess computations for all interior ROIs, displayed as an interfacial excess mapping. Triangles outside the dataset or without an associate ROI inside the dataset are drawn as a wireframe mesh.}
\label{FigDcomLeoben}
\end{figure}

Figures \ref{FigDcomLeoben} summarize the results of letting \nanochem{} construct a quality triangulated surface mesh based on the joint point cloud of boron and carbon atoms in the dataset. The initial planar interface model was computed with principal component analysis (PCA). Subsequently this model was iteratively refined via DCOM. During each iteration the mesh from the previous iteration was refined and mesh-smoothened to equilibrate interior angles of the triangles, followed by a DCOM step, with a subsequent check for mesh self intersections. Specifically, the mesh was refined from an initial triangle edge length and DCOM radius of $\SI{10}{\nano\meter}$ in steps of $\SI{2}{\nano\meter}$ to a target edge length of $\SI{2}{\nano\meter}$. 


Using sequential computing the interface modeling took $\SI{0.3}{\second}$ wall clock time including I/O. For this dataset it was successful to take naively the entire boron and carbon ion point cloud. The algorithm locates the interface initially and successive refined the model fully automated.

With a triangulated surface mesh generated, we equipped \nanochem{} with support for the automated ROI analyses of the previous case studies. For the exemplar dataset here, we  performed completely automated and multi-threaded analyses using four cores on a laptop. Details are available in the supplementary material. ROIs with a height of $\SI{20}{\nano\meter}$ and a radius of $\SI{5}{\nano\meter}$ were taken. From the total of 4934 triangles/ROIs, 1354 were detected as fully-contained ROIs. Composition analyses were performed for all of them. This analysis took $\SI{18.2}{\second}$ of which $\SI{5}{\percent}$ was spent in I/O operations. 


With additional less than \SI{10}{\second} of sequential Python processing the results yield interfacial excess maps Fig. \ref{FigDcomLeoben}a. This map and exemplar composition profiles Fig. \ref{FigDcomLeoben}d and Fig. \ref{FigDcomLeoben}e confirm that boron and carbon have segregated to a joint atomic fraction within the interface of approximately \SI{1.5}{} at.-\%.

With an average width of more than $\SI{5}{\nano\meter}$ the profiles document that the chemical decoration, as it displays in the reconstructed volume, is wider than typically observed for structural units in grain boundaries when these are characterized with high-resolution transmission electron microscopy. Such experiments reported \SIrange{0.5}{2}{\nano\meter} or explored with atom-resolving simulations \cite{Leitner2017,Leitner2018,Raabe2014,Wei2019}. Calculation of segregation profiles at grain boundaries within the coincident site lattice (CSL) approach display widths which are consistently smaller than \SI{0.5}{\nano\meter} \cite{Scheiber2015,Han2016,Lorenz2021}. Literature results from early APFIM measurements \cite{Seidman1991,Hu1992} which relied on a different methodological approach than compared to modern APT measurements also report such narrow segregation profiles for W-Re solid solutions. The here proposed protocol for automated characterization of gradients at interfaces can support research on disentangling the individual contributions of inaccuracies  due to reconstruction models used or the eventual existence of defect phases \cite{KorteKerzel2021}, and/or precipitates at the interface. Whether such a disentangling is possible with experiments alone or requires the support of computer simulations is a topic of current research \cite{Jenkins2020b}. This warrants a detailed analysis in its own right that is beyond the scope of this multi-method-reporting paper.

Motivated by this success, we applied the tool to five other datasets. Each was measured for a similar experimental design: characterizing a single eventually curved interface, in a site-specifically specimen to yield a reconstructed volume with an interface that is decorated with solutes. Studies for these other datasets are parts of ongoing work; and therefore the following findings are so far preliminary. We learned that it is the specific location of the interface plane, its inclination, the spatial arrangement (relative to the interface patch) and the number of solute atoms on either side of the interface which defines if creating an interface model without manual interaction is successful.

For one of the five datasets this was successful using also naively the entire dataset. In the other cases working with all ions of the decorating species was unsuccessful. Especially for strong noise the PCA just splitted the dataset approximately in half uncorrelated to the chemical decoration. In three other cases, using the naive approach delivered a mesh of only a partial patch of the interface. However, with simply constraining the input to operate only on those solute ions inside a thin rotated bounding box about the interface, it was possible to mesh all cases across the entire interface.

Only in one of the five cases the spatial distribution of the decorating ions was so heterogeneous that local curvature built up during DCOM operations so that the algorithm warned and stopped when it detected mesh self-intersections. We learned that here is potential for a further improvement of the method which is to use automated remeshing procedures which are commonly used in the field of finite-element-based interface evolution solvers. Alternatively one could interpret the DCOM-relocated vertex positions as a predictor step and apply a subsequent shape smoothening operation which effectively constraints excessive vertex migration. Algorithms for such tasks have been proposed in the computational geometry community and are used in the continuum mechanics and the annealing microstructure evolution modeling communities.

In summary, we consider the work on this use case and DCOM a success because if such a simple spatial filtering suffices it is likely that also other datasets can profit from our faster, more automated, and more comprehensively documenting tool. Thanks to open-source software, it is possible to containerize the tool, and make it thus accessible for cloud-based computing in services like \nomad{} \cite{Scheffler2022}. This substantiates that much collectively performed research with sharing atom probe datasets is left to become cooperatively harvested. This would also enable even more careful analyses of the subject to pinpoint in which cases there is really no alternative to manual mesh processing.


\subsection{Automated high-throughput analyses of object ensembles}
\label{CaseStudy5}
In the fifth case study we apply the above-mentioned methods in automated high-throughput analyses for characterizing microstructural objects, precipitates, via computing and segmenting triangulated iso-composition surfaces. The tomographic reconstruction was taken from a measured atom probe specimen of a nickel base super alloy. This material contains two different types of precipitates. One type are \gammaxx{} (${Ni}_{3}Nb$) which are niobium-rich precipitates. The other type are \gammax{} precipitates (${Ni}_3X$) which contain aluminium or titanium atoms representing $X$. Many of these precipitates show coprecipitation, i.e. (spatial) configurations where one or multiple precipitates lie in close proximity or make contact to other precipitates. To rationalize their effect on material properties, it is not only relevant to quantify each precipitate individually with respect to its volume, shape, and atom composition, but also to understand the detailed spatial arrangement and relative frequency of different coprecipitation configurations. 

The dataset was measured, reconstructed, and ranged by Rielli et al. \cite{Rielli2021}. The reconstruction contains a total of $\SI{103.2e6}{}$ ions. Experiments were performed on a \cameca{} local electrode atom probe (LEAP) 3000 Si instrument with a detector efficiency of \SI{57}{\percent} in pulsed voltage mode. A target evaporation of \SI{1}{\percent}, a pulse fraction of \SI{20}{\percent}, and a pulse rate of \SI{200}{\kilo\hertz} were used. The specimen was measured while maintaining a temperature of \SI{50}{\kelvin}. The tomographic reconstruction was performed with commercial software \ivas{} (v3.8.4) and crystallographically calibrated \cite{Gault2012a,Ceguerra2019}. 

In the preprocessing steps we imported the dataset representing the reconstruction and ranging definitions from \ivas{} and applied these (\transcoder{}, \ranger{}). Next, we computed an iso-surface of the total atom count per voxel (\nanochem{}) and evaluated the shortest Euclidean distance of each ion to the iso-surface at $\varphi = 0.1$ atoms threshold (\distancer{}). Next, two sets of high-throughput studies were executed: One for characterizing \gammaxx{} and another one for characterizing \gammax{} precipitates. Previous high-resolution transmission electron microscopy studies \cite{Theska2020} on a similar material and processing conditions supports our assumption to restrict the analyses in this work on \gammaxx{} and \gammax{} as the most important precipitate types. We assume that closed objects segmented from niobium iso-composition surfaces (${\varphi}_{Nb}$) represent \gammaxx{} precipitates while closed objects segmented from aluminium plus titanium iso-composition surfaces (${\varphi}_{Al+Ti}$) are \gammax{} precipitates. Having made the same assumption in previous work \cite{Theska2019,Rielli2020,Rielli2021}, we continue to use the brown (\gammaxx{}) and turquoise (\gammax{}) coloring scheme of \cite{Theska2019} to distinguish these precipitates.

Each of the two high-throughput studies was instructed as a set of \nanochem{} runs with different parameterization. Both probe the sensitivity of the object representations as a function of delocalization settings (grid resolution $l$, kernel variance $\sigma$, and threshold values $\varphi_{Nb}$ and $\varphi_{Al+Ti}$) respectively. Specifically, 3D discretized grids within $l =\,\, $\SIrange{0.50}{2.0}{\nano\meter} with \SI{0.25}{\nano\meter} step were probed. For each edge length $l$, anisotropic Gaussian smoothing kernels with $\sigma = \sigma_x = \sigma_y = 2\sigma_z$ were used and $\sigma$ was varied between \SIrange{0.25}{1.0}{\nano\meter} (with \SI{0.25}{\nano\meter} step). A $7^3$ voxel delocalization kernel was centered at the closest voxel covering each ion. The result of each parameter combination yields a set of discretized elemental composition fields (for niobium, and aluminium plus titanium, respectively, reporting atomic fraction \SI{}{\atpercent}). Iso-composition surfaces were computed on the interval \SIrange{1.0}{30.0}{\atpercent} with \SI{0.1}{\atpercent} step. In total, each high-throughput study thus probed $28$ different combinations of delocalization settings with a total of $291$ iso-surfaces analyzed per delocalization each.

Compared to classical analyses it would be necessary that an experimentalist performs $8148$ runs via manual, GUI-based analyses and characterizes each object for these runs. This is a highly impractical task already for a single run as most runs have a complex set of triangulated iso-surfaces \cite{Rielli2020,Rielli2021}. 

By contrast, \nanochem{} performs automated analyses with post-processing for each iso-surface including all its microstructural objects to extract quantities of interest for materials engineers (number density, volume of each object, object shape via fitting approximate oriented bounding boxes, and computing which ions lie inside each object for quantification of ion-species- and element-specific/atom-type-specific compositions). We assume closed objects can qualify as so-called interior objects if no point on their surface mesh lies closer to an edge model of the dataset than a threshold distance $d_{prx} \leq \SI{2.0}{\nano\meter}$. 

\begin{figure}[h!]
\centering
	\includegraphics[draft=false,width=1.0\textwidth]{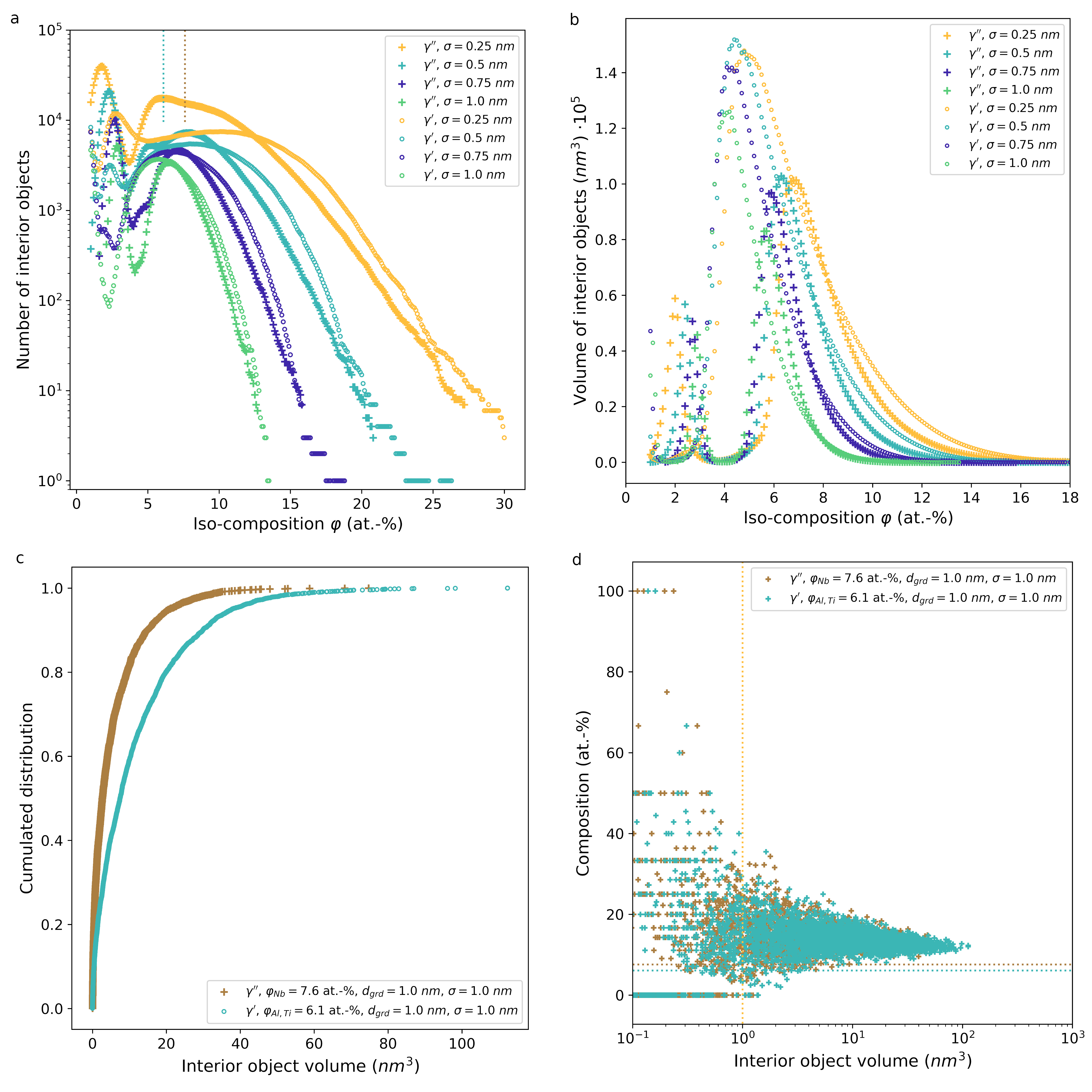}
\caption{High-throughput quantification with \nanochem{} reveals the parameter sensitivity of iso-surface-based descriptors, here exemplified for the number of interior objects a) and the total volume, respectively b) of \gammaxx{} and \gammax{} precipitates as a function of $\sigma$ and $\varphi$ at fixed $l = \SI{1.0}{\nano\meter}$. The volume of the convex hull about the entire dataset is \SI{2.056e6}{\nano\meter^3}. The upper dotted lines mark those runs, which are discussed exemplarily in more detail. For \gammaxx{} and \gammax{} the respective $\varphi$ values are \SI{7.6}{\atpercent} and \SI{6.1}{\atpercent}. The results show there are regions in parameter space for which the descriptors are least sensitive to $\varphi$.
Paraprobe-nanochem can compute detailed statistics with each run. Sub-figure c) exemplifies this with the distribution of volume of interior objects. Sub-figure d) exemplifies statistics of average object composition, here as an example for niobium for \gammaxx{} and aluminium plus titanium for \gammax{}, respectively, as a function of the object volume.}
\label{FigSydneyEng}
\end{figure}

One key result of the study is that all descriptors show a sensitivity on the iso-surface parameterization. The results also quantitatively document a substantial effect which the delocalization settings have on the resulting descriptors. These results clearly support previously made comments \cite{Larson2013} that such settings should not only be always reported but also their effect ideally be quantified in a reproducible and routine manner given now the availability of automated tools. 

For example, Figs. \ref{FigSydneyEng} display the sensitivity of the number of precipitates Fig. \ref{FigSydneyEng}a and total volume of such precipitates Fig. \ref{FigSydneyEng}b as a function of iso-composition $\varphi$, i.e. $\varphi_{Nb}$ and $\varphi_{Al+Ti}$ respectively, for increasingly higher variance of the delocalization kernel $\sigma$. The curve is practically smooth with two maxima. Different qualitative surfaces are obtained: Below $\varphi \leq \SI{3}{\atpercent}$ the iso-surfaces represent a single complex which encloses the majority of the dataset. With increasing $\varphi$ the complex begins to fragment into pieces with a complicated dependency on the threshold value. Evidently, these low threshold values result in a percolating network of complexes, which indicates a threshold that most experimentalists would qualitatively consider as a too small one. With increasing threshold the complexes fragment further, which is more representative of a dataset that hosts a set of individually separated precipitates. The corresponding $\varphi$ range can be interpreted as a region of the parameter space (descriptor $\varphi$) with a lower sensitivity to $\varphi$. The resulting objects are in addition sensitive to the delocalization settings. Beyond $\varphi \ge \SI{8}{\atpercent}$ the number density and corresponding volume of \gammaxx{} or \gammax{} precipitates reduces until eventually no interior objects remain. This is an effect of the successively stronger erosion of the precipitates with increasing $\varphi$. Traditionally one would ask at this stage what is the ``right'' threshold to choose. Our methods enable to quantify the sensitivity to either support specific user choices or supplement these with the required sensitivity. 

Thanks to automation, it was possible to compute further statistics for each run. As these are part of the supplementary material, it suffices to zoom into key results. Figures \ref{FigSydneyEng}c and \ref{FigSydneyEng}d exemplify for the specific runs with $l = \SI{1.0}{\nano\meter}, \sigma = \SI{1.0}{\nano\meter}$, $\varphi_{Nb} = \SI{7.6}{\atpercent}$ for \gammaxx{}, and $\varphi_{Al+Ti} = \SI{6.1}{\atpercent}$ for \gammax{} respectively. We find that the compositional variance of the objects is significantly lower the larger it is the object volume (Fig. \ref{FigSydneyEng}d). This sensitivity originates from at least two effects: Finite ion counting statistics and a relatively stronger effect of the delocalization for small objects. The results support that analyses of nanoscale precipitates requires substantial care if not sometimes a clearly set limit as to what is reliably characterizable with iso-surfaces and APM \cite{Degeuser2020,Gault2021Nat} in general. Especially when no correlative electron microscopy data are available. Observing a quantization of compositions for objects with less than a few voxels (here $\SI{1}{\nano\meter^3}$) is evidence of finite counting effects. 


\label{CaseStudy4b}
To further support our argumentation that a high-throughput approach is not only applicable to other materials but also offers significant value for scientists, we add another short case study. Specifically, an example of research on characterizing the composition of metastable phases in a metastable titanium alloy. The challenge here is that thermodynamics enable eventually for a range of object compositions, which can oftentimes be very similar, especially when defect phases \cite{KorteKerzel2021} are involved and finite counting effects are relevant \cite{Degeuser2020}. Specifically, the user story summarizes practical research experiences from a recent analysis by Zheng and coworkers \cite{Zheng2020,Stoichkov2021}. They studied different transient phases forming during heating to aging temperature in a $\beta$ Ti-5Al-5Mo-5V-3Cr alloy. The reconstructed dataset contained a total of \SI{45e6}{} ions. The authors performed APT experiments, analyzed these via GUI-based processing in commercial software and studied individual precipitates via high-resolution electron microscopy.


Initially, only one phase (Ti-rich) was identified in the APT dataset when exploring titanium iso-composition surfaces. This was in contrast to the electron microscopy work, which clearly showed a high density of two types of precipitates (with different crystal structures) within the matrix. After two weeks manual exploring the APT data, it was possible to identify two types of precipitates. These showed very similar compositions (both Ti rich but with slight variation in the Al content) and overlap. Subsequently, it took approximately \SI{5}{\hour} of GUI work to sort through the precipitates  based on their composition and relative shape (spherical vs. slightly more elliptical)  and assign them different colors for visualizing each precipitate type. This manual approach introduces significant bias into the analysis as the user has to ultimately make a semi-quantitative judgement which precipitate in the reconstruction corresponds to which phase. Finally, the three most representative precipitates were used to estimate the average composition of each phase. However, unintentional user bias can skew the results as the most representative precipitates which are taken for the composition analysis are in reality usually the most extreme cases in terms of composition.

As an example how this user story could be supported, we performed two exemplar high-throughput runs for titanium ($\varphi_{Ti} = $ \SIrange{50}{100}{\atpercent}) and aluminium ($\varphi_{Al} = $ \SIrange{1}{15}{\atpercent}) iso-composition surfaces (\SI{0.1}{\atpercent} step for both runs) for the same dataset. For each run the titanium, aluminium, vanadium, and molybdenum composition was computed for each closed interior object (using a convex hull model for the edge of the dataset). The high-throughput runs yield sets of tables of atom-type-resolved compositions in an \hdf{} file. Using a laptop and four threads the results were available after \swisswatch{1}{39} wall-clock time without further manual interaction. With all computational geometry and compositional data accessible and deterministically reproducible in an automated manner there is a clear benefit and complementary value of using automated analyses for initial assessing a dataset. Such assessment can support users with offering guidance where making traditional GUI-based analyses are of additional value.

\subsection{3D characterization of co-precipitating phases}
\label{CaseStudy6}
As the sixth and last case study, we will discuss how the results for the surface meshes of \gammaxx{} and \gammax{} precipitates (from the previous case study) can be further processed via \intersector{} to yield automated and rigorous quantitative analyses of the precipitates' spatial arrangement and eventual inclusion in (sets of) so-called coprecipitate configurations. In the first and especially the second case study we discussed how \intersector{} enables a quantification of an object's shape and composition dependency as a function of e.g. $\varphi$. Now, we will extend such analyses to entire sets of precipitates for collision and proximity respectively. We quantify all cases where \gammaxx{} precipitates are lying in specific proximity or are having contact with other \gammax{} precipitates. We should mention that the example of studying surface meshes which here are \gammaxx{} and \gammax{} precipitates is again an example that could equally be applied for objects representing other microstructural features such as other phases, or fragments of grains, crystals or polyhedra-composite-based descriptions of crystal defect ensembles.

Two high-throughput runs were performed: The first took all surface meshes of \gammaxx{} precipitates while the second run took all meshes of \gammax{} precipitates. Exemplarily we discuss the configurations for objects within two specific of the runs from the fifth case study. Those for parameter values $l = \sigma = \SI{1.0}{\nano\meter}, \varphi_{Nb} = \SI{7.6}{\atpercent}$, $\varphi_{Al+Ti} = \SI{6.1}{\atpercent}$, and $d_{prx} = \SI{1.0}{\nano\meter}$ as the maximum proximity threshold distance between objects to qualify still as neighbors.

\begin{figure}[h!]
\centering
	\includegraphics[draft=false,width=1.0\textwidth]{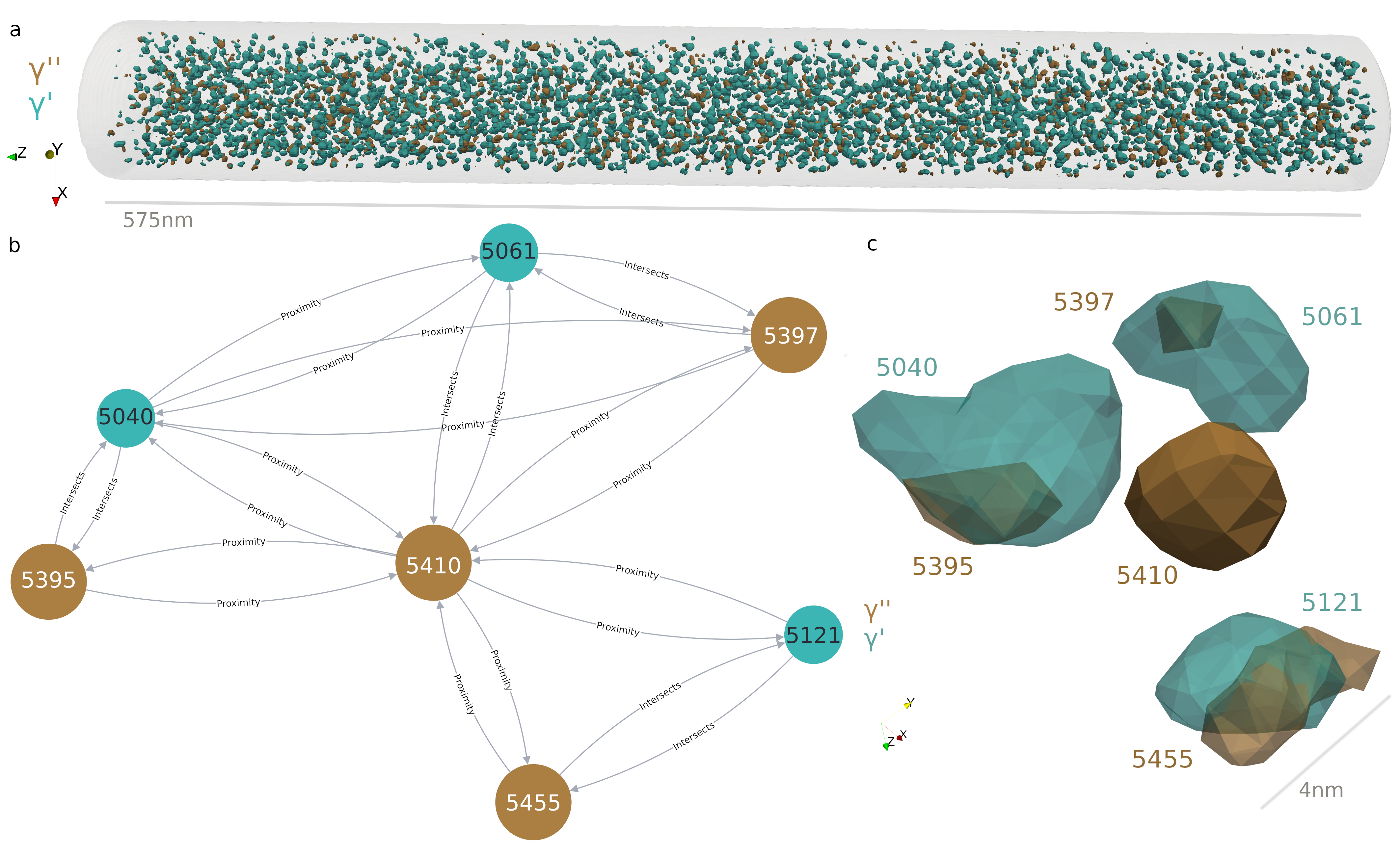}
\caption{Spatial analyses of object collisions and proximity with \intersector{} yield a graph-based quantification of coprecipitate configurations. Sub-figure a) shows a composite of two 3D surface mesh ensembles for one exemplar pair of runs from one of the high-throughput studies ($l = \sigma = \SI{1.0}{\nano\meter}, \varphi_{Nb} = \SI{7.6}{\atpercent}, \varphi_{Al+Ti} = \SI{6.1}{\atpercent}$, $d_{prx} \leq \SI{1.0}{\nano\meter}$). The composite shows the precipitates of all \gammaxx{} (brownish color) and \gammax{} (turquoise color) precipitates in the dataset. Sub-figure b) shows an exemplar sub-graph which resolves all collision and proximity relations between an exemplar chosen target object (\gammaxx{}, $ID = 5410$) to its neighbors. This is a coprecipitate configuration. Sub-figure c) visualizes the target and its neighbors via their surface meshes.}
\label{FigSydneyExemplarCaseOverview}
\end{figure}

Figures \ref{FigSydneyExemplarCaseOverview} display examples for the results of such \intersector{} analyses. The figures report how graph analytics and 3D visualization can support getting a detailed understanding of coprecipitation (configurations). For the selected parameters the analyses returned $2384$ \gammaxx{} and $3634$ \gammax{} interior objects Fig. \ref{FigSydneyExemplarCaseOverview}a, respectively. 

The first case study on edge models of irradiated steel inspected collisions between convex hulls with themselves. Now, however, pairings of objects from different sets were processed. Specifically, three such set combinations were processed: First, the \gammaxx{} meshes against themselves, second the \gammax{} meshes against themselves, and third the \gammaxx{} meshes were compared against the \gammax{} meshes (all for the above-mentioned parameterization $l, \sigma, {\varphi}_{Nb}, {\varphi}_{Al+Ti}, d_{prx}$). 

Individually, each analysis yields a list of disjoint collision pairs and disjoint proximity pairs. Each list of such pairs can be interpreted as a set of bidirectional relations between object pairs. For example, Fig. \ref{FigSydneyExemplarCaseOverview}b and Fig. \ref{FigSydneyExemplarCaseOverview}c reveal that \gammaxx{} object 5395 intersects with \gammax{} object 5040. Note that for a given $l, \sigma, {\varphi}_{Nb}, {\varphi}_{Al+Ti}$ the list of proximity relations depends only on the value of $d_{prx}$ in this example. The larger it is the proximity distance the less restrictive one is with how close two objects are to still be considered as a member of the (same) coprecipitate configuration. 

The most interesting case is when all three lists are combined into a single graph Fig. \ref{FigSydneyExemplarCaseOverview}b because such a graph describes all collision or proximity relations between all members of a given pair of object sets for a specific proximity threshold $d_{prx}$. Subsequently, sub-graphs can be extracted to characterize each coprecipitate configuration via graph analytics.

We should explore the benefits and potential limitations of such a description of coprecipitation: A clear benefit of the graph-based description is that qualifying a configuration is possible through counting disjoint relations. For instance, an object without any collision or proximity relations is a monolith. An object with only one relation (collision or proximity) is a partner in a duplet. Evidently, sub-graph analyses are a viable approach to disentangle all possible intricate configurations and thus avoid ambiguities or inconsistencies such as double counting of relations.

Figure \ref{FigSydneyExemplarCaseOverview} substantiates that this is beneficial in particular for studying the more complicated configurations with multiple objects. However, Fig. \ref{FigSydneyExemplarCaseOverview}c pinpoints that there are also challenges: Namely, one can attribute multiple qualitative types to a given sub-graph: In fact, does object 5410 for instance qualify as a monolith because it does not collide on its neighbors or does it qualify as a quadruplet, or even septet, given that it has six disjoint objects in proximity? We learn that qualitative configuration analyses like those in \cite{Rielli2020,Rielli2021} can be formulated more rigorously with our approach.

Furthermore, we learn that analyzing coprecipitation via iso-surface-based meshes faces the challenge that neighboring objects can show internal collision relations. Examples are shown in Fig. \ref{FigSydneyExemplarCaseOverview}c ($\{5395, 5040\}, \{5397, 5061\}, \{5455, 5121\}$). These internal collisions are a consequence that the analysis logically relates multiple interpretations of iso-surfaces (here of a region in the dataset to represent either a portion of \gammaxx{} or of \gammax{} precipitate volume). This is problematic because if crystal structure information were available, one would expect that there can be only one thermodynamic phase at the same location, i.e. inside an overlapping contour about a set of atoms. Evidently, researchers have to make specific assumptions how they treat and distinguish such configurations when all information available is composition based. For the example here, possible solutions are to consider only collisions, or to treat colliding objects with neighbors of the target like the above three pairs as a single object; and thus assume object 5410 is of quadruplet type.

In summary, we have substantiated how our approach and \intersector{}, combined with graph analytics, give researchers additional novel opportunities for making assessments - in an automated manner. We should note that it is possible to inject further qualitative (or quantitative) pieces of information into the graph, if available. This input can be based on detailed computational geometry analyses: For instance, we could combine the approach in this case study with the tessellation-based segmentation of the first case study to tessellate the ions in the vicinity of the above-discussed precipitates.

By inspecting the topology of the resulting labelled 3D Voronoi cell network, it is possible then not only to detect how objects are arranged but also in which cases they form specific configurations. Equipped with such capabilities it is possible to answer if a triplet \{\gammaxx{}/\gammax{}/\gammaxx{}\} is a sandwich, i.e. a multilayered precipitate and to clarifying how the adjoining \gammaxx{} objects collide. The Voronoi cell composite would then give a 3D description of the hetero-phase junctions as accurate as it is possible for point cloud data without having crystal structure information available and when using experiments with less than ab-initio accuracy and/or precision for resolving true atom positions.

\begin{figure}[h!]
\centering
	\includegraphics[draft=false,width=1.0\textwidth]{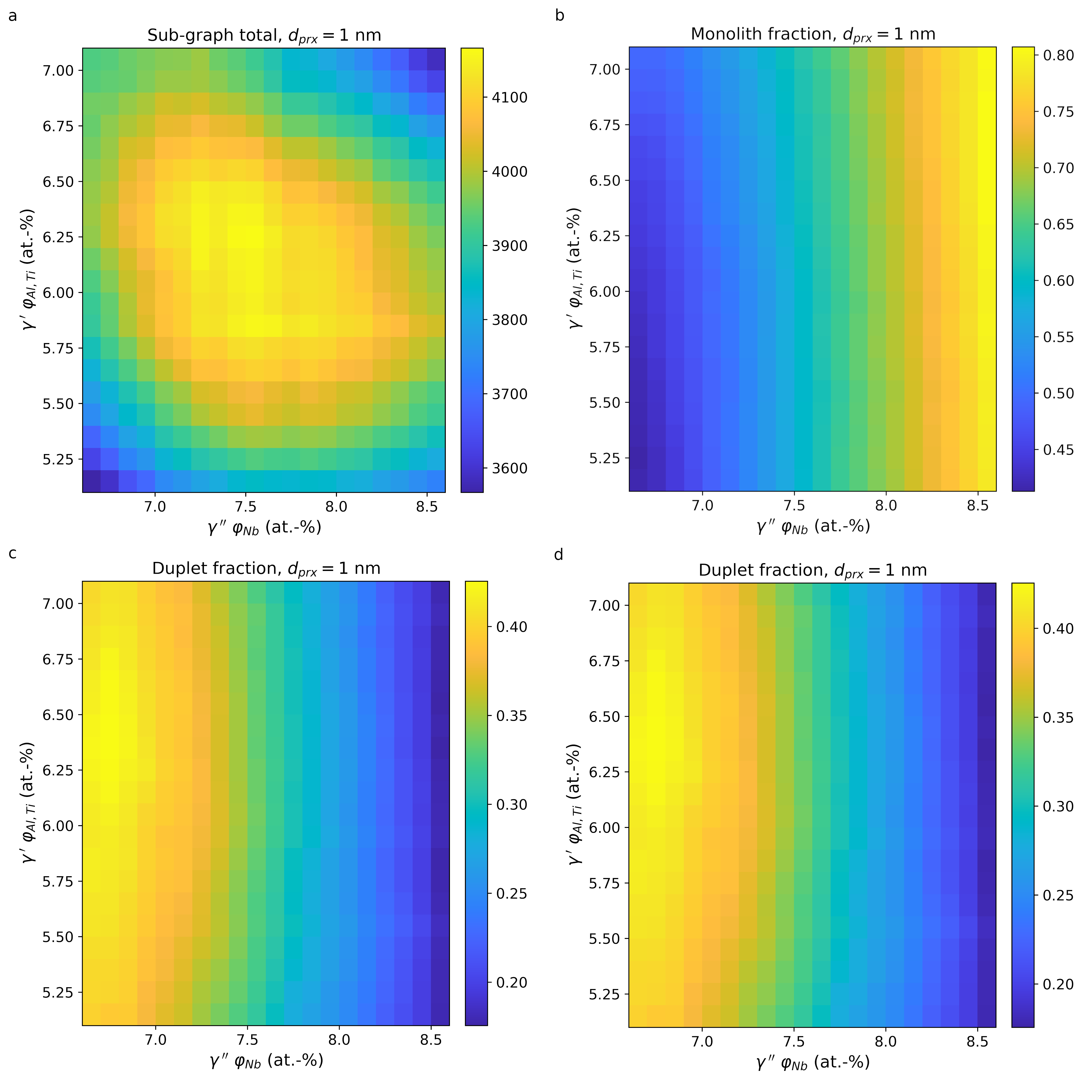}
\caption{With high-throughput quantification workflow using \intersector{} and \autoreporter{}, coprecipitate configurations can be analyzed to inspect how different pairings of runs result in distinguishable collisions and proximity relations between objects. By counting the number of disjoint relations, it is possible to categorize different configurations. Each such configuration is a sub-graph representing a group of objects with no collisions on members of other sub-graphs, i.e. distances $>d_{prx}$ to neighbors. The sub-figures show exemplar results for objects which either collide or lie within $d_{prx} = \SI{1.0}{\nano\meter}$ proximity. Specifically, sub-figure a) shows the total number of disjoint configurations. Each configuration is a disconnected sub-graph (one of which is shown in Fig. \ref{FigSydneyExemplarCaseOverview}). Sub-figures b), c) and d) report the fraction how many of these sub-graphs qualify as monoliths, duplets, and triplets, respectively. }
\label{FigSydneyExemplarCoprecipitation}
\end{figure}

Finally, we want to exemplify that automation of these detailed configuration analyses via scripting and combining analysis yields high-throughput quantification of the sensitivity of how many configurations (monoliths, duplets, and others) exist for a given pairing of object sets as a function of parameterization. The idea is to extend the above analyses of two object sets to combinations of pairs of object sets. Figures \ref{FigSydneyExemplarCoprecipitation} show the results as heat maps:  Each combination of ${\varphi}_{Nb}$ and ${\varphi}_{Al+Ti}$ yields an own collision and proximity graph which can be post-processed to quantify how many disjoint configurations exist (monoliths, duplets, and triplets).

Figure \ref{FigSydneyExemplarCoprecipitation}a shows an island of stability that matches with the low parameter sensitivity of \gammax{} and \gammaxx{} for their respective least sensible regions - compare with Figs. \ref{FigSydneyEng}a, and \ref{FigSydneyEng}b. The stronger the combination of ${\varphi}_{Nb}$ and ${\varphi}_{Al+Ti}$ deviates from the reference combination (${\varphi}_{Nb} = \SI{7.6}{\atpercent}, {\varphi}_{Al+Ti} = \SI{6.1}{\atpercent}$) the more likely the respective objects are in different locations in the tomographic reconstruction. In effect, the ensemble is increasingly poorer in registration which results in a lower number of detectable collision or proximity sub-graphs. Without a rigorous automation such analyses would be even more tedious to perform than the already very tedious high-throughput characterization of iso-surfaces.



\subsection{Scalability}
\label{CaseStudyBenchmark}
We close the study with exploring the numerical performance of the tools and probe their functioning for large datasets. All tools are equipped with functionalities for monitoring how much wall-clock time a tool spends in different code sections. The first benchmarks are with a dataset which is a tomographic reconstruction of a measured Sm-Co-Fe-Zr-Cu hard magnet specimen of Polin et al. \cite{Polin2021,Saxena2021}. The point cloud has \SI{104.9e6}{} ions. 

Profiling data were collected for \ranger{} (get ion type for each ion), \surfacer{} (compute convex hull of the dataset), \distancer{} tool (compute shortest Euclidean distance of each ion to the triangle mesh of the convex hull), and \nanochem{}. Ion positions were delocalized ($l = \sigma = \SI{1}{\nano\meter}$, $7^3$ kernel) before collecting iso-composition surfaces for $\varphi_{Zr} = $ \SIrange{5}{70}{\atpercent} with \SI{1}{\atpercent} step. Microstructural objects were reconstructed for each iso-surface, checked for closure and intersection with the edge of the dataset, characterized for their shape, volume, and which ions they contain. The analyses were executed on a computing node of the \talos{} computer cluster (see e.g. \cite{Kuehbach2020MSMSE} for details). Runs were scheduled exclusively with one MPI process which spawned between \SIrange{1}{40}{} OpenMP threads. This represents a typical scenario of using a workstation rather than a laptop.

\begin{figure}[h!]
\centering
    \quad
	\includegraphics[draft=false,width=0.5\textwidth]{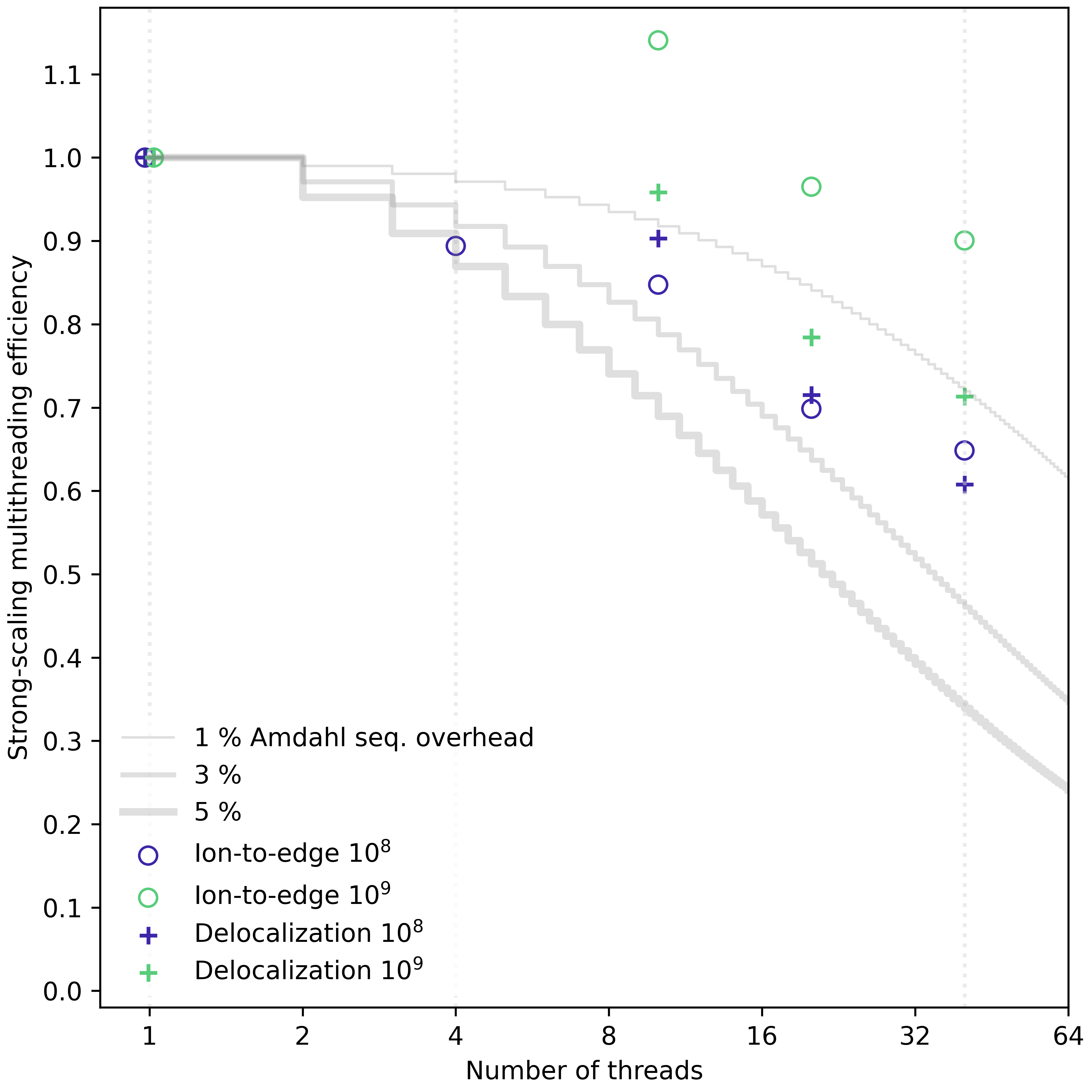} 
\caption{Strong-scaling multithreading efficiency for the \distancer{} and \nanochem{} tool when processing datasets with $10^8$ and $10^9$ ions respectively. Grey staircase curves report which efficiency would be naively expected according to Amdahl's law.}
\label{FigBenchmarks}
\end{figure}

The total wall-clock timings for sequential execution as the reference were as follows (including I/O): \ranger{} took less than \SI{9}{\second}, the \hdf{} file with the convex hull (5200 triangles) was ready after \SI{2}{\minute} and \SI{26}{\second}. Computing the distances for all ions to the hull took \SI{12}{\minute}. Object reconstruction and related pre-processing took \swisswatch{1}{42}; of which the delocalization was the most time-consuming task accounting for \SI{92}{\percent} of the total time. Figure \ref{FigBenchmarks} documents the strong-scaling multithreading performance achieved when solving the same tasks with multiple cores. Using 40 threads, the object reconstruction is finished in less than \SI{6}{\minute}. Ion distances are ready after \SI{30}{\second}.


Finally, we would like to report the successful application of the tools for analyzing a dataset containing one billion ions (\SI{1.074e9}{}), which is one among the largest datasets so far measured with APM worldwide. The atom probe specimen to this dataset was prepared from a Fe-12Cr-0.3C steel sample \cite{Belde2016}. We used the same four tools and above-mentioned questions when analyzing this dataset, but quantified iron instead of zirconium iso-composition surfaces (\SIrange{5}{70}{\atpercent} with \SI{1}{\atpercent} step). \talos{} was used to machine off the tasks with one MPI process which spawned at most 40 threads. Nodes were used exclusively. 


The reference wall-clock timings for sequential execution were as follows (including I/O): \ranger{} took less than \SI{2}{\minute}, the \hdf{} file with the convex hull (21836 triangles) was ready after \SI{34}{\minute}. Computing the distances for all ions to the hull took \swisswatch{4}{20}. Object reconstruction and related pre-processing took \SI{18}{\hour}. The delocalization took \SI{97}{\percent} of the total time. Like in previous benchmarks, we see again an order of magnitude improvement thanks to multithreading: When using 40 threads, ranging was accomplished in less than \SI{1}{\minute}, all distances were available after \SI{7}{\minute}, and the entire analysis with \nanochem{} reduced to \SI{1}{\hour}. 

We should mention that the main reason for the reduction of the computing time was due to time consuming tasks like delocalization and distance computations. Figure \ref{FigBenchmarks} summarizes that both tasks are efficiently strong-scaling with multithreading. Comparing the execution of the same analysis with 40 threads to sequential analyses indicates \SI{70}{\percent} efficiency for delocalization and more than \SI{90}{\percent} efficiency for distancing. We point the interested reader to the supplementary material where we provide a more detailed discussion about the results of these benchmarks.



In summary, we exemplified how open-source software tools can be combined to deliver a toolbox which gives atom probers and scientists interested in studying point cloud data efficient, automated, and scriptable opportunities for reproducible analyses in hitherto seldom explored parameter spaces. The sensitivity of atom probe numerical results can now be addressed by taking advantage of open-source tools and transparent data management practices to motivate future work towards a more detailed quantitative understanding and better interoperability between open-source and commercial software tools.


\section{Data availability}
The workflows of the use cases are shared. Exemplar notebooks for case studies \ref{CaseStudy2}, \ref{CaseStudy4}, \ref{CaseStudy5}, \ref{CaseStudy6}, and a generic one for an oxide-dispersion-strengthened steel \cite{Wang2019,Kuehbach2021MM} are available as \jupyters{}. These can be used as templates for getting started with the tools. Notebooks for the other use cases and numerical results in addition to those in the supplementary material are available from MK upon serious request. Readers who are interested in getting access to specific reconstructed point cloud and associated range files are kindly asked to contact the respective co-authors directly: BJ (case study \ref{CaseStudy1}), \cite{Kuehbach2021MM} (case study \ref{CaseStudy2}), DM (case study \ref{CaseStudy3}), AR/SK/LR (case study \ref{CaseStudy4}), VR/SP (nickel base super alloy part of case studies \ref{CaseStudy5}, \ref{CaseStudy6}), SA/AS (titanium part of case study \ref{CaseStudy5}), and AS (benchmarks \ref{CaseStudyBenchmark}) respectively. A table of SHA256 checksums for all range and reconstruction files is provided in the supplementary material (Tab. \ref{tab:sha256sums}).

\section{Code availability}
\sloppy
This work modified previously reported tools \cite{Kuehbach2021NPJ,Kuehbach2021JAC} and delivers new tools to the paraprobe-toolbox. \\
The source code and tutorial-style \jupyters{} are available via the following repository:
\begin{itemize}
    \item \href{http://gitlab.com/paraprobe/paraprobe-toolbox.git}{http://gitlab.com/paraprobe/paraprobe-toolbox.git}
\end{itemize}
Only these repositories will be maintained in the future.
\fussy

\section{Acknowledgements}
MK gratefully acknowledges the funding and computing time grants (on \talos{}) through BiGmax, the Max-Planck-Society's Research Network on Big-Data-Driven Materials Science, the Fritz-Haber-Institute of the Max-Planck-Society (not only for offering a laptop), and the FAIRmat team (for discussions and exchange of ideas on building software tools for FAIR research data management that motivated the \nexus{} application definitions). The work on implementing software tools for FAIR experimental microscopy methods is funded by the Deutsche Forschungsgemeinschaft (DFG, German Research Foundation) – project 460197019. MK and SB are funded by the FAIRmat consortium.

SP's research is funded by the Australian Research Council Linkage Project LP190101169 and the UNSW Scientia Fellowship scheme. SP and VR thank Dr. Takanori Sato at the Australian Centre for Microscopy and Microanalysis at the University of Sydney and acknowledge use of their facilities via Microscopy Australia. BJ would like to acknowledge funding from the EPSRC program grant MIDAS (EP/S01702X/1) for the study of irradiation damage in zirconium alloys. Part of this work was supported by Rolls-Royce Plc. and the atom probe facilities at the University of Oxford are funded by the EPSRC (EP/M022803/1). SA would like to acknowledge the financial support by the Alexander von Humboldt Foundation. AR, SK, and LR's research was funded in whole, or in part, by the Austrian Science Fund (FWF) [P 34179-N].

We gratefully acknowledge the contribution of the industry partners who enabled the exchange of original datasets in the course of the study. MK would like to thank Nikita Polin (MPIE) for offering the hard magnet benchmark dataset, AS for the inspiring discussions during the testing of the toolbox, Christoph Freysoldt (MPIE) for making editorial suggestions on the manuscript and exchanging experiences on software benchmarking, Sherjeel Shabih for support with containerizing the toolbox for \nomad{}, Christoph Koch for support (both currently at the Humboldt-Universit\"at zu Berlin), and Alisson Kwiatkowski da Silva, Leigh Stephenson, and Shyam Katnagallu (all three also MPIE) for offering access to the billion ion benchmark dataset. MK thanks . 


\section{Author contributions}
MK leads the \paraprobe{} project. He designed the tools, implemented the code, performed the analyses, and wrote the majority of the manuscript. The co-authors inspired the use cases and contributed to the work by offering access to their datasets, discussing the results, and contributing suggestions as well as editorial advice to the respective case studies: BJ (case study \ref{CaseStudy1}), DM (case study \ref{CaseStudy3}), AR/SK/LR (case study \ref{CaseStudy4}), VR/SP (nickel base super alloy part of case studies \ref{CaseStudy5}, \ref{CaseStudy6}), and SA/AS (titanium part of case study \ref{CaseStudy5}. AS contributed to tool testing and ran all benchmarks on \talos{} (case study \ref{CaseStudyBenchmark}). VR contributed additionally by successfully testing the tools on Windows 11 using WSL2, the Windows Subsystem for Linux.

\section{Competing interests statement}
The authors declare they have no competing interests.

\newpage

\section*{Supplementary Material}
\subsection*{Results and discussion}
\paragraph{2.2 Automated composition profiling for closed objects}
We implemented two different approaches for instructing the marching cube algorithm to process triangles at the edge of the dataset. These decisions have quantitative effects for the representation of microstructural features close to and at the edge of the dataset. With the first method we process each voxel of the delocalization grid with the marching cubes algorithm. With the second method we exclude voxels that are close to the edge of the dataset to spatially restrict the placing of triangles.

The most frequent case is that scientists are interested in composition quantification. In this case the composition $c$ is $c := \frac{N_{trg}}{N_{total}}$ with a number of target atoms $N_{trg}$ over the total number of atoms $N_{total}$ for each voxel (\SI{}{\atpercent}). Delocalization converts counts into intensities $\frac{a}{b}$ with $a \geq 0, b > 0$, $a, b \in \mathcal{R}$ and $a \leq b$, with $a$ representing $N_{trg}$ and $b$ representing $N_{total}$, respectively.

For voxels inside the dataset $a \land b \geq 1$ holds from which $\frac{a}{b} < a$ follows. For voxels at the edge or close to holes in the dataset, though, the opposite holds ($\frac{a}{b} \geq a$). Consequently, using the first strategy an object becomes also triangulated towards the edge of the dataset (see Fig. 3c). This effectively closes in many cases the triangle surface patch of an objects towards the edge. Thereby, the respective iso-surface facets protrude out of the edge and point cloud (see blue wireframe in Fig. 3c) though. These parts of an object's mesh are biased and especially sensitive to the delocalization parameters. An alternative is to evaluate the intensity ($b \leq 1$) for each voxel; and thus inform the marching cubes algorithm about which voxels are so close to the edge of the dataset that they better should not be processed. This correction method yields open objects at the edge of the dataset like the surface patch which is shown in Fig. 5a. These object representations match qualitatively closer to the results from commercial software (\ivas{}, \apsuite{}).



\paragraph{2.7 Scalability}
We learn that the delocalization is the primary target for future code optimization. There are several possible strategies: We should note that we have not implemented the two-step fast-Fourier-transformation-based approach that is presumably used in \ivas{} and \apsuite{} \cite{Larson2013}. Load imbalance is a key reason for limited scalability especially for the delocalization because we distribute blocks of voxel slabs along the longest axis of the specimen. This requires compromises, though, as to how equally the ions are distributed across the threads. Load imbalance is especially relevant when benchmarking the processing of microstructural objects. We found that not only the total number of objects but also the distribution of geometrical primitives (triangles, points) differs strongly between iso-surfaces. We learned that further work is necessary before substantiated statements about the strong-scaling efficiency of these code portions can be made. The processing time for each object and task has to be assessed in relation to the cardinality of the object set, the individual shape and size of the objects and their spatial arrangement. Finally, it should also be mentioned that modern processors use adaptive protocols for setting the clock cycling to values which guard the processor against overheating. This can affect with which effective clock cycle especially a series of short running processes gets executed.

\subsection*{Methods}
\paragraph{A comment on the usage of graphical user interface (GUI)-based software in atom probe}
We respect that GUI-based workflows are important for scientists, for most because the usage of such tools is often intuitive. What is conceptually behind GUI-based software is a set of predefined algorithms and functionalities which the users compose into a workflow as they interact with the software. A common limitation with how this user experience is currently implemented in many commercial software tools, though, is that the documentation of these, indeed often very flexible workflow compositing system, is that many of the intermediate numerical results and associated metadata are not exportable if at all tracked. This is relevant, though, for FAIR-compliant research especially with respect to the repeatability and/or reproducibility of such workflows. There are several likely reasons why such limitations of on-the-fly documentation systems of GUI-based software exist. There are technical intricacies and, given a GUIs flexibility, the need to have a solution that works ideally for arbitrary usage pattern. Both challenges represent implementation efforts, and thus increase development costs.

We envision that the role of open-source software tools like the paraprobe-toolbox is in the supporting of established GUI-based workflows within the e.g. atom probe microscopy community and related disciplines. Specifically, it is the offering of automation capabilities for making data-driven assessments. These can support scientists with identifying where investing further time and detail into e.g. traditional GUI-based workflows is worthwhile which are then eventually more conveniently performed via GUIs. Ideally, though, our work is understood in such a way that with making research more compliant with the FAIR data stewardship principles, there is also an increasing interest, a tremendous set of opportunities, and need to offer scientists tools which also substantially stronger automate the collecting of metadata. Ideally, this can incentivize a stronger working together between vendors of proprietary software and developers of (open-source) software to build effective and efficient tools for a FAIR research data infrastructure for the materials science communities \cite{Scheffler2022}.


\paragraph{Geometrical models for the edge of a dataset and distance-based filtering}
To enable fast distancing of point clouds to arbitrarily spatially arranged triangle sets, we modified the tool with adding an iterative preprocessing stage: First, the point cloud is coarsely discretized. Second, an extremal distance is computed for each voxel. This distance is successively refined to an eventually shorter extremal distance per voxel. These distances can then guide the subsequent distancing of each ion through offering on average shorter querying distances so that the bounded volume hierarchy queries result in less triangle candidates; and thus faster distancing.

Added mesh statistics and mesh inspection functionalities of \surfacer{}, including closure tests delivered also a more detailed understanding of algorithms for filtering point clouds such as the one reported in \cite{Kuehbach2021NPJ}. This algorithm reduces numerical costs when computing \ashapes{} for large datasets. The key idea is to compute the \ashape{} for a version of the point cloud where interior points are filtered out. The resulting \ashapes{} pick up differently curved regions in the dataset but have the disadvantage that the set of (exterior) triangles is neither necessarily watertight nor necessarily free of regions with double surface patches or other internal structure. 

Details depend on the $\alpha$ value. It is possible that the \ashape{} has double surface patches, i.e. regions with inner and (oftentimes approximately parallel) outer surface patches. For ions in the proximity, or in between such double surface patches, the closest triangle of the \ashape{} can be a triangle of an inner patch. In this case, though, setting a minimum distance of the ion to the edge of the dataset can cause a stronger than necessary removal of ions. This reduces the significance of the analysis in e.g. spatial statistics applications.

In this work, we learned from inspecting the exported interior tetrahedra for each \ashape{} as a function of $\alpha$ that this spatial filtering technique needs improvement. Namely, an additional filtering step so that the eventually created interior portions of double surface patches are removed. One opportunity is to evaluate the distance of the triangles to the ion density field that gets computed during the spatial filtering. Interior portions of double surface patches have usually a larger distance to the edge of the dataset than their exterior portions. This could be explored in the future to further improve the spatial filtering algorithm and thus improve the quality of \ashape{}-based models for serving as models for the edge of a point cloud.

\paragraph{Delocalization}
Paraprobe-nanochem implements a multi-threaded kernel density estimation. Specifically, a truncated anisotropic 3D Gaussian delocalization kernel is centered at each reconstructed ion position and Eq. \ref{EquationDelocalization} evaluated

\begin{equation}
A\cdot\int_{-aL}^{+aL}\int_{-aL}^{+aL}\int_{-aL}^{+aL}e^{-[(x^2/(2\sigma_x^2))+(y^2/(2\sigma_x^2))+(z^2/(2(0.5\sigma_x)^2))]}dxdydz = 1
\label{EquationDelocalization}
\end{equation}

to quantify the respective delocalized ion intensities in a ${[(2a)+1]}^3$ kernel for each voxel. We set $\mu_x = \mu_y = \mu_z = 0$ and assume that $a \in \mathcal{N}, a > 0$ is the kernel half-size, $L \in \mathcal{R}, L > 0$ the edge length of (cubic) voxel, and $\sigma_z = 0.5\sigma_x$ with $\sigma_x = \sigma_y := \sigma \in \mathcal{R}$, $\sigma > 0$ the kernel width, and $x, y, z$ the reconstructed ion positions. Using the symmetry of the error function yields analytical solutions for the kernel contributions to each voxel that can be evaluated numerically using double precision. The implementation could be replaced in the future by a fast-Fourier-transformation-based approach to improve numerical efficiency. The renormalization constant $A$ accounts for ion intensity contributions beyond the tails of the kernel to prevent a leakage of the total ion intensity.


The implementation enables different intensities to be computed, including total number of atoms per voxel, element composition (\SI{}{\atpercent}), or concentration (in atoms \SI{}{\per\nano\meter^3}). A preprocessing identifies the location of the ions based on which the voxel grid is spatially partitioned along the specimen main axis ($z$). Subsequently, the delocalization is computed in such a manner that each thread processes approximately the same total number of ions. Using overlapping halo regions assures a correct accounting over adjacent regions of the voxel grid.

\paragraph{Spatial correlation analyses - implementation details}
The above analysis capabilities require robust computational geometry methods to detect if triangulated polyhedra collide or intersect volumetrically. For convex polyhedra the Gilbert–Johnson–Keerthi (GJK) algorithm \cite{Gilbert2007} is one such. This algorithm is implemented for instance in the \pqp{} library \cite{Larsen2020} and elsewhere \cite{Hornus2015,Magalhaes2017a,Magalhaes2019}. For non-convex polyhedra it is possible to use a tessellation of the polyhedra into sets of individually all convex smaller polyhedra. Subsequently, these can be evaluated for collisions via e.g. the GJK algorithm to infer eventual collisions of the original non-convex polyhedra.

Eventually not as efficient as the GJK implementation in \pqp{} or the study of \cite{Magalhaes2019}, we tetrahedralize each surface mesh first using \tetgen{} \cite{Si2015,Kuehbach2020MSMSE}. Subsequently, we evaluate tetrahedron-tetrahedron intersections \cite{Si2015,Ganovelli2002,Hornus2015}. Evidently, using such an approach requires that each object (triangulated surface mesh) gets successfully tetrahedralized as a piecewise-linear complex \cite{Si2015}. Occasionally, we detected cases, though, where portions of the mesh were locally connected via non-trivial, almost point-like contacts. Although these situations are known to occur for MC \cite{Lewiner2003} and are characterizable, they can occasionally result in non-trivial-to-debug situations during tetrahedralization. Therefore, we implemented another strategy for detecting object collisions which checks if two objects share at least one ion with the same evaporation ID.

To compute the intersection volume of arbitrary polyhedra we use a strategy of tetrahedralizing each polyhedron and accumulate the volume of individual tetrahedron-tetrahedron-intersections. A numerical robust yet efficient algorithm for evaluating a de facto analytical intersection volume between two arbitrarily-shaped and -oriented tetrahedra is a challenging mathematical problem. To the best of our knowledge this has not yet been fully solved \cite{McCoid2021}.

Therefore, we use a more pragmatic approach which is to evaluate NEF tetrahedra intersections using functionalities of the \cgal{} library \cite{Hachenberger2007,Hachenberger2021}. We are aware that this approach has limitations which are not central to our work but worth to become improved in future work: One is that computing NEF polyhedra can be very costly. Another one is that meshed-based processing using floating point numbers, especially when decomposing a polyhedron into tetrahedra, can result in numerical inaccuracies, even though the NEF-based approach is conceptually capable of being much more accurate (usage of arbitrary floating point arithmetics provided).

For the \intersector{} tool we detect intersections between polyhedra via performing object collision tests rather than proximity analyses \cite{Hornus2015}. Proximity analyses were evaluated for objects which do not collide by using the analytical triangle-triangle distance computation capabilities of \pqp{} \cite{Larsen2020} and compare the shortest distance to $d_{prx}$. A hierarchy of axis-aligned bounding boxes was used to prune the search space of objects, triangles, and tetrahedra respectively. The post-processing of graph data was implemented in \python{}.

For this work we made a few assumptions during the implementation of \intersector{}: First, surface meshes can represent convex or non-convex polyhedra but must not have holes, self-intersections, or be numerically degenerated. Second, we accept a collection of sets only when it represents results from a linearly sampled sequence of $k$ values. For instance, a collection of objects from an iso-surface high-throughput characterization (via \nanochem{}) with increasing values of $\varphi$ in steps of $\Delta\varphi$ on $[\varphi_{min}, \varphi_{max}]$ is a suited candidate collection. To begin with, we implemented pair-wise comparisons between sets $k$ and $k+1$ using linearly space probing of parameter values. One could change this in the future for a non-linear spacing and still use the same $k\pm1$ traversing. Although we have not implemented it, allowing for larger jumps in $k$, to cover for instance relations between the state of an object between $k$ and $k\pm n$ with $n > 1$ is a possible generalization. 

We store all processed directed graphs in the \hdf{} output file of the \intersector{} run. This offers several benefits, such as programmatic options for automating the visualization of said analyses to support a detailed understanding, rendering, debugging, and counting of relations between nodes via graph analytics. Such analytics yield robust descriptors to qualify which objects are monoliths, duplets, triplets, or arbitrarily complex configurations as the paper reports. Recalling our motivation, \intersector{} delivers an automated alternative to otherwise very tedious manual analyses of coprecipitation phenomena \cite{Rielli2021} or two-dimensional slicing through images of discretized datasets \cite{Theska2019}, thus offering rigorous qualification and quantification. 

\paragraph{Implementation details}
The above-described tools were implemented in \cxx{} and a set of convenience \python{} tools. Third-party specialists libraries are called which have open-source licenses for academic usage. These were \cgal{} (v5.2.1) \cite{CGAL2021}, which we compiled with the Eigen (v3.3.9) \cite{Eigen2021} linear algebra, the Boost \cxx{} (v1.76.0) \cite{Schling2011} template library, and the GMP (v6.2.1) and MPFR (v4.0.2) numerical libraries. Additionally, we used code from \pqp{} (v1.3) \cite{Gottschalk1996,Larsen2000,Larsen2020} for computing triangle-triangle intersections. Surface meshes were tessellated with \tetgen{} (v1.5.1) \cite{Si2015,Kuehbach2020MSMSE}. Point clouds were tessellated with \voroxx{} (v0.4.6) \cite{Rycroft2009}. Bounded volume hierarchies were used to reduce the number of location and intersection queries for geometric primitives  \cite{Gottschalk1996,Kuehbach2021NPJ}. Graph analytics were instructed via \cypher{} scripting using \neofourj{} \cite{Neo4j2021} (v4.3.1) and \python{}. Data were written with the \hdf{} library (v1.12.0) \cite{HDF52018}.


In continuation of our previous work \cite{Kuehbach2021NPJ,Kuehbach2021JAC}, we parallelized the tools via an MPI/OpenMP work partitioning strategy. At the coarse layer work packages are distributed via the Message Passing interface (MPI) library. At the fine layer these work packages are split further to be solved via Open Multi-Processing (OpenMP) multithreading. We configured \paraprobe{} with cmake build tools (v3.19) and compiled with the GNU \cxx{} compiler (v7.5) using in most cases -O3 -march=native optimization. 

Except for the benchmarks and the molybdenum-hafnium case study, all case studies were executed on a laptop. Specifically, this was a Dell Latitude 5480 i7-782 with a four hyper-threading core pair CPU with \SI{32}{\gibi\byte} main memory plus a SanDisk X400 M.2 2280 SSD drive with \SI{512}{\gibi\byte}, all instructed by Ubuntu 18.04.5. We used four OpenMP threads, OMP\_PLACES=cores mapping, and one MPI process. We explore the strong-scaling multithreading efficiency with executing runs on the \talos{} system (see \cite{Kuehbach2020MSMSE,Kuehbach2021JAC} for details). Here, nodes were used exclusively with OMP\_PLACES=threads mapping. We compare wall-clock timing data of individual functions (I/O monitored separately) while processing the same datasets using from \SIrange{1}{40}{} OpenMP threads.

The molybdenum-hafnium case study was processed with a Dell XPS 15 9510 laptop with an Intel i9-11900H eight-core CPU with \SI{32}{\gibi\byte} main memory plus a SK Hynix 1TB PC711 NVMe disk, all instructed by Windows11 Pro. We used the Windows Subsystem for Linux (WSL2) running Ubuntu 20.04.4 LTS. Four of the eight cores and \SI{20}{\gibi\byte} main memory were allocated for the WSL2 part. Thereby the work also documents that the toolbox can be used not only on Linux. We would be happy to be contacted by interested individuals who would like to run the toolbox on Mac OS systems so that we can extend the range of applicable systems for the atom probe community.

\newpage
\paragraph{SHA256 checksums}
\begin{center}
\begin{table}[!htb]
\scriptsize
\caption{SHA256 checksums for the point cloud and range files for the datasets of each case study. Benchmarks are marked as (B) cases.}
\centering
\begin{tabular}{lll}
	\addlinespace[0.2em]
	\toprule
    \bf{File}          & \bf{Case study} & \bf{SHA256} \\ 
    \midrule
\textup{R33\_07490-v02.pos} & 2.1 & 322a24d780b54bdcea132a0014c152aa8a03b408d33ea25fc88ecfa379f2bd12 \\
\textup{R33\_07490-v02.cluster.rrng} & 2.1 & 289153bb0bc29679fd9a362e9acfd856b49b3795c58f8a0c9045d01ed442e621 \\
\textup{R33\_07490-v02.cluster.indexed.rrng} & 2.1 & 4b934716e274eb4b9a152aaf2f2452357dae0f1da10e55ac9862408241f6f1bc \\
\textup{R33\_07490-v02.cluster.indexed.pos} & 2.1 & 74cdde57f0dd34dd153f377898a67e35b62103f76fa7c053193833b548dfac73 \\
\textup{R21\_07575-v02.rrng} & 2.2 & fee7128f30ed88f7e115ec30fbbd1f3c0962d55772849d702cc3917056f61c93 \\
\textup{R21\_07575-v02.pos} & 2.2 & 66bab07cc649236bc63cec29bb67375e226edc4f92fde00313c6f203d43bf4ca \\
\textup{R5096\_41455-v02.pos} & 2.3 & b1f05fe36c0acee0ad91b46ba5f7b490638ed51f0455ada22df01bbb25b26950 \\
\textup{R5096\_41455-v02.rrng} & 2.3 & 154b4dac9aa8e86191c4746dd92f7faa88af2105494a24d3e3b9fdd04a9ace13 \\
\textup{R21\_08680-v02.pos} & 2.4 & be6287c45233de97f0376fd8e1e078f7cfa029bac8b1a4087936076a2ea0de51 \\
\textup{R21\_08680.rrng} & 2.4 & c75aead0e82d7e4149cf07192e734ee9765259259b9e7cba96f284d4a7ec7cea \\
\textup{R04\_22071.RRNG} & 2.5, 2.6 & 4cb6dc8da15412e33104fc64c6b2e6480484d08f8ca8641895fd00ea848e9f8f \\
\textup{R04\_22071.pos} & 2.5, 2.6  & 8040adffed5c1f9761cdc2da6b881625c9f9335f3e96c557374f8bae0dc89078 \\
\textup{R5006\_29110\_Top\_Level\_ROI.apt} & 2.5 & 614fdfd8b7dbb3e24decea5493a24df3e7b468062e854682f620a440c8e7dcd0 \\
\textup{R5006\_29110\_Top\_Level\_ROI.RRNG} & 2.5 & 30f11b1cfa5db8f1b3b22fffed9f922ae8a0c885d4c50d76538d452b8a19c6e9 \\
\textup{D1\_High\_Hc\_R5076\_52126.apt} & 2.7 (B) & b803a34aa6f3b87acbf9db152c17453789084278cef37237b898acb2ae405403 \\
\textup{D1\_High\_Hc\_R5076\_52126.RRNG} & 2.7 (B) & 3107ae9a46357267eb294d6df5b3819d42c4fe604a819eaa86864f3b7d067d61 \\
\textup{R76\_20169-v01.pos} & 2.7 (B) & 9a0c71984acb5290c5cb31e98ddd29e06d08da078817baa305b038cef278c237 \\
\textup{20169.RRNG} & 2.7 (B) & db9ad704afe141a1040e72698702bd1ef656c26b4b390aad4d6c6e4d20f58de8 \\
    \bottomrule
	\addlinespace[1em]
\end{tabular}
\label{tab:sha256sums}
\end{table}
\end{center}

\newpage
\section*{References}
\sloppy
\printbibliography[heading=none]{}

@Article{Hyde2000,
  author = {J. M. Hyde and C. A. English},
  title  = {{An Analysis of the Structure of Irradiation induced Cu-enriched Clusters in Low and High Nickel Welds}},
  doi    = {10.1557/proc-650-r6.6},
  pages  = {6-12},
  year = {2000},
  volume = {650},
  journal={MRS Proceedings},
}

@InProceedings{Ester1996,
  author = {M. Ester and H.-P. Kriegel and J. Sander and X. Xu},
  title  = {A Density-Based Algorithm for Discovering Clusters in Large Spatial Databases with Noise},
  pages  = {226-231},
  year = {1996},
  publisher = {AAAI Press},
  booktitle = {},
}

@InProceedings{Marquis2017,
author = {E. A. Marquis and V. Araullo-Peters and Y. Dong and A. Etienne and S. Fedotova and K. Fujii and K. Fukuya and E. Kuleshova and A. Lopez and A. London and S. Lozano-Perez and Y. Nagai and K. Nishida and B. Radiguet and D. Schreiber and N. Soneda and M. Thuvander and T. Toyama and F. Sefta and P. Chou},
title = {{On the Use of Density-Based Algorithms for the Analysis of Solute Clustering in Atom Probe Tomography Data}},
booktitle = {{Proceedings of the 18th International Conference on Environmental Degradation of Materials in Nuclear Power Systems – Water Reactors}},
pages = {2097-2113},
year = {2017},
doi = {10.1007/978-3-030-04639-2_141},
publisher = {Springer,
Cham},
}

@Article{Theska2020,
author = {F. Theska and K. Nomoto and F. Godor and B. Oberwinkler and A. Stanojevic and S. P. Ringer and S. Primig},
title = {{On the early stages of precipitation during direct ageing of Alloy 718}},
journal = {Acta Mater.},
volume = {188},
year  = {2020},
pages = {492–503},
doi = {10.1016/j.actamat.2020.02.034},
}

@article{Hornbuckle2015,
journal = {Ultramicroscopy},
volume = {159},
year = {2015},
pages = {346–353},
title = {{A procedure to create isoconcentration surfaces in
low-chemical-partitioning,
high-solute alloys}},
author = {B.C. Hornbuckle and M. Kapoor and G.B. Thompson},
doi = {10.1016/j.ultramic.2015.03.003},
}

@Article{Kuehbach2021NPJ,
  author  = {M. K\"uhbach and P. Bajaj and H. Zhao and M. H. \c{C}elik and E. A. J\"agle and B. Gault},
  title   = {{On strong-scaling and open-source tools for analyzing atom probe tomography data}},
  doi     = {10.1038/s41524-020-00486-1},
  pages   = {{21}},
  volume  = {7},
  journal = {npj Comput. Mater.},
  year    = {2021},
}

@Article{Wilkinson2016,
  author  = {M. D. Wilkinson and M. Dumontier and I. J. Aalbersberg and G. Appleton and M. Axton and A. Baak and N. Blomberg and J.-W. Boiten and L. B. {da Silva Santos} and P. E. Bourne and J. Bouwman and A. J. Brookes and T. Clark and M. Crosas and I. Dillo and O. Dumon and S. Edmunds and C. T. Evelo and R. Finkers and A. Gonzalez-Beltran and A. .G. Gray and P. Groth and C. Goble and J. S. Grethe and J. Heringa and P. A.C. {'}t Hoen and R. Hooft and T. Kuhn and R. Kok and J. Kok and S. J. Lusher and M. E. Martone and A. Mons and A. L. Packer and B. Persson and P. Rocca-Serra and M. Roos and R. {van Schaik} and S.-A. Sansone and E. Schultes and T. Sengstag and T. Slater and G. Strawn and M. A. Swertz and M. Thompson and J. {van der Lei} and E. {van Mulligen} and J. Velterop and A. Waagmeester and P. Wittenburg and K. Wolstencroft and J. Zhao and B. Mons},
  title   = {{The FAIR Guiding Principles for scientific data management and stewardship}},
  journal = {Sci. Data},
  volume = {3},
  year    = {2016},
  pages = {{160018}},
  doi     = {10.1038/sdata.2016.18},
}

@article{Barton2019,
author = {D. Barton and B. Hornbuckle and K. Darling and G. Thompson},
year = {2019},
title = {{The Influence of Isoconcentration Surface Selection in Quantitative Outputs from Proximity Histograms}},
journal = {Microsc. Microanal.},
volume = {25},
issue = {2},
pages = {401-409},
doi = {10.1017/S143192761900014X},
}

@article{Martin2015,
journal = {Mater. Sci. Technol.},
volume = {32},
issue = {3},
title = {{Insights into microstructural interfaces in aerospace alloys characterised by atom probe tomography}},
author = {T. L. Martin and A. Radecka and L. Sun and T. Simm and D. Dye and K. Perkins and B. Gault and M. Moody and P. A. J. Bagot},
pages = {232-241},
year = {2015},
doi = {10.1179/1743284715Y.0000000132},
}

@article{Hellman2000,
title = {{Analysis of Three-dimensional Atom-probe Data by the Proximity Histogram}},
year = {2000},
journal = {Microsc. Microanal.},
volume = {6},
issue = {5},
pages = {437-444},
doi = {10.1007/s100050010051},
author = {O. C. Hellman and J. Vandenbroucke and J. R\"using and D. Isheim and D. N. Seidman},
}

@article{Keutgen2020,
author = {J. Keutgen and A. London and O. Cojocaru-Mir\'{e}din},
year = {2020},
title = {{Solving Peak Overlaps for Proximity Histogram Analysis of Complex Interfaces for Atom Probe Tomography Data}},
volume = {27},
issue = {1},
journal = {Microsc. Microanal.},
pages = {28-35},
doi = {10.1017/S1431927620024800},
}

@Article{Ceguerra2019,
author = {A. V. Ceguerra and A. C. Day and S. P. Ringer},
title = {{Assessing the Spatial Accuracy of the
Reconstruction in Atom Probe Tomography and a New Calibratable Adaptive Reconstruction}},
journal = {Microsc. Microanal.},
volume = {25},
year = {2019},
pages = {309-319},
doi = {10.1017/S1431927619000369},
}

@Electronic{Hornus2015,
url = {http://hal.inria.fr/hal-01157239v1},
year = {2015},
author = {S. Hornus},
title = {A review of polyhedral intersection detection and new techniques},
note = {RR-8730, Inria Nancy - Grand Est (Villers-l\`{e}s-Nancy, France)},
}

@Article{Lorenz2021,
journal = {Acta Mater.},
volume = {221},
year = {2021},
pages = {{117393}},
title = {{Impact of the segregation energy spectrum on the enthalpy and entropy of segregation}},
author = {D. Scheiber and L. Romaner},
doi = {10.1016/j.actamat.2021.117393},
}

@Article{Han2016,
journal = {Acta Mater.},
volume = {104},
year = {2016},
pages = {259-273},
title = {{Grain-boundary metastability and its statistical properties}},
author = {J. Han and V. Vitek and D. J. Srolovitz},
doi = {10.1016/j.actamat.2015.11.035},
}

@Article{Scheiber2015,
journal = {Acta Mater.},
volume = {88},
year = {2015},
pages = {180-189},
title = {{Ab initio description of segregation and cohesion of grain boundaries in W–25 at.\% Re alloys}},
author = {D. Scheiber and V. I. Razumovskiy and P. Puschnig and R. Pippan nd L. Romaner},
doi = {10.1016/j.actamat.2014.12.053},
}

@Misc{Neo4j2021,
author = {{Neo4j}},
url = {http://neo4j.com/docs/},
year = {2021},
note = {(last accessed May 19, 2022)},
}

@Article{Hachenberger2007,
journal = {Comput. Geometry},
volume = {38},
issue = {1-2},
year = {2007},
pages = {64-99},
title = {{
Boolean operations on 3D selective Nef complexes: Data structure, algorithms, optimized implementation and experiments}},
author = {P. Hachenberger and L. Kettner and K. Mehlhorn},
doi = {10.1016/j.comgeo.2006.11.009},
}

@Article{Leitner2017,
title = {{On grain boundary segregation in molybdenum materials}},
author = {K. Leitner and P. J. Felfer and D. Holec and J. Cairney and W. Knabl and A. Lorich and H. Clemens and S. Primig},
journal = {Mater. Des.},
volume = {135},
pages = {204-212},
doi = {10.1016/j.matdes.2017.09.019},
year = {2017},
}

@Article{Ghamarian2020,
journal = {Ultramicroscopy},
year = {2020},
volume = {215},
pages = {{112996}},
doi = {10.1016/j.ultramic.2020.112996},
title = {{Morphological classification of dense objects in atom probe tomography data}},
author = {I. Ghamarian and L.-J. Yu and E. A. Marquis},
}

@Article{Mueller1956a,
author = {E. W. M\"uller},
journal = {J. Appl. Phys.},
title = {{Resolution of the Atomic Structure of a Metal Surface by the Field Ion Microscope}},
year = {1956},
volume = {27},
pages = {474-476},
doi = {10.1063/1.1722406},
}

@Book{Miller2000,
title = {{Atom Probe Tomography: Analysis at the Atomic Level}},
author = {M. K. Miller},
edition = {1},
year = {2000},
publisher = {Springer, New York},
doi = {10.1007/978-1-4615-4281-0},
}

@Book{Larson2013,
author = {D. J. Larson and T. J. Prosa and R. M. Ulfig and B. P. Geiser and T. F. Kelly},
title = {Local Electrode Atom
Probe Tomography},
year = {2013},
edition = {1},
publisher = {Springer,
New York},
doi = {10.1007/978-1-4614-8721-0},
}

@Article{Jain2013,
author = {A. Jain and S.P. Ong and G. Hautier and W. Chen and W. D. Richards and S. Dacek and S. Cholia and D. Gunter and D. Skinner and G. Ceder and K. A. Persson},
title = {{The Materials Project: A materials genome approach to accelerating materials innovation}},
journal = {APL Mater.},
year = {2013},
volume = {1},
number = {1},
pages = {{011002}},
doi = {10.1063/1.4812323},
}

@Article{Wang2019,
author = {J. Wang and D. K. Schreiber and N. Bailey and P. Hosemann and M. B. Toloczko},
title = {{The Application of the OPTICS Algorithm to Cluster Analysis in Atom Probe Tomography Data}},
journal = {Microsc. Microanal.},
year = {2019},
volume = {25},
pages = {338-348},
doi = {10.1017/S1431927618015386},
}

@Misc{Felfer2021IonList,
author = {P. Felfer},
title = {Ion list APT},
year = {2021},
url = {http://github.com/peterfelfer/Atom-Probe-Toolbox/blob/master/IonenlisteAPT.xlsx},
note = {(last accessed May 19, 2022)},
}

@Misc{Felfer2022b,
author = {P. Felfer and V. Dalbauer and B. Ott and M. Heller and C. Macauley and M. Weiser and {et} {al.}},
title = {Atom-Probe-Toolbox},
year = {2022},
url = {http://github.com/peterfelfer/Atom-Probe-Toolbox},
note = {(last accessed May 19, 2022)},
}

@Article{Still2021,
journal = {Microsc. Microanal.},
title = {{Alpha Shape Analysis (ASA) Framework for Post- Clustering Property Determination in Atom Probe Tomographic Data}},
author = {E. K. Still and D. K. Schreiber and J. Wang and P. Hosemann},
year = {2021},
doi = {10.1017/S1431927620024939},
pages = {1-21},
volume = {},
}

@Misc{Hyperspy2021,
title = {hyperspy},
author = {F. {de la Pe\~{n}a} and E. Prestat and V. T. Fauske and P. Burdet and J. L\"ahnemann and T. Furnival and P. Jokubauskas and M. Nord and T. Ostasevicius and K. E. MacArthur and D. N. Johnstone and M. Sarahan and T. aarholt and J. Taillon and V. Migunov and A. Eljarrat and J. Caron and T. Poon and S. Mazzucco and C. Francis and B. Martineau and S. Somnath and T. Slater and N. Tappy and M. Walls and N. Cautaerts and F. Winkler},
year = {2021},
doi = {10.5281/zenodo.5608741},
note = {(last accessed May 19,
2022)},
}

@Misc{AbTem2021,
title = {The abTEM code: transmission electron microscopy from first principles},
author = {J. Madsen and T. Susi},
year = {2021},
doi = {10.12688/openreseurope.13015.2},
note = {(last accessed May 19,
2022)},
}

@Misc{LiberTem2021,
title = {LiberTEM},
author = {A. Clausen and D. Weber and K. Ruzaeva and V. Migunov and A. Baburajan and A. Bahuleyan and M. Bryan and J. Caron and R. Chandra and S. Dey and S. Halder and D. S. katz and B. D. A. Levin and M. Nord and C. Ophus and S. Peter and L. Puskas and J. {van Schyndel} and J. Shin and S. Sunku and K. M\"uller-Caspary and R. E. Dunin-Borkowski},
year = {2021},
doi = {10.5281/zenodo.5547992},
note = {(last accessed May 19,
2022)},
}

@Article{Draxl2019,
author = {C. Draxl and M. Scheffler},
title = {{The NOMAD laboratory: from data sharing to artificial intelligence}},
journal = {J. Phys. Mater.},
year = {2019},
pages = {{036001}},
volume = {2},
doi = {10.1088/2515-7639/ab13bb},
}

@Article{Zhou2021a,
  author  = {X. Zhou and J. R. Mianroodi and A. Kwiatkowski da Silva and T. Koenig and G. B. Thompson and P. Shanthraj and D. Ponge and B. Gault and B. Svendsen and D. Raabe},
  title   = {{The hidden structure dependence of the chemical life of dislocations}},
  doi     = {10.1126/sciadv.abf0563},
  number  = {16},
  volume  = {7},
  journal = {Sci. Adv},
  year    = {2021},
}

@Article{Brehm2020,
author = {M. Brehm and M. Thomas and S. Gehrke and B. Kirchner},
title = {{TRAVIS – A Free Analyzer for Trajectories from Molecular Simulation}},
journal = {J. Chem. Phys.},
year = {2020},
volume = {152},
pages = {{164105}},
doi = {10.1063/5.0005078},
}

@Article{Casalino2021,
title = {{AI-driven multiscale simulations illuminate mechanisms of SARS-CoV-2 spike dynamics}},
author = {L. Casalino and A. C. Dommer and Z. Gaieb and E. P. Barros and T. Sztain and S.-H. Ahn and A. Trifan and A. Brace and A. T. Bogetti and A. Clyde and H. Ma and H. Lee and M. Turilli and S. Khalid and L. T. Chong and C. Simmerling and D. J. Hardy and J. D. C. Maia and J. C. Phillips and T. Kurth and A. C. Stern and L. Huang and J. D. McCalpin and M. Tatineni and T. Gibbs and J. E. Stone and S. Hah and A. Ramanathan and R. E. Amaro},
year = {2021},
doi = {10.1177/10943420211006452},
volume = {35},
pages = {432-451},
journal = {Int. J. High Perform. Comput. Appl.},
}

@Book{Kirkland2020,
author = {E. J. Kirkland},
year = {2020},
edition = {3},
doi = {10.1007/978-3-030-33260-0},
title = {{Advanced Computing in Electron Microscopy}},
publisher = {Springer, Cham},
}

@Book{Williams1996,
author = {D. B. Williams and C. B. Carter},
title = {Transmission Electron Microscopy},
year = {1996},
doi = {10.1007/978-1-4757-2519-3},
edition = {1},
publisher = {Springer, New York},
}

@Misc{Rielli2020,
author = {V. V. Rielli and F. Theska and F. Godor and A. Stanojevic and B. Oberwinkler and S. Primig},
title = {{High-Throughput Analysis of Precipitates in a Ni-based superalloy}},
url = {http://www.youtube.com/watch?v=YAKjdaxxlZk&list=PL-0gZJOh94SfNCVFaZm6WNGLc2BDkvRex&index=17},
year = {2020},
note = {{(last accessed May 19,
2022)}},
}

@Misc{Felfer2022a,
author = {P. Felfer and M. Weiser},
title = {{APT Winterschool day 3 - lecture 4 - (local) interfacial excess and interfacial excess mapping}},
url = {http://www.youtube.com/watch?v=gOwD8z_ZY5Y},
year = {2022},
note = {{(last accessed March 12, 2022)}},
}

@Misc{APTTC2020,
author = {D. Haley and A. Ceguerra and M. K\"uhbach and A. London and C. Oberdorfer and B. Radiguet and D. Reinhard and D. Isheim},
title = {{List of software tools for atom probe}},
year = {2021},
url = {http://docs.google.com/spreadsheets/d/1pGKUYOadY8vtEtvb9cMMZOWg7lHHkJ0SL8ew4CsSDVQ/edit#gid=0},
note = {(last accessed May 19, 2022)},
}

@article{Elzoka2020,
title = {{Enabling near-atomic–scale analysis of frozen water}},
author = {A. A. El-Zoka and S.-H. Kim and S. Deville and C. Newman and L. T. Stephenson and B. Gault},
journal = {Sci. Adv.},
year = {2020},
volume = {6},
number = {49},
doi = {10.1126/sciadv.abd6324},
}

@Article{London2017,
title={Single-Ion Deconvolution of Mass Peak Overlaps for Atom Probe Microscopy}, 
volume={23}, 
doi = {10.1017/S1431927616012782}, 
number={2}, 
journal={Microsc. Microanal.}, 
author={A. J. London and D. Haley and M. Moody},
year={2017}, 
pages={300–306},
}

@Article{Stephenson2007,
  author  = {L. T. Stephenson and M. P. Moody and P. V. Liddicoat and S. P. Ringer},
  title   = {{New Techniques for the Analysis of Fine-Scaled Clustering Phenomena within Atom Probe Tomography (APT) Data}},
  journal = {Microsc. Microanal.},
  year    = {2007},
  volume  = {13},
  number  = {6},
  pages   = {448-463},
  doi     = {10.1017/s1431927607070900},
}

@article{Cerjak2017,
title = {{Atom probe investigations on temper embrittlement and reversible temper embrittlement in S 690 steel weld metal}},
author = {H. Cerjak and F. Mendez-Martin and M. Domakova},
journal = {Sci. Technol. Weld. Joining},
year = {2017},
pages = {140-147},
volume = {23},
issue = {2},
doi = {10.1080/13621718.2017.1346771},
}

@Article{Peng2019,
author = {Z. Peng and Y. Lu and C. Hatzoglou and A. K. {da Silva} and F. Vurpillot and D. Ponge and D. Raabe and B. Gault},
title = {An Automated Computational Approach for Complete In-Plane Compositional Interface Analysis by Atom Probe Tomography},
journal = {Microsc. Microanal.},
year = {2019},
volume = {25},
pages = {389-400},
doi = {10.1017/S1431927618016112},
}

@Article{Leitner2018,
journal = {Mater. Des},
volume = {142},
year = {2018},
pages = {36-43},
title = {How grain boundary chemistry controls the fracture mode of molybdenum},
author = {K. Leitner and D. Scheiber and S. Jakoba and S. Primig and H. Clemens and E. Povoden-Karadeniz and L. Romaner},
doi = {10.1016/j.matdes.2018.01.012},
}

@Article{Raabe2014,
journal = {Curr. Opin. Solid State Mater. Sci.},
volume = {18},
year = {2014},
pages = {253-261},
title = {Grain boundary segregation engineering in metallic alloys: A pathway to the design of interfaces}}

@Article{Seidman1991,
journal = {Mater. Sc. Eng. A},
volume = {137},
year = {1991},
pages = {57-67},
title = {{Solute-atom segregation at internal interfaces on an atomic scale: atom-probe experiments and computer simulations}},
author = {D. N. Seidman},
doi = {10.1016/0921-5093(91)90318-H},
}

@Misc{Collins2021,
title = {ExPaNDS ontologies v1.0},
author = {S. P. Collins and S. {da Gra\c{c}a Ramos} and D. Iyayi and H. G\"orzig and A. {Gonz\'{a}lez Beltr\'{a}n} and A. Ashton and S. Egli and C. Minotti},
url = {https://zenodo.org/record/4806026#.YnwBz1RBy3_},
year = {2021},
note = {(last accessed May 19, 2022)},
}

@Electronic{Goerzig2022,
title = {NeXusOntology},
author = {H. G\"orzig and P. R. Jemian and B. Watts},
url = {https://github.com/nexusformat/NeXusOntology},
year = {2022},
note = {(last accessed May 19, 2022)},
}

@Article{Hu1992,
journal = {Scr. Metall. et Mater.},
volume = {27},
year = {1992},
pages = {693-698},
title = {{}Relationship of chemical composition and structure on an atomic scale for metal/metal interfaces: The W(Re) system}}

@Misc{FAIRmatNeXusProposal2022,
title = {{NeXus FAIRmat proposal}},
author = {M. Aeschlimann and M. Albrecht and S. Brockhauser and C. Emminger and R. Ernstorfer and M. Greiner and M. Grundmann and T. Haraszti and M. Hartelt and L. Himanen and C. Koch and M. K\"uhbach and T. Pincelli and L. Rettig and M. Scheidgen and S. Shabih and C. Sturm
and {the} {FAIRmat team members}},
url = {https://fairmat-experimental.github.io/nexus-fairmat-proposal},
year = {2022},
note = {(last accessed May 19, 2022)},
}

@Article{Wei2019,
title = {3D nanostructural characterisation of grain boundaries in atom probe data utilising machine learning methods},
author = {Y. Wei and Z. Peng and M. K\"uhbach and A. breen and M. Legros and M. Larranaga and F. Mompiou and B. Gault},
journal = {PLOS One},
year = {2019},
doi = {10.1371/journal.pone.0225041},
}

@Article{Lorensen1987,
author = {W. E. Lorensen and H. E. Cline},
journal = {ACM SIGGRAPH Computer Graphics},
title = {{Marching Cubes: A High Resolution 3D Surface Construction Algorithm}},
volume = {21},
year = {1987},
pages = {163-169},
doi = {10.1145/37401.37422},
}

@Article{Lorensen2020,
title = {{History of the Marching Cubes Algorithm}},
author = {W. E. Lorensen},
journal = {IEEE Comput. Graph.},
year = {2020},
pages = {8-15},
volume = {40},
doi = {10.1109/MCG.2020.2971284},
}

@Article{Lewiner2003,
author = {T. Lewiner and H. Lopes and A. Wilson Vieria and G. Tavares},
title = {{Efficient implementation of Marching Cubes’ cases with topological guarantees}},
journal = {J. Graph. Tools},
year = {2003},
volume = {8},
pages = {},
doi = {10.1080/10867651.2003.10487582},
}

@Article{ORourke1985,
author = {J. O\'{}Rourke},
title = {{Finding Minimal Enclosing Boxes}},
journal = {Int. J. Comput. Inf.},
year = {1985},
volume = {14},
doi = {10.1007/BF00991005},
pages = {183-199},
}

@Article{Chang2011,
author = {C.-T. Chang and B. Gorissen and S. Melchior},
journal = {ACM Trans. Graph.},
pages = {{122}},
doi = {10.1145/2019627.2019641},
volume = {30},
title = {Fast oriented bounding box optimization on the rotation group {$SO(3,\mathcal{R})$}},
year = {2011},
}

@Article{Barequet2001,
author = {G. Barequet and S. Har-Peled},
journal = {J. Algorithms},
volume = {38},
year = {2001},
pages = {91-109},
title = {{Efficiently Approximating the Minimum-Volume Bounding Box of a Point Set in Three Dimensions}},
doi = {10.1006/jagm.2000.1127},
}

@Misc{Melchior2018,
author = {B. Gorissen and S. Melchior},
title = {{OptimalOBB}},
year = {2018},
url = {http://github.com/chadogome/OptimalOBB},
note = {(last accessed May 19, 2022)},
}

@Misc{Engwirda2020,
author = {D. Engwirda},
title = {{Source code repository to construct generalised Voronoi-type dual meshes in MATLAB.}},
year = {2020},
url = {http://github.com/dengwirda/dual-mesh},
note = {(last accessed May 19, 2022)},
}

@Article{Lazar2017,
journal = {Model. Simul. Mater. Sci. Eng.},
title = {{VoroTop: Voronoi cell topology visualization and analysis toolkit}},
author = {E. A. Lazar},
year = {2017},
volume = {26},
pages = {{015011}},
doi = {10.1088/1361-651X/aa9a01},
}

@Article{Lazar2015,
title = {{Topological framework for local structure analysis in condensed matter}},
author = {E. A. Lazar and J. Han and D. J. Srolovitz},
year = {2015},
volume = {112},
pages = {E5769-E5776},
number = {43},
doi = {10.1073/pnas.1505788112},
journal = {PNAS},
}

@Article{KorteKerzel2021,
journal = {Int. Mater. Rev.},
title = {Defect phases – thermodynamics and impact on material properties},
author = {S. Korte-Kerzel and T. Hickel and L. Huber and D. Raabe and S. Sandl\"obes-Haut and M. Todorova and J. Neugebauer},
doi = {10.1080/09506608.2021.1930734},
volume = {0},
number = {0},
pages = {1-29},
year = {2021},
}

@Article{Degeuser2020,
doi = {10.1016/j.actamat.2020.02.023},
author = {F. {de Geuser} and B. Gault},
journal = {Acta Mater.},
title = {{Metrology of small particles and solute clusters by atom probe tomography}},
volume = {188},
pages = {406-415},
year = {2020},
}

@Article{Krakauer1993,
author = {B. W. Krakauer and D. N. Seidman},
title = {{Absolute atomic-scale measurements of the Gibbsian interfacial excess of solute at internal interfaces}},
journal = {Phys. Rev. B},
year = {1993},
pages = {6724-6728},
volume = {48},
doi = {10.1103/PhysRevB.48.6724},
}

@Article{Felfer2015a,
  author  = {P. Felfer and B. Scherrer and J. Demeulmeester and W. Vandervoorst and J. M. Cairney},
  title   = {{Mapping interfacial excess in atom probe data}},
  journal = {Ultramicroscopy},
  year    = {2015},
  volume  = {{159}},
  pages   = {438-444},
  doi     = {10.1016/j.ultramic.2015.06.002},
}

@Article{Sarker2020,
journal = {Appl. Phys. Lett.},
volume = {116},
issue = {15},
doi = {10.1063/5.0002049},
year = {2020},
author = {J. Sarker and S. Broderick and A. F. M. {Anhar Uddin Bhuiyan} and Z. Feng and H. Zhao and B. Mazumder},
title = {{}A combined approach of atom probe tomography and unsupervised machine learning to understand phase transformation in ${({\textnormal{Al}}_{\textnormal{x}}{\textnormal{Ga}}_{1-x})}_{2}{\textnormal{O}}_{3}$}}

@InProceedings{Kazhdan2006,
  author    = {M. Kazhdan and M. Bolitho and H. Hoppe},
  title     = {Poisson Surface Reconstruction},
  booktitle = {Eurographics Symposium on Geometry Processing},
  year      = {2006},
  editor    = {K. Polthier and A. Sheffer},
}

@Article{Momma2011,
  author  = {K. Momma and F. Izumi},
  title   = {VESTA 3 for three-dimensional visualization of crystal,
volumetric and morphology data},
  year    = {2011},
  volume  = {44},
  pages   = {1272-1276},
  doi     = {10.1107/s0021889811038970},
  journal = {J. Appl. Crystallogr.},
}

@Article{Edelsbrunner1983,
  author  = {H. Edelsbrunner and D. G. Kirkpatrick and R. Seidel},
  title   = {{On the Shape of a Set of Points in the Plane}},
  journal = {IEEE Trans. Inf. Theory},
  year    = {1983},
  volume  = {29},
  number  = {4},
  pages   = {551-559},
  doi     = {10.1109/tit.1983.1056714},
}

@Article{Edelsbrunner1994,
  author  = {H. Edelsbrunner and E. P. M\"ucke},
  title   = {{Three-Dimensional Alpha Shapes}},
  journal = {ACM Trans. Graph.},
  year    = {1994},
  volume  = {13},
  number  = {1},
  pages   = {43-72},
  doi     = {10.1145/174462.156635},
}

@InProceedings{Hellman2000a,
  author    = {O. C. Hellman and J. A. Vandenbroucke and J. R\"using and D. Isheim and D. N. Seidman},
  title     = {{Identification of 2D Boundaries from 3D Atom Probe Data,
and Spatial Correlation of Atomic Distributions with Interfaces}},
  booktitle = {{Proceedings of the MRS Fall Meeting 1999: Symposia A/C – Multiscale Phenomena in Materials - Experiments in Modeling}},
  year      = {2000},
  editor    = {B. Devincre and D. H. Lassila and R. Phillips and I. M. Robertson},
  volume    = {578},
  pages     = {395-400},
  doi = {10.1557/PROC-578-395},
}

@Article{Jenkins2020b,
journal = {Microsc. Microanal.},
volume = {26},
issue = {2},
year = {2020},
title = {{Reflections on the Analysis of Interfaces and Grain Boundaries by Atom Probe Tomography}},
author = {B. M. Jenkins and F. Danoix and M. Goun\'{e} and P. A. J. Bagot and Z. Peng and M. P. Moody and B. Gault},
pages = {247-257},
doi = {10.1017/S1431927620000197},
}

@Article{Belde2016,
journal = {Acta Mat.},
volume = {113},
year = {2016},
pages = {19-31},
title = {{Vessel microstructure design: A new approach for site-specific core-shell micromechanical tailoring of TRIP-assisted ultra-high strength steels}},
author = {M. Belde and H. Springer and D. Raabe},
doi = {10.1016/j.actamat.2016.04.051},
}

@Article{Jenkins2020c,
journal = {Materialia},
volume = {11},
year = {2020},
pages = {{100717}},
title = {{The effect of composition variations on the response of steels subjected to high fluence neutron irradiation}},
author = {B. M. Jenkins and J. O. Douglas and N. Almirall and N. Riddle and P. A. J. Bagot and J. M. Hyde and G. R. Odette and M. P. Moody},
doi = {10.1016/j.mtla.2020.100717},
}

@Article{Mayweg2021b,
journal = {Acta Mat.},
volume = {215},
year = {2021},
pages = {{117048}},
title = {{Correlation between grain size and carbon content in white etching areas in bearings}},
author = {D. Mayweg and L. Morsdorf and Y. Li and M. Herbig}}

@Book{Mayweg2021a,
title = {{Microstructural characterization of white etching cracks in 100Cr6 bearing steel with emphasis on the role of carbon}},
author = {D. Mayweg},
edition = {1},
publisher = {Dissertation, RWTH Aachen University},
school = {10.18154/RWTH-2021-02467},
year = {2021},
}

@Article{Zheng2020,
journal = {Scr. Mater.},
volume = {176},
year = {2020},
pages = {7-11},
title = {Shuffle-induced modulated structure and heating-induced ordering in the metastable $\beta$-titanium alloy, Ti-5Al-5Mo-5V-3Cr},
author = {Y. Zheng and A. Stoichko and Q. Feng and B. Rajarshi and B. Dipankar F. L. Hamish},
doi = {10.1016/j.scriptamat.2019.09.027},
}

@Article{Kuehbach2021JAC,
title = {{On Open and Strong-Scaling Tools for Atom Probe Crystallography: High-Throughput Methods for Indexing Crystal Structure and Orientation}},
author = {M. K\"uhbach and  M. Kasemer and B. Gault and A. Breen},
journal = {J. Appl. Crystallogr.},
year = {2021},
volume = {54},
pages = {1490-1508},
doi = {10.1107/S1600576721008578},
}

@Article{Xu2015,
author = {R. Xu and C.-C. Chen and L. Wu and M. C. Scott and W. Theis and C. Ophus and M. bartels and Y. Yang and H. Ramezani-Dakhel and M. R. Sawaya and H. Heinz and L. D. Marks and P. Ercius and J. Miao},
year = {2015},
title = {Three-dimensional coordinates of individual atoms in materials revealed by electron tomography},
journal = {Nat. Mater.},
volume = {14},
pages = {1099-1103},
issue = {11},
doi = {10.1038/nmat4426},
}

@article{Morgado2021,
title = {Revealing atomic-scale vacancy-solute interaction in nickel},
journal = {Scr. Mater.},
volume = {203},
pages = {114036},
year = {2021},
doi = {10.1016/j.scriptamat.2021.114036},
author = {F. F. Morgado and S. Katnagallu and C. Freysoldt and B. Klaes and F. Vurpillot and J. Neugebauer and D. Raabe and S. Neumeier and B. Gault and L. T. Stephenson},
}

@Article{Hellman2003,
  author  = {O. C. Hellman and J. B. {du Rivage} and D. N. Seidman},
  title   = {{Efficient sampling for three-dimensional atom probe microscopy data}},
  journal = {Ultramicroscopy},
  year    = {2003},
  volume  = {95},
  pages   = {199-205},
  doi     = {10.1016/S0304-3991(02)00317-0},
}

@Article{Newman2006,
author = {T. S. Newman and H. Yi},
title = {{A survey of the marching cubes algorithm}},
journal = {Comput. Graph.},
volume = {30},
year = {2006},
pages = {854-879},
doi = {10.1016/j.cag.2006.07.021},
}

@Article{Gilbert2007,
  author  = {Gilbert,
M. and Vurpillot,
F. and Vella,
A. and Bernas,
H. and Deconihout,
B.},
  title   = {Some aspects of the silicon behaviour under femtosecond pulsed laser field evaporation},
  journal = {Ultramicroscopy},
  year    = {2007},
  volume  = {107},
  number  = {9},
  pages   = {767-772},
  doi     = {10.1016/j.ultramic.2007.02.027},
  type    = {Journal Article},
}

@Article{Karnesky2007,
  author  = {R. A. Karnesky and D. Isheim and D. N. Seidman},
  title   = {Direct measurement of two-dimensional and three-dimensional interprecipitate distance distributions from atom-probe tomographic reconstructions},
  journal = {Appl. Phys. Lett},
  year    = {2007},
  volume  = {91},
  number  = {1},
  doi     = {10.1063/1.2753097},
}

@Article{Ulfig2009b,
  author  = {R. M. Ulfig and E. Oltman and D. J. Larson and V. S. Smentkowski},
  title   = {{Improvements in Three-Dimensional Compositional Analysis of Complex Alloys}},
  journal = {Microsc. Microanal.},
  year    = {2009},
  volume  = {15},
  pages   = {294-295},
  doi     = {10.1017/s1431927609098791},
}

@Article{Haley2009,
  author  = {D. Haley and T. Petersen and G. Barton and S. P. Ringer},
  title   = {{Influence of field evaporation on Radial Distribution Functions in Atom Probe Tomography}},
  journal = {Philos. Mag.},
  year    = {2009},
  volume  = {89},
  number  = {11},
  pages   = {925-943},
  doi     = {10.1080/14786430902821610},
}

@Article{Gault2012a,
  author  = {Gault,
B. and Moody,
M. P. and Cairney,
J. M. and Ringer,
S. P.},
  title   = {Atom probe crystallography},
  journal = {Mater. Today},
  year    = {2012},
  volume  = {15},
  number  = {9},
  pages   = {378-386},
  doi     = {10.1016/s1369-7021(12)70164-5},
  type    = {Journal Article},
}

@Article{Torres2011,
journal = {Ultramicroscopy},
volume = {111},
issue = {6},
year = {2011},
pages = {464-468},
title = {{The influence of voxel size on atom probe tomography data}},
author = {K. L. Torres and M. Daniil and M. A. Willard and G. B. Thompson},
doi = {10.1016/j.ultramic.2011.01.001},
}

@Article{Felfer2012b,
  author  = {P. Felfer and C. R. Killmore and J. G. Williams and K. R. Carpenter and S. P. Ringer and J. M. Cairney},
  title   = {A quantitative atom probe study of the Nb excess at prior austenite grain boundaries in a Nb microalloyed strip-cast steel},
  journal = {Acta Mater.},
  year    = {2012},
  volume  = {60},
  pages   = {5049-5055},
  doi     = {10.1016/j.actamat.2012.06.013},
}

@Article{Felfer2013,
  author  = {P. Felfer and A. Ceguerra and S. Ringer and J. Cairney},
  title   = {{Applying computational geometry techniques for advanced feature analysis in atom probe data}},
  journal = {Ultramicroscopy},
  year    = {2013},
  volume  = {132},
  pages   = {100-106},
  doi     = {10.1016/j.ultramic.2013.03.004},
}

@Book{Botsch2010,
title = {{Polygon Mesh Processing}},
author = {M. Botsch and L. Kobbelt and M. Pauly and P. Alliez and B. Levy},
edition = {1},
year = {2010},
publisher = {CRC Press, New York},
doi = {10.1201/b10688},
}

@Article{Ceguerra2013,
  author  = {A. V. Ceguerra and A. J. Breen and L. T. Stephenson and P. J. Felfer and V. J. Araullo-Peters and P. V. Liddicoat and X. Y. Cui and L. Yao and D. Haley and M. P. Moody and B. Gault and J. M. Cairney and S. P. Ringer},
  title   = {{The rise of computational techniques in atom probe microscopy}},
  journal = {Curr. Opin. Solid. State. Mater. Sci.},
  year    = {2013},
  volume  = {17},
  pages   = {224-235},
  doi     = {10.1016/j.cossms.2013.09.006},
}

@Article{Kontis2021,
journal = {Scr. Mater.},
volume = {194},
year = {2021},
pages = {{113626}},
title = {{Interactions of solutes with crystal defects: A new dynamic design parameter for advanced alloys}},
author = {P. Kontis},
doi = {10.1016/j.scriptamat.2020.11.011},
}

@Article{Si2015,
  author  = {H. Si},
  title   = {TetGen,
a Delaunay-Based Quality Tetrahedral Mesh Generator},
  journal = {ACM Trans. Math. Softw.},
  year    = {2015},
  volume  = {41},
  number  = {2},
  pages   = {{}},
  doi     = {10.1145/2629697},
}

@Article{Oberdorfer2015,
  author    = {C. Oberdorfer and S. M. Eich and M. L\"utkemeyer and G. Schmitz},
  title     = {Applications of a versatile modelling approach to 3D atom probe simulations},
  journal   = {Ultramicroscopy},
  year      = {2015},
  volume    = {159},
  pages     = {184-194},
  doi       = {10.1016/j.ultramic.2015.02.008},
}

@Article{Felfer2015b,
  author  = {P. Felfer and A. V. Ceguerra and S. P. Ringer and J. M. Cairney},
  title   = {{Detecting and extracting clusters in atom probe data: A simple,
automated method using Voronoi cells}},
  journal = {Ultramicroscopy},
  year    = {2015},
  volume  = {150},
  pages   = {30-36},
  doi     = {10.1016/j.ultramic.2014.11.015},
}

@Article{Kontis2018,
  author = {P. Kontis and Z. Li and D. M. Collins and J. Cormier and D. Raabe and B. Gault},
  title = {{The effect of chromium and cobalt segregation at dislocations on nickel-based superalloys}},
  journal = {Scr. Mater.},
  year = {2018},
  volume = {145},
  pages = {76-80},
  doi = {10.1016/j.scriptamat.2017.10.005},
}

@Book{Lefebvre2016,
authors = {W. Lefebvre-Ulrikson and F. Vurpillot and X. Sauvage},
title = {{Atom Probe Tomography Put Theory Into Practice}},
year = {2016},
publisher = {Elsevier,
Amsterdam},
edition = {1},
doi = {10.1016/B978-0-12-804647-0.01001-9},
}

@Article{Felfer2016,
  author  = {P. Felfer and J. Cairney},
  title   = {{A computational geometry framework for the optimisation of atom probe reconstructions}},
  journal = {Ultramicroscopy},
  year    = {2016},
  volume  = {169},
  pages   = {62-68},
  doi     = {10.1016/j.ultramic.2016.07.008},
}

@Article{Katnagallu2018,
  author  = {S. Katnagallu and B. Gault and B. Grabowski and J. Neugebauer and D. Raabe and A. Nematollahi},
  title   = {{Advanced data mining in field ion microscopy}},
  journal = {Mater. Charact.},
  volume = {146},
  pages = {307-318},
  year    = {2018},
  doi     = {10.1016/j.matchar.2018.02.040},
}

@Article{Miller1986,
  author  = {M. K. Miller and J. A. Horton},
  title   = {{An Atom Probe Field-Ion Microscope Study of Boron Decorated Boundaries in Ni$_\textnormal{3}$Al}},
  journal = {Scr. Metall.},
  year    = {1986},
  volume  = {20},
  pages   = {789-792},
  doi     = {10.1016/0036-9748(86)90513-2},
}

@Online{HDF52018,
  author = {{The HDF Group}},
  title  = {{Hierarchical Data Format,
version 5}},
  year   = {2018},
  note   = {http://www.hdfgroup.org/HDF5/},
}

@Article{Janssen2019,
author = {J. Jan{\ss}en and S. Surendralal and Y. Lysogorskiy and M. Torodova and T. Hickel and R. Drautz and J. Neugebauer},
title = {{pyiron: An integrated development environment for computational materials science}},
year = {2019},
journal = {Comput. Mater. Sc.},
volume = {163},
pages = {24-36},
doi = {10.1016/j.commatsci.2018.07.043},
}

@Article{Scheffler2022,
author = {M. Scheffler and M. Aeschlimann and M. Albrecht and T. Bereau and H.-J. Bungartz and C. Felser and M. Greiner and A. Gro{\ss} and C. T. Koch and K. Kremer and W. E. Nagel and M. Scheidgen and C. W\"oll and C. Draxl},
year = {2022},
title = {{FAIR data enabling new horizons for materials research}},
journal = {Nature},
pages = {635-642},
volume = {604},
doi = {10.1038/s41586-022-04501-x},
}

@Inbook{Draxl2020,
author={C. Draxl and M. Scheffler},
editor= {W. Andreoni and S. Yip},
title = {{Big Data-Driven Materials Science and Its FAIR Data Infrastructure}},
booktitle={Handbook of Materials Modeling: Methods: Theory and Modeling},
year={2020},
publisher={Springer International Publishing, Cham},
pages={49-73},
doi = {10.1007/978-3-319-44677-6_104},
}

@Misc{Haley2018b,
  author = {D. Haley and A. Ceguerra},
  urldate   = {2018},
  title  = {{3Depict - Visualisation \& Analysis for Atom Probe}},
  url    = {http://threedepict.sourceforge.net},
  note = {(last accessed on May 19, 2022)},
}

@Article{Stukowski2010a,
  journal = {Model. Simul. Mat. Sci. Eng.},
  title = {{Extracting dislocations and non-dislocation crystal defects from atomistic simulation data}},
  author = {A. Stukowski and K. Albe},
  year = {2010},
  volume = {18},
  pages = {{085001}},
  doi = {10.1088/0965-0393/18/8/085001},
}

@Article{Rycroft2009,
  author  = {C. Rycroft},
  title   = {{VORO$++$: A three-dimensional Voronoi cell library in C$++$}},
  journal = {Chaos},
  year    = {2009},
  volume  = {19},
  pages = {},
  doi     = {10.2172/946741},
}

@Article{Kuehbach2020MSMSE,
  author  = {M. K\"uhbach and F. Roters},
  title   = {{Quantification of 3D spatial correlations between state variables and distances to the grain boundary network in full-field crystal plasticity spectral method simulations}},
  journal = {Model. Simul. Mat. Sci. Eng.},
  year    = {2020},
  volume = {28},
  pages = {{055005}},
  doi  = {10.1088/1361-651X/ab7f8},
}

@Article{Spencer2017,
title = {Reconciling grain growth and shear-coupled grain boundary migration},
author = {L. T. Spencer and K. Chen and J. Han and P. K. Purohit and D. J. Srolovitz},
journal = {Nat. Commun.},
year = {2017},
pages = {{1764}},
volume = {8},
doi = {10.1038/s41467-017-01889-3},
}

@Electronic{Magalhaes2017a,
  author = {S. V. Magalh{\~{a}}es},
  title  = {{MeshIntersection}},
  url    = {http://github.com/sallesviana/MeshIntersection.git},
  year   = {2017},
}

@Misc{Jenkins2019,
author = {A. J. London and B. Jenkins},
title = {Cluster-Alpha-Edge},
url = {https://github.com/andyroo101/Cluster-Alpha-Edge.git},
year = {2019},
note = {(last accessed May 19, 2022)},
}

@Article{Bolte2006,
  author       = {S. Bolte and F. P. Cordeli{\`{e}}res},
  journaltitle = {J. Microsc.},
  title        = {A guided tour into subcellular colocalization analysis in light microscopy},
  doi          = {10.1111/j.1365-2818.2006.01706.x},
  number       = {3},
  pages        = {213-232},
  volume       = {224},
  year         = {2006},
}

@Article{Zinchuk2008,
  author       = {V. Zinchuk and O. Zinchuk},
  journaltitle = {Curr. Protoc. Cell Biol.},
  title        = {Quantitative Colocalization Analysis of Confocal Fluorescence Microscopy Images},
  doi          = {10.1002/0471143030.cb0419s39},
  number       = {1},
  volume       = {39},
  year         = {2008},
}

@Article{Manders1992,
  author       = {E.M. Manders and J. Stap and G.J. Brakenhoff and R. van Driel and J.A. Aten},
  date         = {1992-11},
  journaltitle = {J. Cell Sci.},
  title        = {Dynamics of three-dimensional replication patterns during the S-phase,
analysed by double labelling of {DNA} and confocal microscopy},
  doi          = {10.1242/jcs.103.3.857},
  number       = {3},
  pages        = {857-862},
  volume       = {103},
}

@Article{Gilles2017,
  author       = {J.-F. Gilles and M. {Dos Santos} and T. Boudier and S. Bolte and N. Heck},
  journaltitle = {Methods},
  title        = {{DiAna},
an {ImageJ} tool for object-based 3D co-localization and distance analysis},
  doi          = {10.1016/j.ymeth.2016.11.016},
  pages        = {55-64},
  volume       = {115},
  year         = {2017},
}

@Article{Costes2004,
  author       = {S. V. Costes and D. Daelemans and E. H. Cho and Z. Dobbin and G. Pavlakis and S. Lockett},
  journaltitle = {Biophys. J.},
  title        = {Automatic and Quantitative Measurement of Protein-Protein Colocalization in Live Cells},
  doi          = {10.1529/biophysj.103.038422},
  number       = {6},
  pages        = {3993-4003},
  volume       = {86},
  year         = {2004},
}

@InCollection{Magalhaes2019,
  author    = {S. V. G. {de Magalh{\~{a}}es} and W. R. Franklin and M. V. A. Andrade},
  booktitle = {Lecture Notes in Computational Science and Engineering},
  date      = {2019},
  title     = {Exact Fast Parallel Intersection of Large 3-D Triangular Meshes},
  doi       = {10.1007/978-3-030-13992-6_20},
  edition   = {1},
  pages     = {365-383},
  publisher = {Springer International Publishing},
}

@Misc{Lafarge2020,
  author = {T. Lafarge and B. Pateiro-Lopez},
  title  = {{CRAN R Project Package "alphashape3d"}},
  url    = {http://cran.r-project.org/web/packages/alphashape3d/index.html},
  year   = {2020},
  note = {(last accessed May 19, 2022)},
}

@Electronic{McCoid2021,
  author = {C. McCoid and M. J. Gander},
  date   = {2021},
  title  = {{Intersection of tetrahedra}},
  url    = {http://www.unige.ch/~mccoid/ongoing/mccoid2020tetrahedra.pdf},
}

@Article{Ganovelli2002,
  author       = {Fabio Ganovelli and Federico Ponchio and Claudio Rocchini},
  date         = {2002-01},
  journaltitle = {J. Graph. Tools},
  title        = {Fast Tetrahedron-Tetrahedron Overlap Algorithm},
  doi          = {10.1080/10867651.2002.10487557},
  number       = {2},
  pages        = {17--25},
  volume       = {7},
  publisher    = {Informa {UK} Limited},
}

@Article{Kuehbach2021MM,
  author       = {M. K\"uhbach and A. J. London and J. Wang and D. K. Schreiber and F. Mendez-Martin and I. Ghamarian and H. Bilal and A. V. Ceguerra},
  date         = {2021},
  journaltitle = {Microsc. Microanal.},
  title        = {Community-Driven Methods for Open and Reproducible Software Tools for Analyzing Datasets from Atom Probe Microscopy},
  doi          = {10.1017/s1431927621012241},
  pages        = {1-16},
}

@Article{Baerentzen2005,
  author       = {J.A. Baerentzen and H. Aanaes},
  date         = {2005-05},
  journaltitle = {{IEEE} Transactions on Visualization and Computer Graphics},
  title        = {Signed Distance Computation Using the Angle Weighted Pseudonormal},
  doi          = {10.1109/tvcg.2005.49},
  number       = {3},
  pages        = {243--253},
  volume       = {11},
  publisher    = {Institute of Electrical and Electronics Engineers ({IEEE})},
}

@Book{CGAL2021,
  author    = {The {CGAL Project}},
  title     = {{CGAL} User and Reference Manual},
  edition   = {5.3},
  publisher = {CGAL Editorial Board},
  url       = {http://doc.cgal.org/5.3/Manual/packages.html},
  year      = {2021},
}

@InCollection{Hachenberger2021,
author =  {P. Hachenberger and L. Kettner},
title =   {{3D} Boolean Operations on Nef Polyhedra},
publisher =  {CGAL Editorial Board},
edition =    {5.3},
booktitle =   {{CGAL} User and Reference Manual},
url = {https://doc.cgal.org/5.3/Manual/packages.html#PkgNef3},
year =   {2021},
}

@InCollection{Kai2021,
author =  {T. K. F. Da and S. Loriot and M. Yvinec},
title =   {{3D} Alpha Shapes},
publisher =  {CGAL Editorial Board},
edition =     {5.3},
booktitle =   {{CGAL} User and Reference Manual},
url = {http://doc.cgal.org/5.3/Manual/packages.html#PkgAlphaShapes3},
year = {2021},
}

@InCollection{Loriot2021,
author =  {S. Loriot and M. Rouxel-Labb{\'{e}} and J. Tournois and I. O. Yaz},
title =   {Polygon Mesh Processing},
publisher =  {CGAL Editorial Board},
edition =     {5.3},
booktitle =   {{CGAL} User and Reference Manual},
url = {http://doc.cgal.org/5.3/Manual/packages.html#PkgPolygonMeshProcessing},
year =      {2021},
}

@InCollection{Katrioplas2021,
author =  {K. Katrioplas and M. Rouxel-Labb\'{e}},
title =   {Optimal Bounding Box},
publisher =  {CGAL Editorial Board},
edition =    {5.3},
booktitle =   {{CGAL} User and Reference Manual},
url = {http://doc.cgal.org/5.3/Manual/packages.html#PkgOptimalBoundingBox},
year =        {2021},
}

@InCollection{Fischer2021,
author =  {K. Fischer and B. G\"artner and T. Herrmann and M. Hoffmann and S. Sch{\"o}nherr},
  title = {Bounding Volumes},
  publisher = {CGAL Editorial Board},
  edition = {5.3},
  booktitle = {{CGAL} User and Reference Manual},
  url = {http://doc.cgal.org/5.3/Manual/packages.html#PkgBoundingVolumes},
  year = {2021},
}

@Article{Zepedaruiz2020,
  author       = {L. A. Zepeda-Ruiz and A. Stukowski and T. Oppelstrup and N. Bertin and N. R. Barton and R. Freitas and V. V. Bulatov},
  journaltitle = {Nat. Mater.},
  title        = {Atomistic insights into metal hardening},
  doi          = {10.1038/s41563-020-00815-1},
  number       = {3},
  pages        = {315-320},
  volume       = {20},
}

@Article{Huber2018,
year = {2018},
title = {{A machine learning approach to model solute grain boundary segregation}},
author = {L. Huber and R. Hadian and B. Grabowski and J. Neugebauer},
journal = {npj Comput. Mater.},
volume = {4},
pages = {{64}},
doi = {10.1038/s41524-018-0122-7},
}

@Article{Stukowski2010,
  author       = {A. Stukowski},
  journaltitle = {Model. Simul. Mater. Sci. Eng.},
  title        = {Visualization and analysis of atomistic simulation data with {OVITO}{\textendash}the Open Visualization Tool},
  doi          = {10.1088/0965-0393/18/1/015012},
  number       = {1},
  pages        = {{015012}},
  volume       = {18},
  year         = {2010},
}

@Article{Gault2021Nat,
  author       = {B. Gault and A. Chiaramonti and O. Cojocaru-Mir\'{e}din and P. Stender and R. Dubosq and C. Freysoldt and S. K. Makineni and T. Li and M. Moody and J. M. Cairney},
  journaltitle = {Nat. Rev. Methods Primers},
  title        = {Atom probe tomography},
  doi          = {10.1038/s43586-021-00054-x},
  number       = {1},
  volume       = {1},
  year         = {2021},
}

@Article{Gault2021,
title = {{Reflections on the Spatial Performance of Atom Probe Tomography in the Analysis of Atomic Neighborhoods}},
author = {B. Gault and B. Klaes and F. Morgado and C. Freysoldt and Y. Li and F. {De Geuser} and L. T. Stephenson and F. Vurpillot},
  year        = {2021},
  journal = {Microsc. Microanal.},
  pages = {1-11},
  doi = {10.1017/S1431927621012952},
}

@Article{Zhou2022,
  author      = {X. Zhou and Y. Wei and M. K\"uhbach and H. Zhao and F. Vogel and R. D. Kamachali and G. B. Thompson and D. Raabe and B. Gault},
  title       = {Revealing in-plane grain boundary composition features through machine learning from atom probe tomography data},
  year        = {2022},
  volume = {226},
  pages = {{117633}},
  journal = {Acta Mat.},
doi = {10.1016/j.actamat.2022.117633},
}

@Article{Stoichkov2021,
journal = {Scr. Mater.},
volume = {194},
year = {2021},
pages = {{113672}},
title = {{Nucleation and growth of $\alpha$ phase in a metastable $\beta$-Titanium Ti-5Al-5Mo-5V-3Cr alloy: Influence from the nano-scale, ordered-orthorhombic $O^{\prime\prime}$ phase and $\alpha$ compositional evolution}},
author = {S. Antonov and R. Shi and D. Li and Z. Kloenne and Y. Zheng and H. L. Fraser and D. Raabe and B. Gault},
doi = {10.1016/j.scriptamat.2020.113672},
}

@Article{Karnesky2007a,
  author       = {R. A. Karnesky and C. K. Sudbrack and D. N. Seidman},
  journaltitle = {Scr. Mater.},
  title        = {Best-fit ellipsoids of atom-probe tomographic data to study coalescence of {$\gamma\prime$} (L{$_{12}$}) precipitates in Ni{\textendash}Al{\textendash}Cr},
  doi          = {10.1016/j.scriptamat.2007.04.020},
  number       = {4},
  pages        = {353-356},
  volume       = {57},
}

@Article{Theska2019,
  author       = {F. Theska and S. P. Ringer and S. Primig},
  journaltitle = {Microsc. Microanal.},
  title        = {Atom Probe Microscopy of Strengthening Effects in Alloy 718},
  doi          = {10.1017/s1431927618015611},
  number       = {2},
  pages        = {470-480},
  volume       = {25},
  year         = {2019},
}

@Article{Jenkins2020,
  author       = {B. M. Jenkins and A. J. London and N. Riddle and J. M. Hyde and P. A.J. Bagot and M. P. Moody},
  journaltitle = {Mat. Characterization},
  title        = {Using alpha hulls to automatically and reproducibly detect edge clusters in atom probe tomography datasets},
  doi          = {10.1016/j.matchar.2019.110078},
  pages        = {110078},
  volume       = {160},
  year         = {2020},
}

@Article{Rielli2021,
  author       = {V. V. Rielli and F. Theska and F. Godor and A. Stanojevic and B. Oberwinkler and S. Primig},
  journaltitle = {Mat. {\&} Design},
  title        = {Evolution of nanoscale precipitates during common Alloy 718 ageing treatments},
  doi          = {10.1016/j.matdes.2021.109762},
  pages        = {{109762}},
  volume       = {205},
}

@Book{Eberly2020,
  author    = {D. H. Eberly},
  title     = {{Robust and Error-Free Geometric Computing}},
  edition   = {1},
  publisher = {CRC Press},
  year      = {2020},
}

@InProceedings{Gottschalk1996,
  author    = {S. Gottschalk and M. C. Lin and D. Manocha},
  booktitle = {Proceedings of the 23rd annual conference on Computer graphics and interactive techniques - {SIGGRAPH} {\textquotesingle}96},
  date      = {1996},
  title     = {{OBBTree: a hierarchical structure for rapid interference detection}},
  doi       = {10.1145/237170.237244},
}

@InProceedings{Polin2021,
 title = {{Atomic‐scale Insights to Design of High‐Performing Hard Magnets Gained by Atom Probe Tomography}},
 author = {N. Polin and S. H. Kim and A. saxena and S. Giron and E. Adabifiroozjaei and A. Gabay and G. Hadjipanayis and D. Raabe and O. Gutfleisch and B. Gault},
 date = {2021},
 booktitle = {Proceedings of the Atom Probe Tomography \& Microscopy Conference 2021 (Virtual), September 27-30, Portland, U.S.},
}

@InProceedings{Saxena2021,
title = {A materials informatics framework to discover patterns in atom probe tomography data},
author = {A. Saxena and N. Polin and D. Raabe and B. Gault and C. Freysoldt},
 date = {2021},
 booktitle = {Proceedings of the Atom Probe Tomography \& Microscopy Conference 2021 (Virtual), September 27-30, Portland, U.S.},
}

@InProceedings{Heller2021,
title = {An open-source software toolbox for FAIR atom probe data analysis},
author = {M. Heller and B. Ott and V. Dalbauer and T. Fuchs and E. G\"obel and C. Macauley and P. Felfer},
year = {2021},
booktitle = {Proceedings of the Atom Probe Tomography \& Microscopy Conference 2021 (Virtual), September 27-30, Portland, U.S.},
}

@article{Kelly2021,
title={Project Tomo: Toward Atomic-scale Analytical Tomography},
volume={26}, 
doi={10.1017/S1431927620022205}, 
number={S2}, 
journal={Microsc. Microanal.}, 
author={T. Kelly and R. Dunin-Borkowski and J. Meyer}, 
year={2020}, 
pages={2618–2621}
}

@Misc{Larsen2020,
  author = {E. Larsen and S. Gottschalk},
  title  = {GammaUNC/PQP library},
  url    = {http://github.com/GammaUNC/PQP},
  note = {(last accessed May 19, 2022)},
}

@InProceedings{Larsen2000,
  author    = {E. Larsen and S. Gottschalk and M.C. Lin and D. Manocha},
  booktitle = {Proceedings 2000 {ICRA}. Millennium Conference. {IEEE} International Conference on Robotics and Automation. Symposia Proceedings (Cat. No.00CH37065)},
  title     = {Fast distance queries with rectangular swept sphere volumes},
  doi       = {10.1109/robot.2000.845311},
  year      = {2000},
}

@Misc{Eberly2010,
url = {http://www.geometrictools.com/Documentation/IntersectionTriangleCylinder.pdf},
author = {D. Eberly},
title = {Intersection of a Triangle and a Cylinder},
year = {2010},
note = {(last accessed May 19, 2022)},
}

@Electronic{Eigen2021,
  author = {G. Guennebaud and B. Jacob and {{} {others}}},
  title = {Eigen v3},
  url = {http://eigen.tuxfamily.org},
  year = {2010},
}

@Book{Schling2011,
author = {B. Schling},
title = {The Boost C++ Libraries},
year = {2011},
edition = {1},
isbn = {0982219199},
doi = {10.5555/2049814},
publisher = {XML Press},
}

@article{Koennecke2015,
author = {M. K\"onnecke and F. A. Akeroyd and H. J. Bernstein and A. S. Brewster and S. I. Campbell and B. Clausen and S. Cottrell and J. U. Hoffmann and P. R. Jemian and D. M\"annicke and R. Osborn and P. F. Peterson and T. Richter and J. Suzuki and B. Watts and E. Wintersberger and J. Wuttke},
doi = {10.1107/S1600576714027575},
journal = {J. Appl. Crystallogr.},
number = {1},
pages = {301--305},
title = {{The NeXus data format}},
volume = {48},
year = {2015},
}

@incollection{cgal:lty-pmp-22a,
  author = {S{\'e}bastien Loriot and Mael Rouxel-Labb{\'e} and Jane Tournois and Ilker O. Yaz},
  title = {Polygon Mesh Processing},
  publisher = {{CGAL Editorial Board}},
  edition = {{5.4}},
  booktitle = {{CGAL} User and Reference Manual},
  url = {https://doc.cgal.org/5.4/Manual/packages.html#PkgPolygonMeshProcessing},
  year = 2022
}

@Conference{Kluyver2016,
Title = {Jupyter Notebooks -- a publishing format for reproducible computational workflows},
Author = {T. Kluyver and B. Ragan-Kelley and F. P{\'e}rez and B. Granger and M. Bussonnier and J. Frederic and K. Kelley and J. Hamrick and J. Grout and S. Corlay and P. Ivanov and D. Avila and S. Abdalla and C. Willing},
booktitle = {Positioning and Power in Academic Publishing: Players, Agents and Agendas},
Editor = {F. Loizides and B. Schmidt},
Organization = {IOS Press},
Pages = {87-90},
Year = {2016},
doi = {10.3233/978-1-61499-649-1-87},
}

@incollection{cgal:ap-pcad-22a,
  author = {Pierre Alliez and Sylvain Pion and Ankit Gupta},
  title = {Principal Component Analysis},
  publisher = {{CGAL Editorial Board}},
  edition = {{5.4}},
  booktitle = {{CGAL} User and Reference Manual},
  url = {https://doc.cgal.org/5.4/Manual/packages.html#PkgPrincipalComponentAnalysisD},
  year = 2022
}
\fussy















\end{document}